\documentclass{aa}  
\def \eg {e.g.}

\def \ie {i.e.}
\def \cf {cf.}
\def \lcdm {{\hbox{$\Lambda$CDM}}}
\def \omegam {{\hbox{$\Omega_{\rm m}$}}}
\def \omegal {{\hbox{$\Omega_\Lambda$}}}
\def \hzero {{\hbox{$H_0$}}}
\def \arcmin {\hbox{$^\prime$}}
\def \arcsec {\hbox{$^{\prime\prime}$}}
\def \deg {\hbox{$^\circ$}}
\def \nh {\hbox{$N_{\rm H}$}}

\def \mach {{\hbox{$\mathcal{M}$}}}

\def \msun {\hbox{${\rm M_\odot}$}}

\def \mfive {\hbox{$M_{500}$}}

\def \rfive {\hbox{$r_{500}$}}

\newcommand{\ergs }{\mbox{erg s$^{-1}$}}

\newcommand{\kmsmpc }{\mbox{km s$^{-1}$ Mpc$^{-1}$}}

\newcommand{\kev }{\mbox{keV}}

\newcommand{\mujyb }{\mbox{$\mu$Jy beam$^{-1}$}}

\newcommand{\muG }{\mbox{$\mu$G}}
\newcommand{\whz }{\mbox{W Hz$^{-1}$}}
\newcommand{\epic }{EPIC}
\newcommand{\epicE }{European Photon Imaging Camera}
\newcommand{\obsid }{ObsID}

\newcommand{\uv }{\textit{uv}}

\newcommand{\wsclean }{\textsc{WSClean}}
\newcommand{\casa }{\textsc{casa}}

\newcommand{\caracal }{\textsc{caracal}}

\newcommand{\oxkat }{\textsc{oxkat}}

\newcommand{\xspec }{\textsc{xspec}}

\newcommand{\esas }{\textsc{esas}}
\newcommand{\esasE }{Extended Source Analysis Software}
\newcommand{\sas }{\textsc{sas}}
\newcommand{\sasE }{Scientific Analysis System}

\newcommand{\aoflagger }{\textsc{AOFlagger}}
\newcommand{\dysco }{\textsc{Dysco}}
\newcommand{\dppp }{DP3}
\newcommand{\dpppE }{Default PreProcessing Pipeline}

\newcommand{\xmm }{{\em XMM-Newton}}
\newcommand{\chandra }{{\em Chandra}}

\newcommand{\asca }{{\em ASCA}}

\newcommand{\rosat }{{\em ROSAT}}

\newcommand{\gmrt }{GMRT}

\newcommand{\vla }{VLA}

\newcommand{\lofar }{LOFAR}
\newcommand{\lofarE }{LOw Frequency ARray}
\newcommand{\ska }{SKA}
\newcommand{\skaE }{Square Kilometer Array}

\newcommand{\meerkat }{MeerKAT}

\newcommand{\lotss }{LoTSS}

\usepackage{placeins}
\usepackage{graphicx}
\usepackage{amssymb}
\usepackage{multirow,bigdelim}
\usepackage{txfonts}
\usepackage[export]{adjustbox}
\usepackage{hyperref}
\hypersetup{colorlinks,citecolor=blue,filecolor=black,linkcolor=red,urlcolor=black}
\begin{document} 

%\title{The MeerKAT view of Abell 754}
\title{The prototypical major cluster merger Abell 754}
%\subtitle{I. Calibration of MeerKAT data and radio/X-ray spectral study of the halo}
\subtitle{I. Calibration of MeerKAT data and radio/X-ray spectral mapping of the cluster}

\authorrunning{A. Botteon et al.} 
\titlerunning{The prototypical major cluster merger Abell 754. I.}

\author{A. Botteon\inst{\ref{ira}}, R.~J. van Weeren\inst{\ref{leiden}}, D. Eckert\inst{\ref{geneva}}, F. Gastaldello\inst{\ref{iasf}}, M. Markevitch\inst{\ref{goddard}}, S. Giacintucci\inst{\ref{nrl}}, G. Brunetti\inst{\ref{ira}}, R. Kale\inst{\ref{tiff}}, and T. Venturi\inst{\ref{ira}}} 

\institute{
INAF - IRA, via P.~Gobetti 101, I-40129 Bologna, Italy \label{ira} \\
\email{andrea.botteon@inaf.it} 
\and
Leiden Observatory, Leiden University, PO Box 9513, NL-2300 RA Leiden, The Netherlands \label{leiden}
\and
Department of Astronomy, University of Geneva, ch. d'Ecogia 16, 1290 Versoix, Switzerland \label{geneva}
\and
INAF - IASF Milano, via A.~Corti 12, I-20133 Milano, Italy \label{iasf}
\and
NASA/Goddard Space Flight Center, Greenbelt, MD 20771, USA \label{goddard}
\and
U.S. Naval Research Laboratory, 4555 Overlook Avenue SW, Code 7213, Washington, DC 20375, USA \label{nrl}
\and
National Centre for Radio Astrophysics, Tata Institute of Fundamental Research, Pune University, Pune 411007, India \label{tiff}
% \and
% Dipartimento di Fisica e Astronomia, Universit\`{a} di Bologna, via P.~Gobetti 93/2, I-40129 Bologna, Italy \label{unibo}
% \and
% ASTRON, the Netherlands Institute for Radio Astronomy, Postbus 2, NL-7990 AA Dwingeloo, The Netherlands \label{astron}
% \and
% Hamburger Sternwarte, Universit\"{a}t Hamburg, Gojenbergsweg 112, D-21029 Hamburg, Germany \label{hamburg}
% \and
% INAF - Osservatorio Astronomico di Brera, via E.~Bianchi 46, I-23807 Merate, Italy \label{brera}
% \and
% INAF - Osservatorio di Astrofisica e Scienza dello Spazio, via P.~Gobetti 93/3, I-40129, Bologna, Italy \label{oas}
% \and
% INFN, Sezione di Bologna, viale Berti Pichat 6/2, I-40127, Bologna, Italy \label{infn}
%\and
%Th\"{u}ringer Landessternwarte, Sternwarte 5, D-07778 Tautenburg, Germany
%\and
%Mbarara University of Science \& Technology, PO Box 1410 Mbarara, Uganda 
%\and
%Anton Pannekoek Institute for Astronomy, University of Amsterdam, Postbus 94249, NL-1090 GE Amsterdam, The Netherlands
%\and
%European Southern Observatory, Karl-Schwarzschild-Str. 2, D-85748 Garching, Germany 
%\and
%Department of Physical Sciences, The Open University, Milton Keynes MK7 6AA, England \\
%\and
%Space Science and Technology Department, The Rutherford Appleton Laboratory, Chilton, Didcot, Oxfordshire OX11 0NL 
}

\date{Received XXX; accepted YYY}

\abstract
{Abell 754 is a rich galaxy cluster at $z=0.0543$ and is considered the prototype of a major cluster merger. Like many dynamically unrelaxed systems, it hosts diffuse radio emission on Mpc-scales. Extended synchrotron sources in the intra-cluster medium (ICM) are commonly interpreted as evidence that a fraction of the gravitational energy released during cluster mergers is dissipated into nonthermal components.}
{Here, we use new \meerkat\ UHF- and L-band observations to study nonthermal phenomena in Abell 754. These data are complemented with archival \xmm\ observations to investigate the resolved spectral properties of both the radio and X-ray cluster emission.}
{For the first time, we employed the pipeline originally developed to calibrate \lofar\ data to \meerkat\ observations. This allowed us to perform a direction-dependent calibration and obtain highly sensitive radio images in UHF- and L-bands which capture the extended emission with unprecedented detail. By using a large \xmm\ mosaic, we produced thermodynamic maps of the ICM.}
{Our analysis reveals that the radio halo in the cluster center is bounded by the well-known shock in the eastern direction. Furthermore, in the southwest periphery, we discover an extended radio source that we classify as a radio relic which is possibly tracing a shock driven by the squeezed gas compressed by the merger, outflowing in perpendicular directions. The low-luminosity of this relic appears compatible with direct acceleration of thermal pool electrons. We interpret the observed radio and X-ray features in the context of a major cluster merger with a nonzero impact parameter.}
{Abell 754 is a remarkable galaxy cluster showcasing exceptional features associated with the ongoing merger event. The high quality of the new \meerkat\ data motivates further work on this system.}


%\keywords{acceleration of particles -- radiation mechanisms: non-thermal -- radiation mechanisms: thermal -- galaxies: clusters: intracluster medium -- galaxies: clusters: general -- shock waves}

\maketitle
%
%-------------------------------------------------------------------

\section{Introduction}

\begin{table}[t]
 \centering
 \caption{Properties of A754 derived from the literature.}
 \label{tab:a754_summary}
  \begin{tabular}{lr} 
  \hline
  \hline
  $z$ & 0.0543 \\
  Right ascension (h, m, s) & 09 09 08 \\
  Declination (\deg, \arcmin, \arcsec) & $-$09 39 58 \\
  \mfive\ ($10^{14}$ \msun) & $6.85^{+0.12}_{-0.13}$  \\
  $Y_{500}$ ($10^{-3}$ arcmin$^2$) & $40.2\pm2.7$ \\ 
  \rfive\ (kpc) & $1322\pm8$ \\
  $L_{\rm X}$ ($10^{44}$ \ergs) & $5.56\pm0.15$ \\  
  $K_0$ (keV cm$^2$) & $270\pm70$ \\
  $kT_{\rm vir}$ (keV) & $11.1\pm0.4$ \\
  $\sigma_{\rm v}^{\rm turb}$ (km s$^{-1}$) & $676\pm46$ \\
  $\sigma_{\rm v}^{\rm gal}$ (km s$^{-1}$) & $953\pm40$ \\
  \hline
  \end{tabular}
 \tablefoot{Reported quantities are: redshift \citep[$z$;][]{smith04}, equatorial coordinates, mass and integrated Comptonization parameter within the radius inside which the mean mass density is 500 times the critical density at the cluster redshift \citep[\mfive, $Y_{500}$, and \rfive;][]{planck16xxvii}, X-ray luminosity in the 0.1--2.4 \kev\ band \citep[$L_{\rm X}$;][]{chen07}, central entropy \citep[$K_0$;][]{cavagnolo09}, virial temperature \citep[$kT_{\rm vir}$;][]{hudson10}, ICM turbulent velocity dispersion \citep[$\sigma_{\rm v}^{\rm turb}$;][]{eckert17turbulent}, and galaxy velocity dispersion \citep[$\sigma_{\rm v}^{\rm gal}$;][]{christlein03}.}
\end{table}

Mergers between galaxy clusters are among the most energetic events in the Universe \citep[\eg][]{sarazin02rev}. During these collisions, shocks and turbulence are injected into the intra-cluster medium (ICM) and often generate cluster-wide synchrotron emission with steep spectrum \citep[\eg][]{feretti12rev, vanweeren19rev}. Diffuse radio emission in clusters probes a complex hierarchy of novel mechanisms in the ICM that are essentially able to dissipate gravitational energy into relativistic particles and magnetic fields on Mpc-scale \citep[see][for a review]{brunetti14rev}. Exploring such a chain of mechanisms has a fundamental impact on our understanding of the microphysics of the ICM and on the evolution of clusters themselves. The study of nonthermal phenomena in merging galaxy clusters is indeed one of the major scientific drivers of many current and future radio interferometers and X-ray microcalorimeters. \\ 
\indent
Extended synchrotron emission in the ICM is nowadays observed in more than 100 merging clusters of galaxies, and it is broadly classified into radio halos and relics \citep[\eg][]{vanweeren19rev}. It is currently thought that radio halos trace turbulent regions where relativistic particles are trapped and reaccelerated through scattering with turbulence \citep[\eg][]{brunetti01coma, petrosian01, brunetti07turbulence, brunetti16stochastic, miniati15run, nishiwaki22}. Instead, radio relics originate as a consequence of the particle acceleration and magnetic field amplification ongoing at merger shocks located in cluster outskirts \citep[\eg][]{ensslin98relics, roettiger99a3667, kang12relics, kang20}. \\
\indent
In recent years, the advent of the new generation wide-band interferometers - pathfinders and precursors of the \skaE\ (\ska) - has brought a major advance in the discovery of these objects and the characterization of their properties. One of the instruments that is contributing to this advance is \meerkat\ \citep{jonas09}, which enables deep, broadband, wide-field, (sub-)10\arcsec\ resolution continuum and polarimetric observations. Observations carried out with \meerkat\ allow to recover the extended diffuse radio emission with large angular scale of nearby clusters, as well as numerous background sources along the cluster line of sight that can be used as Faraday rotation probes to study the weak ICM magnetic field. Recent studies based on \meerkat\ observations have been focused on the analysis of samples of clusters \citep{knowles21, knowles22, kale22}, radio halos \citep{venturi22, sikhosana23, sikhosana24arx, botteon23}, relics \citep{parekh20meerkat, parekh22, degasperin22, chibueze23, koribalski24double}, mini-halos \citep{riseley22ms1455, riseley23, riseley24, trehaeven23}, and interaction between ICM and radio galaxies \citep{ramatsoku20eso137, chibueze21, rudnick22, giacintucci22, velovic23}. These studies showcase the versatile capabilities of \meerkat\ in investigating nonthermal phenomena in galaxy clusters. \\
\indent
Recent years have also faced important leaps forward in the calibration and imaging techniques of radio interferometric data to account for the wide-band and large field-of-view (FoV) of the new instruments. These developments are crucial to achieve deep images with high dynamic range, which are essential for studying the diffuse and faint radio emission from clusters. As a result, \meerkat\ data reduction can now benefit of semi-automated calibration pipelines, such as \oxkat\ \citep{heywood20} and \caracal\ \citep{jozsa20}, while additional optimization steps, such as the correction for direction-dependent effects, can be performed on user demand to improve the calibration quality even further \citep[\eg][]{parekh21calibration, riseley22ms1455, trehaeven23}. Prior to \meerkat, significant development to correct direction-dependent effects has been carried out for \lofarE\ \citep[\lofar;][]{vanhaarlem13} which, operating at frequencies below 200 MHz, is more affected by ionospheric distortions \citep[\eg][]{vanweeren16calibration, vanweeren21, williams16, tasse18}. \\ 
\indent
Abell 754 (hereafter A754) is a nearby, rich, hot, and massive galaxy cluster in the stage of a violent merger, see Tab.~\ref{tab:a754_summary} for a summary of its main properties. It has been actively studied in the optical and X-ray bands, and is considered the prototype of a major cluster merger. Previous investigations have unveiled its complex galaxy distribution \citep{fabricant86, zabludoff95, godlowski98, okabe08}, X-ray morphology, and gas temperature structure \citep{henry95, henriksen96a754, markevitch03, henry04}. These findings suggest a main collision along an east-west axis, probably with a nonzero impact parameter \citep[\eg][]{roettiger98}. Remarkably, A754 is one of the few galaxy clusters where a candidate shock front, appearing as a clear X-ray surface brightness discontinuity a few arcminutes east of the cluster core, was reported with \rosat\ \citep{krivonos03}. The shock nature of this edge was later confirmed using the \chandra\ measurement of the temperature jump across the discontinuity, corresponding to a weak shock front with Mach number $\mach=1.57^{+0.16}_{-0.12}$ \citep{macario11}. Signs of nonequilibrium ionization plasma in the ICM due to the shock heating during the merger process have also been reported in the system \citep{inoue16}. Early radio observations of A754 provided evidence for ultrarelativistic electrons and magnetic fields within the ICM, making this cluster one of the first known systems hosting extended synchrotron emission \citep{wielebinski77, mills78, harris80a754}. In particular, low-frequency \vla\ observations at 74 and 330 MHz with the \vla\ C-array confirmed the existence of a radio halo and suggested the presence of possible radio relics east and west of the radio halo \citep{kassim01}. Only the east relic was later confirmed with 1.4 GHz \vla\ D-array and \gmrt\ observations \citep{bacchi03, kale09, macario11}. The X-ray detected shock front coincides with the position of the radio relic, which at low frequency appears connected to the centrally located radio halo \citep{kale09, macario11}. It is worth noting that these studies were typically conducted with low-resolution ($\gtrsim$60 arcsec) radio imaging, which was necessary to recover the extended diffuse emission using narrow-band data. \\
\indent
A754 is the closest galaxy cluster after Coma that hosts diffuse nonthermal sources in the ICM and a well-characterized merger shock \citep[see \eg][for recent work]{bonafede21, bonafede22, churazov21, churazov23coma}. Its redshift of $z = 0.0543$ provides a good compromise to perform spatially resolved studies while still allowing to cover its outskirts with a modest number of pointings with different facilities (for reference, $\rfive \simeq 21.2$ arcmin). For these reasons, A754 is an attractive laboratory to investigate in great detail the physical processes leading to the dissipation of kinetic energy during cluster mergers over a wide range of scales. We have recently targeted A754 with \meerkat\ UHF (544--1088 MHz) and L (856--1712 MHz) band observations, and this paper represents the first in a series dedicated to exploiting these data. In the first part, we focus on the new strategy used to calibrate the \meerkat\ data, employing the pipeline originally developed to calibrate low-frequency \lofar\ observations. In the second part, we present and discuss our scientific results, complementing the new \meerkat\ data with archival \xmm\ observations. \\
\indent
Here, we adopt a \lcdm\ cosmology with $\omegal = 0.7$, $\omegam = 0.3$ and $\hzero = 70$ \kmsmpc, in which 1 arcsec corresponds to 1.056 kpc at the cluster redshift and the luminosity distance is $D_{\rm L} = 242.2$ Mpc. We adopt the convention $S_\nu \propto \nu^{-\alpha}$ for radio synchrotron spectrum, where $S_\nu$ is the flux density at frequency $\nu$ and $\alpha$ is the spectral index.

\section{MeerKAT data reduction}

\begin{figure*}
 \centering
 \includegraphics[width=\hsize,trim={0.2cm 0.2cm 0.2cm 0.2cm},clip]{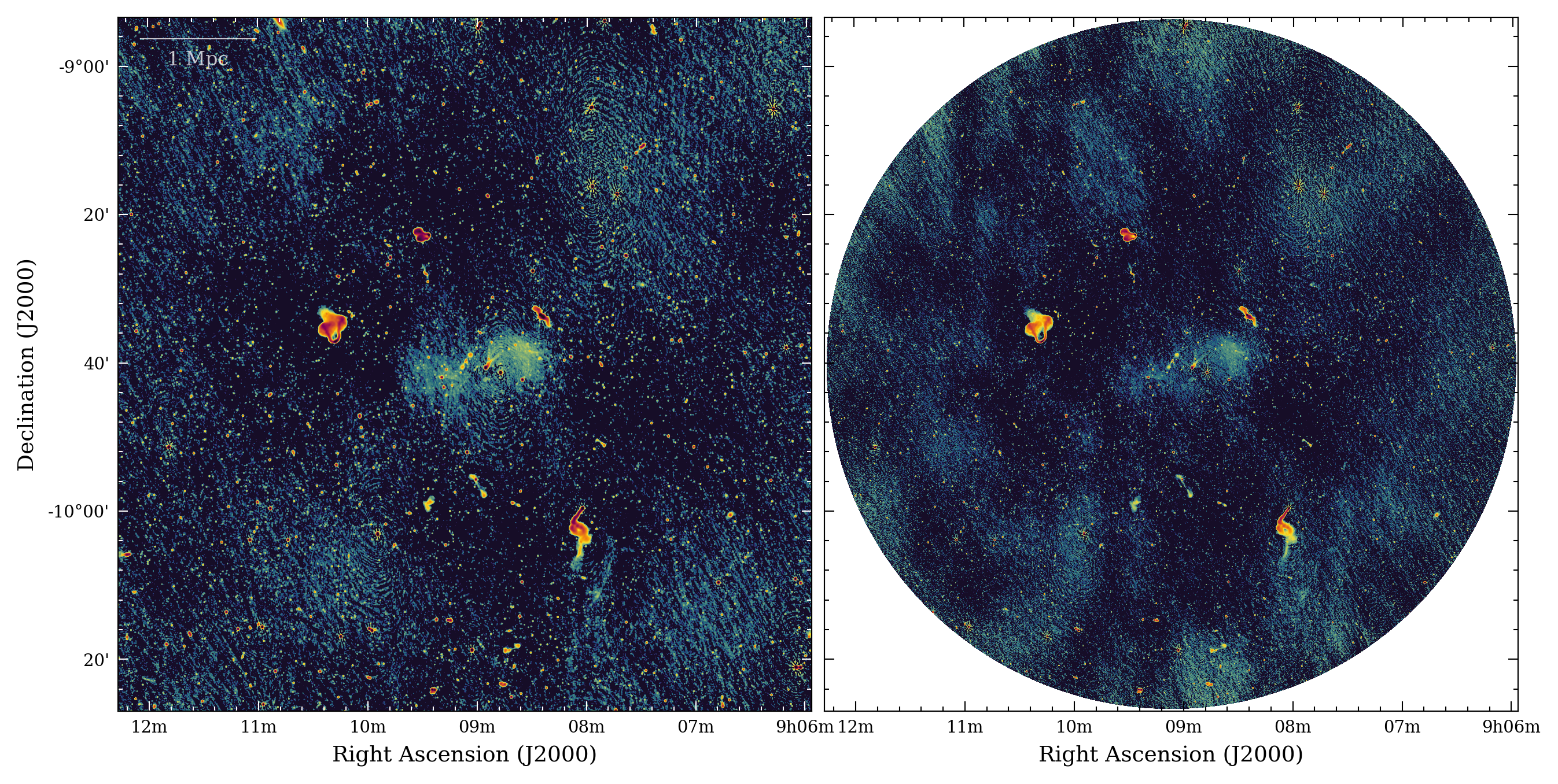}
 \caption{The primary beam corrected images produced by the SDP Continuum pipeline for the rising tracks in UHF (\textit{left}) and L (\textit{right}) band.}
 \label{fig:sdp_images}
\end{figure*}

To process the \meerkat\ observations we use the method of \citet{vanweeren21}, employing the \texttt{facetselfcal.py}\footnote{\url{https://github.com/rvweeren/lofar_facet_selfcal}} pipeline. This calibration pipeline mainly uses the \dpppE\ \citep[\dppp;][]{vandiepen18}, \wsclean\ \citep{offringa14}, and the \texttt{losoto}\footnote{\url{https://github.com/revoltek/losoto}} \citep{degasperin19}, \texttt{ddf-pipeline}\footnote{\url{https://github.com/mhardcastle/ddf-pipeline}} \citep{shimwell19, tasse21}, and \texttt{python-casacore}\footnote{\url{https://github.com/casacore/python-casacore}} \citep{casacore19} packages.

\subsection{Observation setup}

Our observing program on A754 (Proposal ID: SCI-20220822-AB-01) made use of \meerkat\ UHF- and L-band receivers for a total time of 20~hours. The 10~hours observation in each band was split into 5~hour-observing runs, two ``rising`` tracks on 2023 January 6 and 7 (Capture Block IDs: 1673038580 and 1673124676) and two ``setting'' tracks on 2023 March 25 and 26 (Capture Block IDs: 1679769376 and 1679855479), to maximize the \uv-coverage on the target. Data were recorded using 4096 channels covering the frequency ranges 544--1088 MHz (UHF band) and 856--1712 MHz (L band) and adopting an integration time of 8~s. \\
\indent
For each observing run, two 10~min scans separated roughly by 3~hours were performed on the bandpass calibrator PKS B0407--658. The target A754 was observed with 30~min scans, which were bookended by 2~min observations on the compact source 3C237 to track the time-varying instrumental gains. Additionally, two 5~min scans on the polarization calibrators 3C138 (for the rising tracks) and 3C286 (for the setting tracks) were included. The final on-source time on A754 amounts to 8~hours in each band.

\subsection{Download of preliminary calibrated data}

\meerkat\ observations are stored in the archive\footnote{\url{https://archive.sarao.ac.za}} in a data format known as \meerkat\ Visibility Format (MVF). We converted the data into \casa\ \citep{mcmullin07, casa22} Measurement Set (MS) format by using the online tool available in the \meerkat\ archive, which makes use of the \texttt{mvftoms.py} script of the \texttt{katdal}\footnote{\url{https://github.com/ska-sa/katdal}} package. In particular, we used the ``Default Calibrated'' option of the online tool which applies a first round of conservative flags produced by the ingest process and all the calibration solutions (including delay, bandpass, and gain) found by the SARAO Science Data Processor\footnote{\url{https://skaafrica.atlassian.net/wiki/spaces/ESDKB/pages/338723406/SDP+pipelines+overview}} (SDP) Calibration pipeline, which sets the flux density scale according to the model of PKS B0407--658 by \citet{hugo21}. By default, this option keeps channels in the range 163--3885. The resulting full polarization MS file we downloaded was $\sim$1.2 TB in size per observing run. For a quick data quality assessment, we also downloaded the images produced by the SDP Continuum pipeline that were available in the archive. In Fig~\ref{fig:sdp_images} we show the UHF- and L-band images produced by the pipeline for the two rising tracks. The quality of the images for the setting tracks is comparable.

\subsection{Data preparation, averaging, and RFI removal}

\begin{figure*}
 \centering
 \begin{minipage}[b]{0.65\linewidth}
 \includegraphics[width=.49\hsize,trim={0cm 0cm 0cm 0cm},clip]{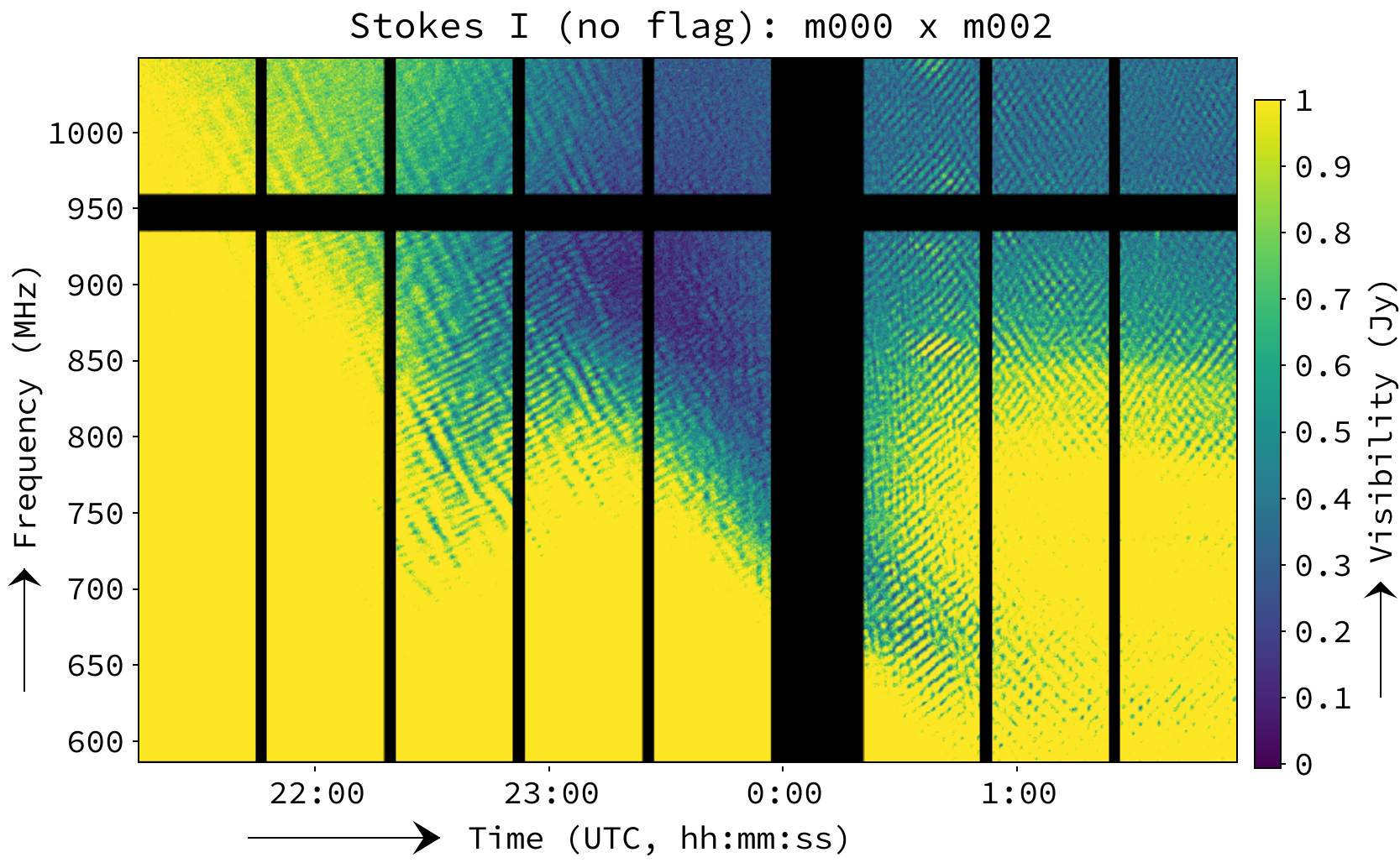} \hspace{0.4cm}
 \includegraphics[width=.49\hsize,trim={0cm 0cm 0cm 0cm},clip]{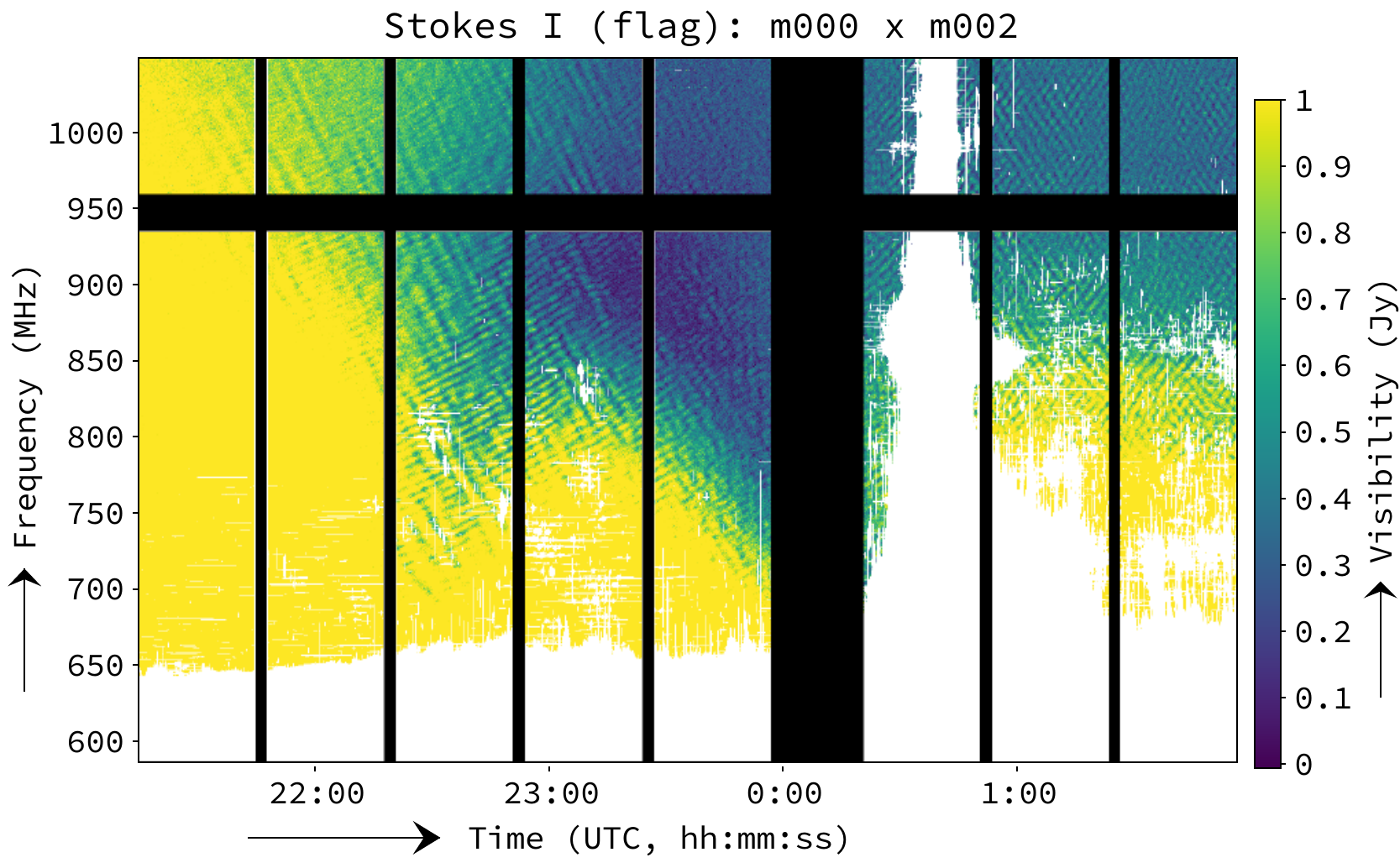} \\
 \includegraphics[width=.49\hsize,trim={0cm 0cm 0cm 0cm},clip]{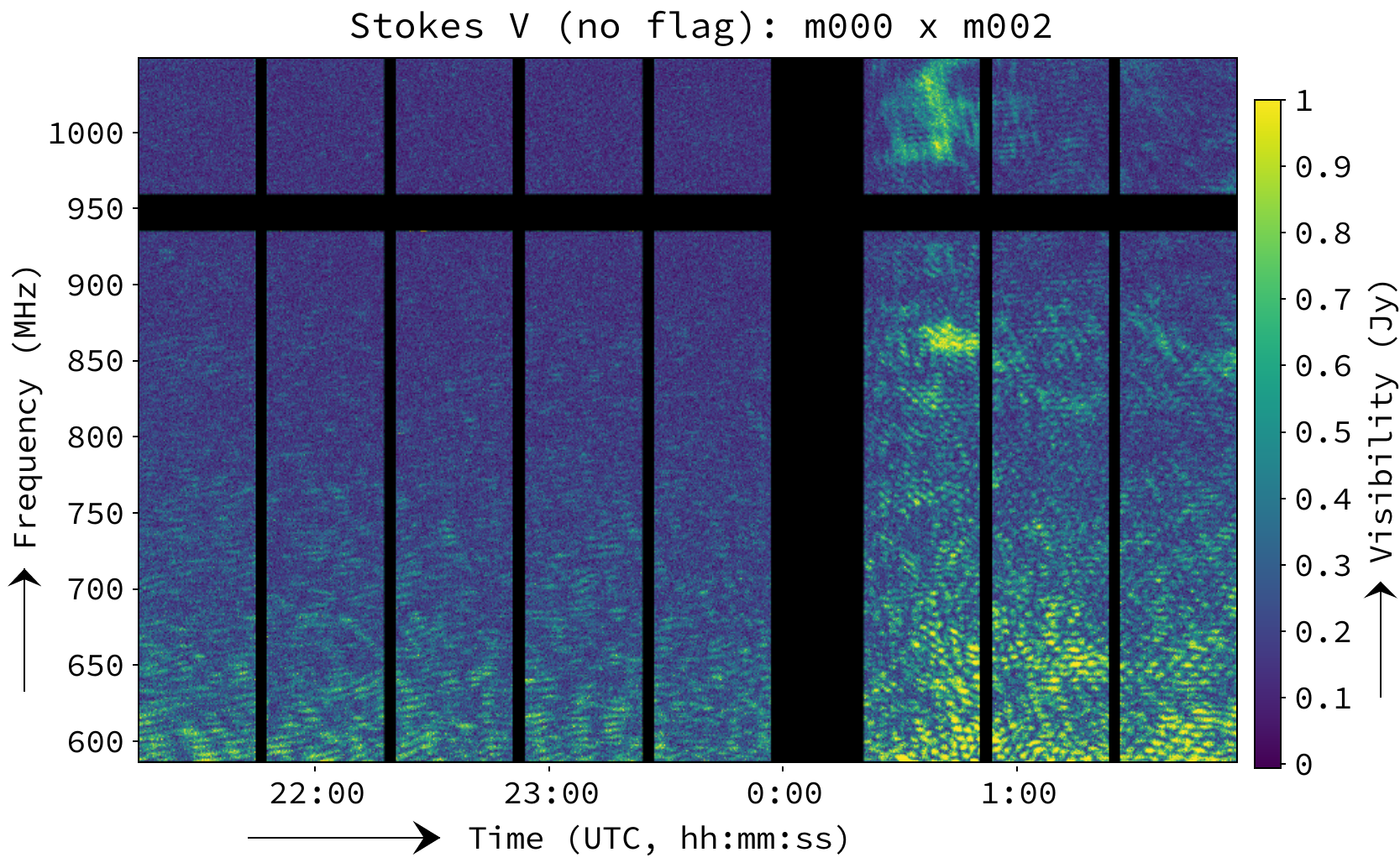} \hspace{0.4cm}
 \includegraphics[width=.49\hsize,trim={0cm 0cm 0cm 0cm},clip]{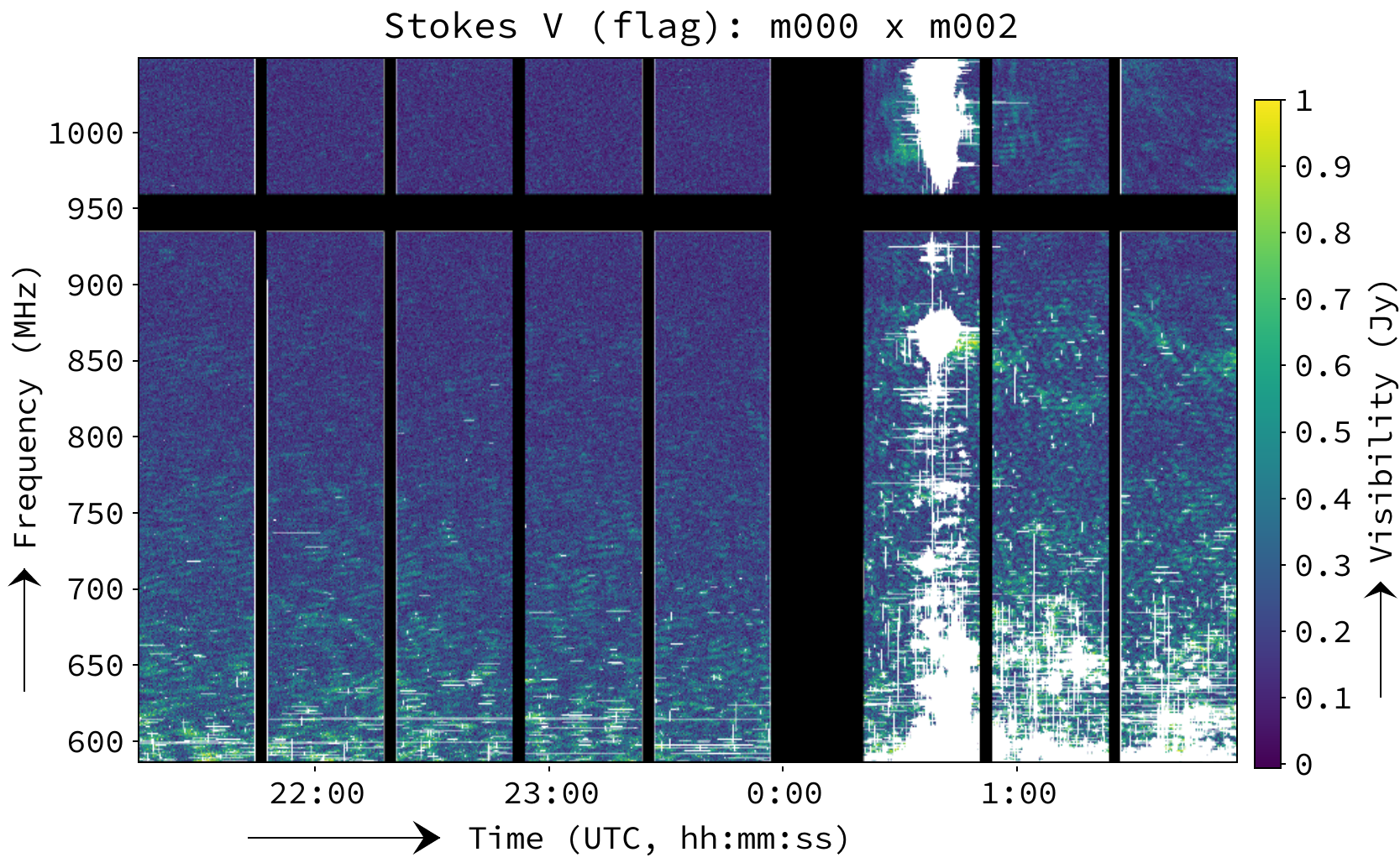} \\
 \includegraphics[width=.49\hsize,trim={0cm 0cm 0cm 0cm},clip]{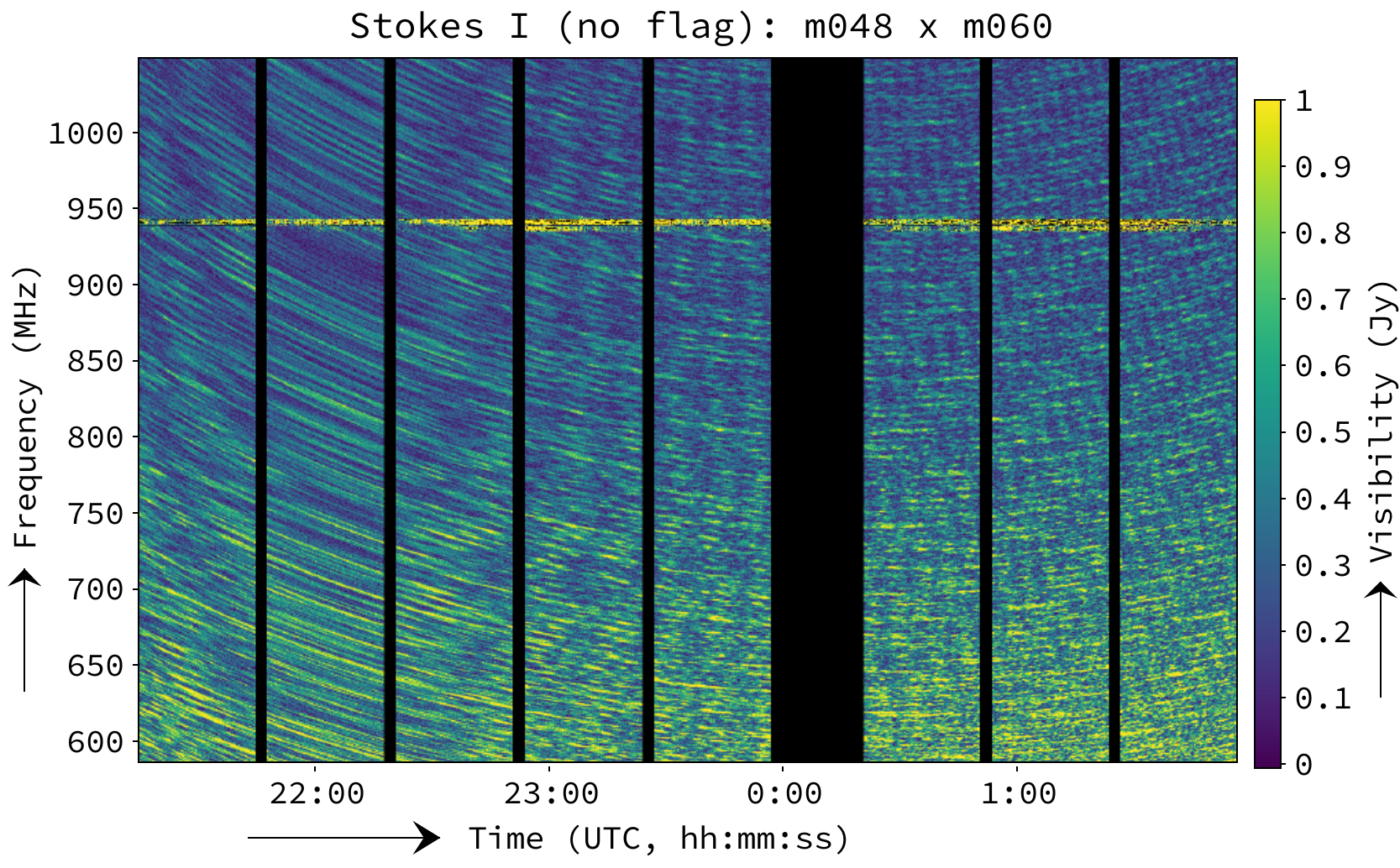} \hspace{0.4cm}
 \includegraphics[width=.49\hsize,trim={0cm 0cm 0cm 0cm},clip]{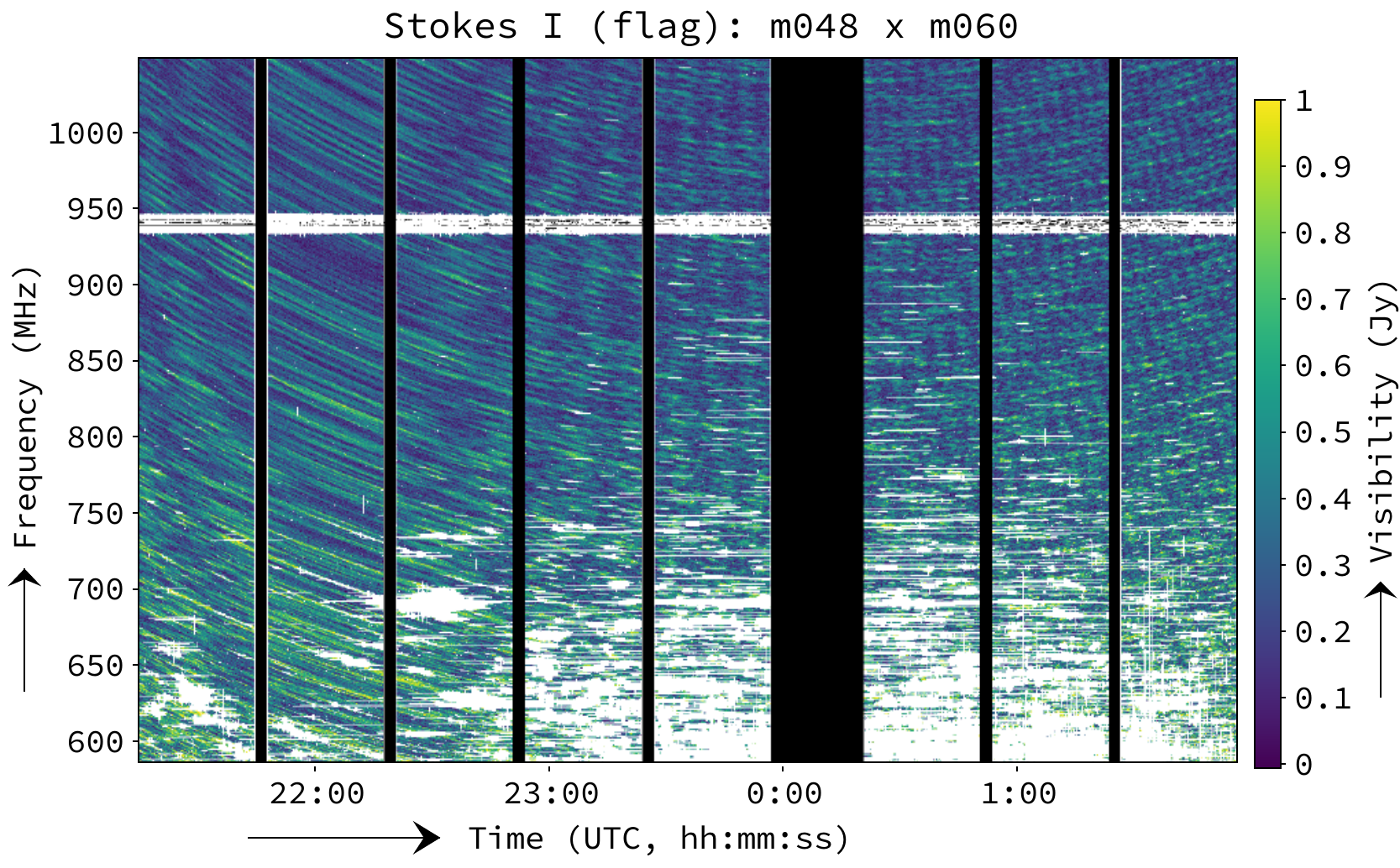} \\
 \includegraphics[width=.49\hsize,trim={0cm 0cm 0cm 0cm},clip]{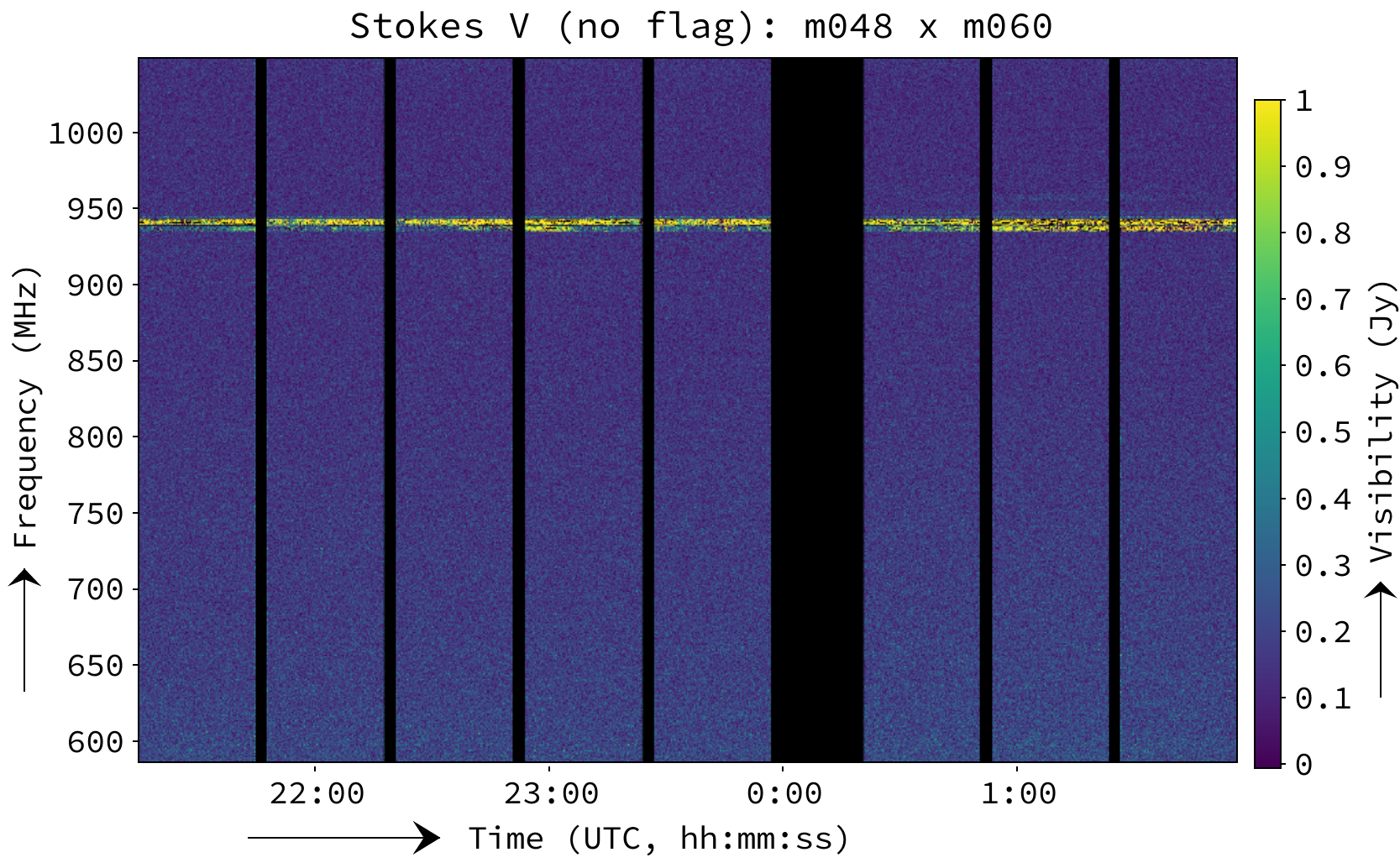} \hspace{0.4cm}
 \includegraphics[width=.49\hsize,trim={0cm 0cm 0cm 0cm},clip]{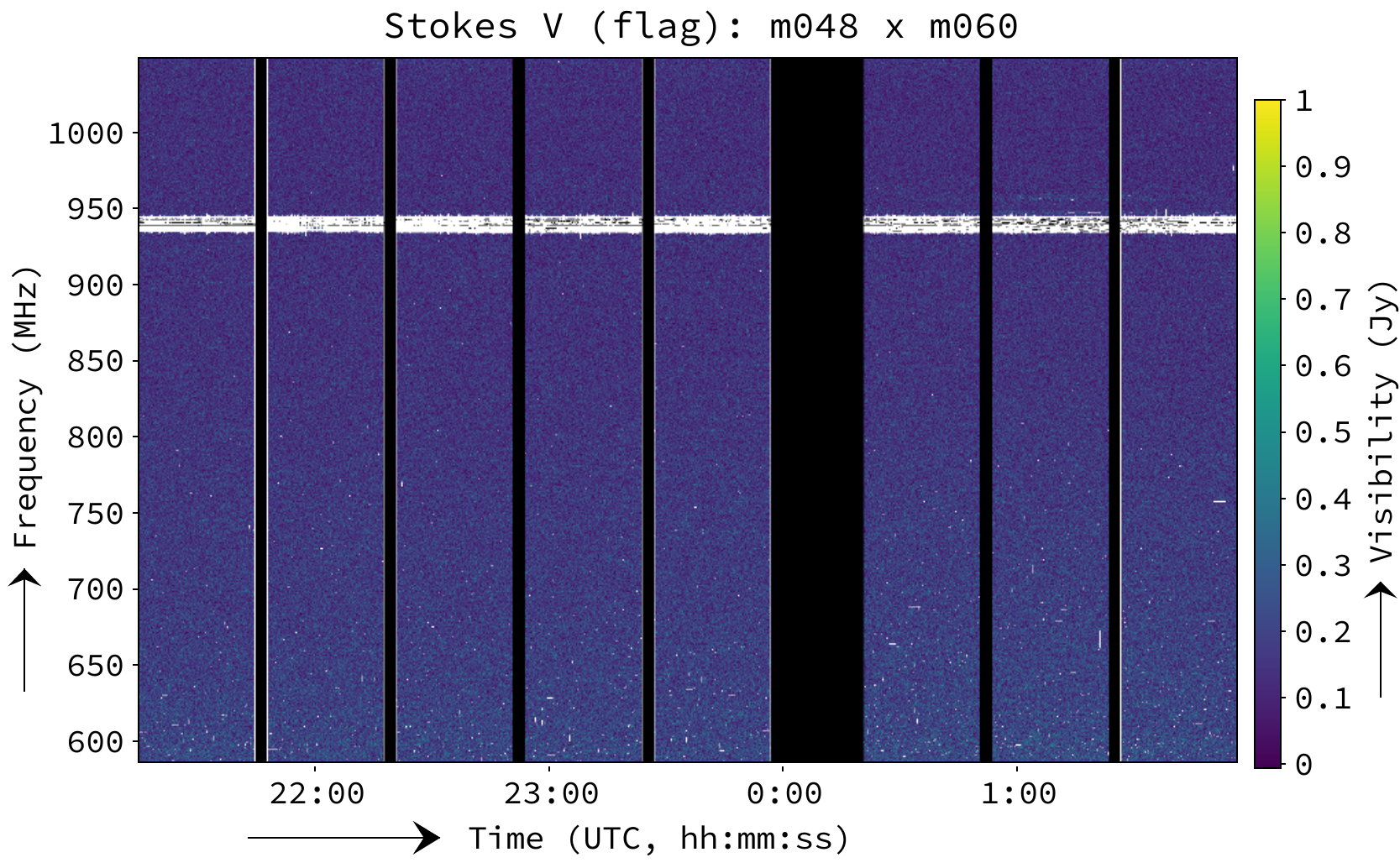}
 \end{minipage}% <- don't forget this %
 \hfill
 \begin{minipage}[b]{0.3\linewidth}
 \caption{Dynamic spectra of the Stokes I and V visibilities of the rising track data of the UHF observation. Plots on the right-hand side show the default flagging strategy applied to Stokes I and Stokes V for the shortest (29~m; m000 $\times$ m002; top panels) and longest (7.7~km; m048 $\times$ m060; bottom panels) baselines (see plot titles). Black bands denote data flagged by the SDP pipeline while white regions are the data flagged by the \aoflagger\ strategy.}
 \label{fig:flagging}
 \end{minipage}
\end{figure*}

We used the \casa\ task \texttt{split} to create new MS files containing only the target field. We then inspected the MS files with the \texttt{rfigui} program of the \aoflagger\ software suite \citep{offringa10, offringa12} to flag data affected by Radio Frequency Interference (RFI) and the edge channels that are within the bandpass roll-off. In particular, we found that applying the default flagging strategy to Stokes V is more efficient in flagging RFI present in \meerkat\ observations compared to its application to Stokes I, which tends to be too aggressive (see Fig.~\ref{fig:flagging} for a comparison between the two). After having tested the Stokes V flagging strategy on different \meerkat\ UHF- and L-band observations, we set it as default strategy directly in \texttt{facetselfcal.py}, if the user enables the option to run \aoflagger\ on input data. Before self-calibration, data were averaged by a factor of 2 in time and by a factor of 4 in frequency while being compressed with \dysco\ \citep{offringa16} to reduce the data volume to a more manageable size of $\sim$20 GB per MS file. As our scientific analysis is focused on the central region of the field, time and bandwidth smearing are not an issue.

\subsection{Self-calibration}

\subsubsection{Full field-of-view calibration}\label{sec:full_fov_cal}

\begin{figure*}
 \centering
 \includegraphics[width=\hsize,trim={0.2cm 0.2cm 0.2cm 0.3cm},clip]{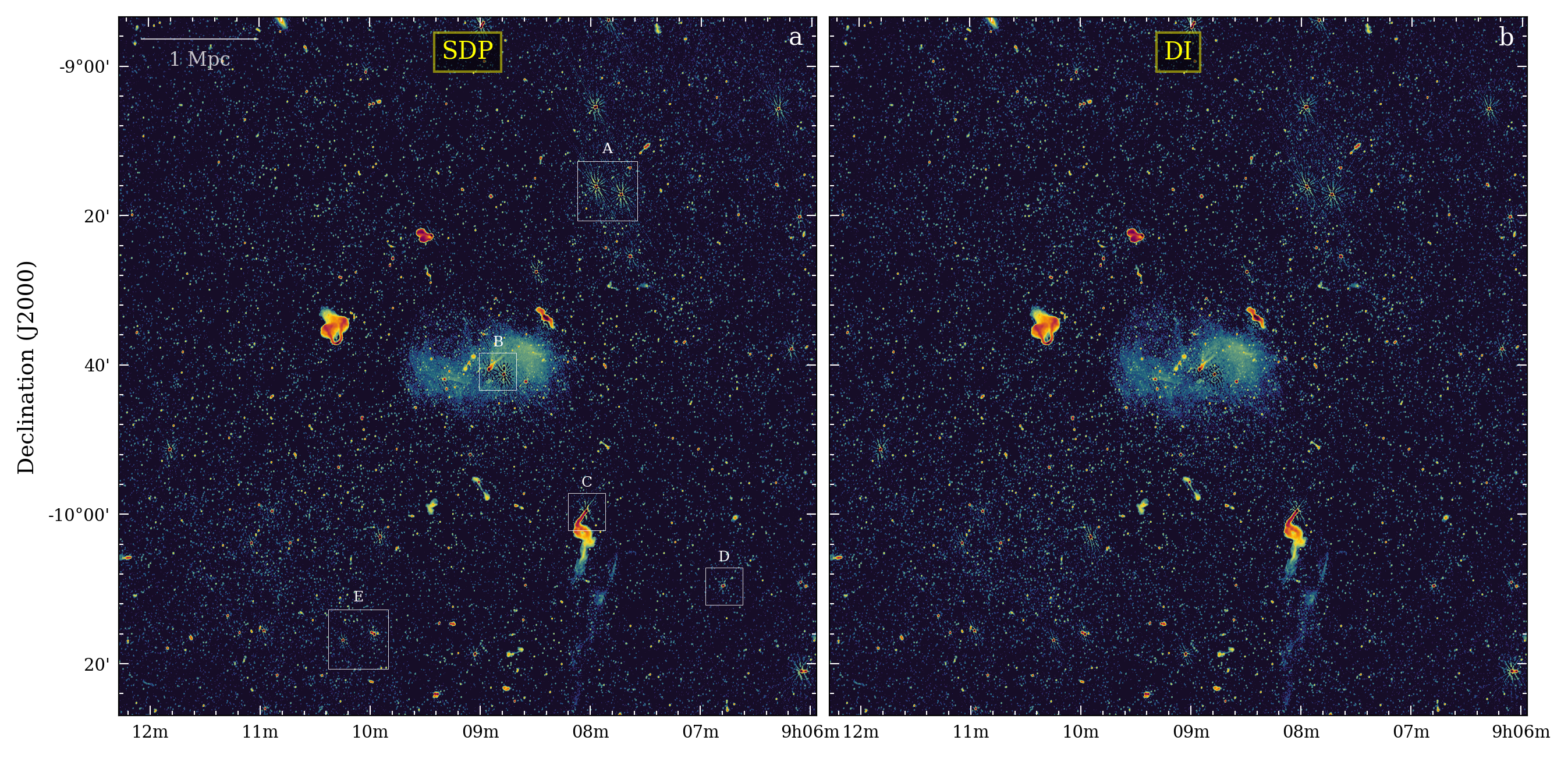} \\ 
 \vspace{0.2cm}
 \hspace{-0.38cm} \includegraphics[width=.985\hsize,trim={0.2cm 0.2cm 0.2cm 0.3cm},clip]{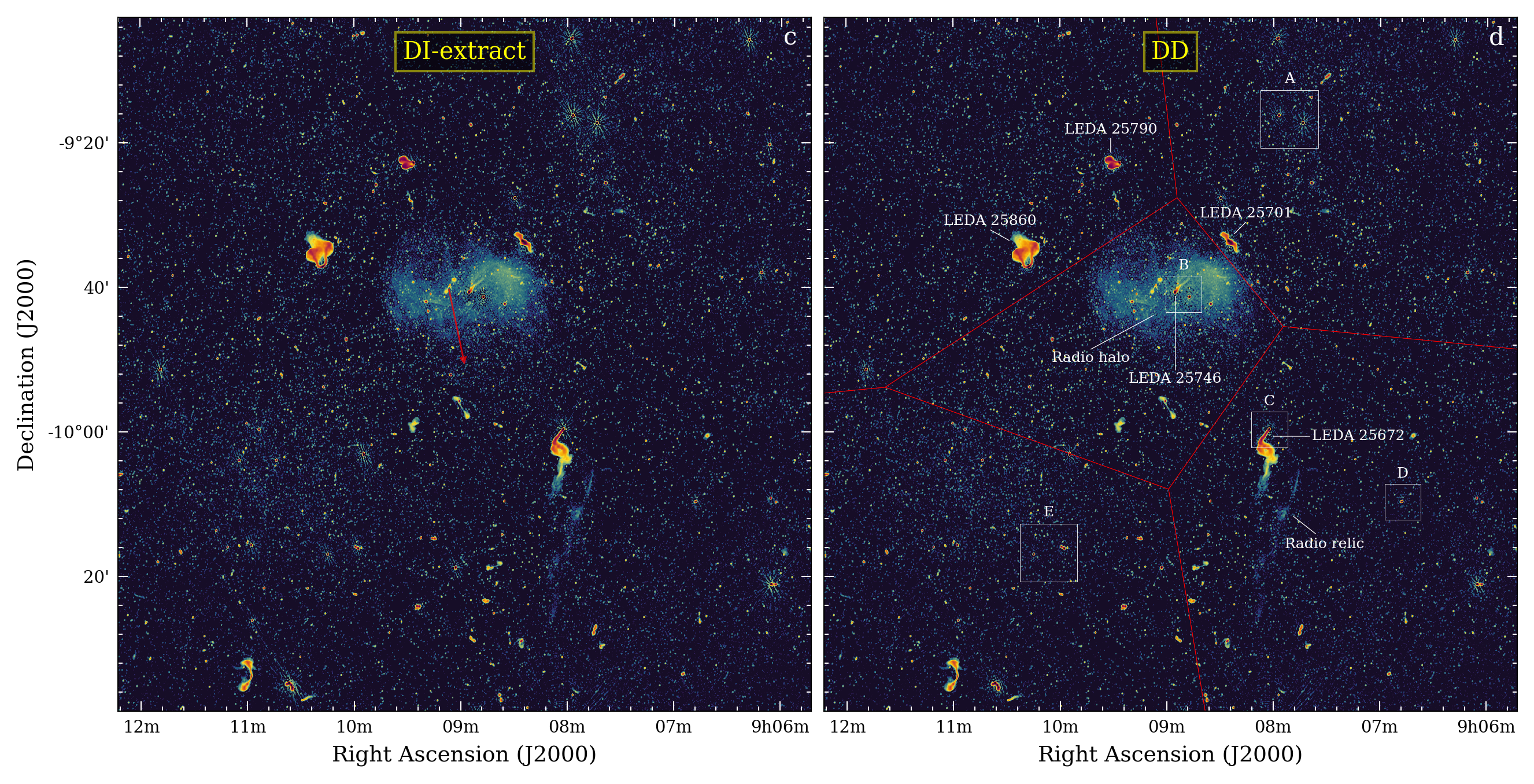}
 \caption{Comparison between UHF images obtained with different calibrations. The color scale is the same in all panels. Images are not corrected for the primary beam attenuation. (\textit{a}) SDP Continuum pipeline calibration. (\textit{b}) Direction-independent calibration on the full FoV (\cf\ Section~\ref{sec:full_fov_cal}). (\textit{c}) Direction-independent calibration on the extracted and phase-shifted dataset (\cf\ Section~\ref{sec:cal_DI}); red arrow shows the shift from the original to the new phase center. (\textit{d}) Direction-dependent calibration (\cf\ Section~\ref{sec:cal_DD}); red polygons show the five facets use for the direction-dependent calibration. The main sources discussed in the paper are labeled.}
 \label{fig:selfcal_compare}
\end{figure*}

\begin{figure*}
 \centering
 \includegraphics[width=.18\hsize,trim={0.2cm 0.2cm 0.3cm 0.3cm},clip]{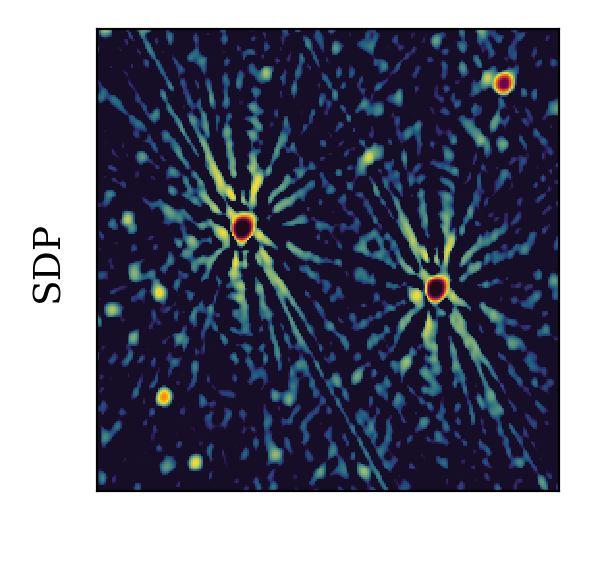} 
 \includegraphics[width=.18\hsize,trim={0.2cm 0.2cm 0.3cm 0.3cm},clip]{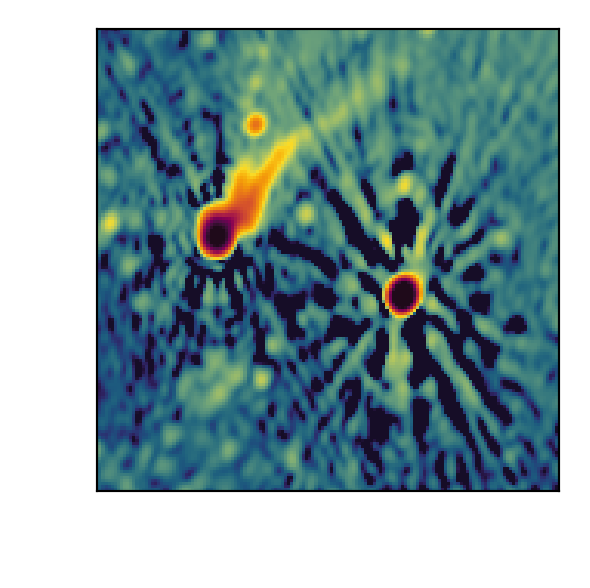} 
 \includegraphics[width=.18\hsize,trim={0.2cm 0.2cm 0.3cm 0.3cm},clip]{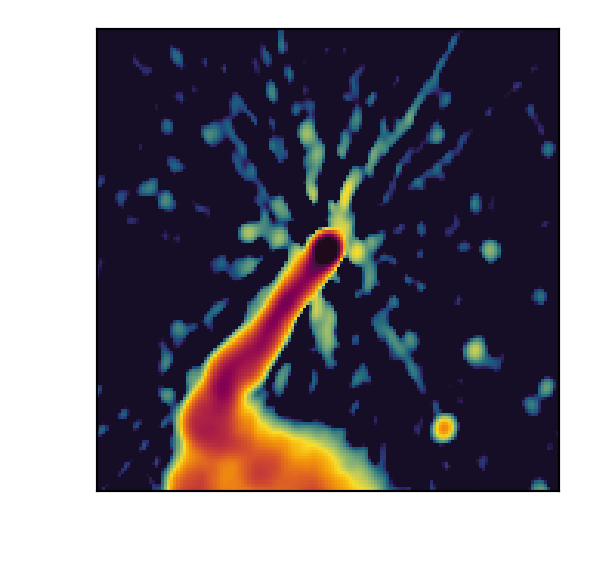} 
 \includegraphics[width=.18\hsize,trim={0.2cm 0.2cm 0.3cm 0.3cm},clip]{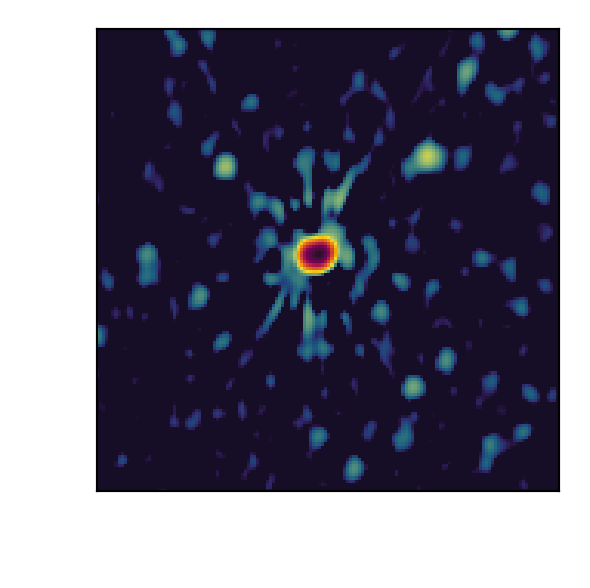} 
 \includegraphics[width=.18\hsize,trim={0.2cm 0.2cm 0.3cm 0.3cm},clip]{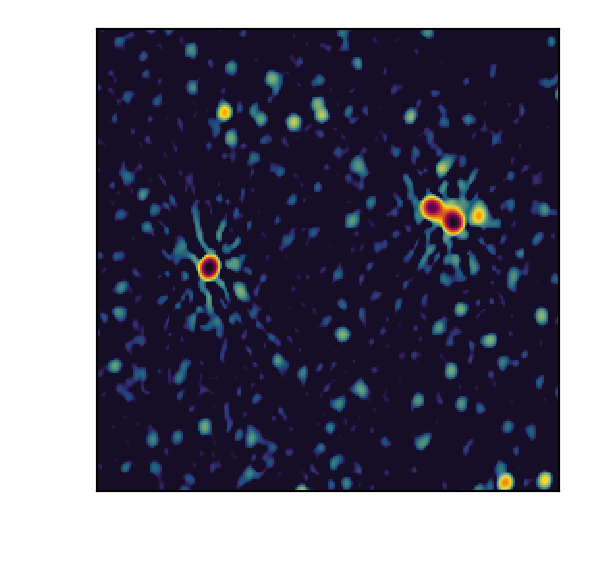} \\
 \includegraphics[width=.18\hsize,trim={0.2cm 0.2cm 0.3cm 0.3cm},clip]{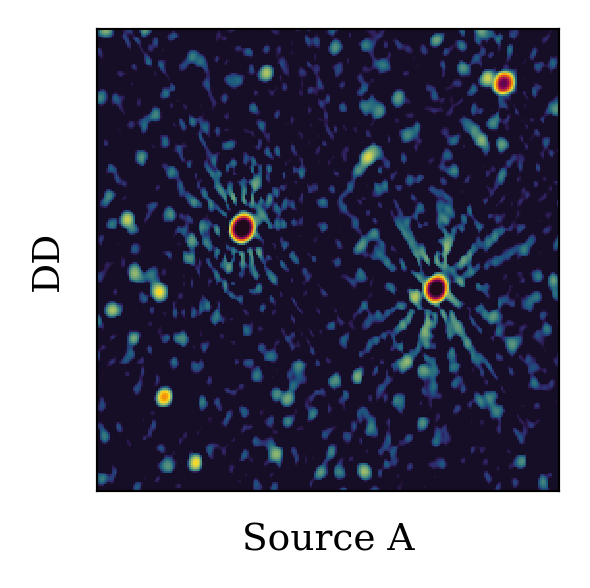} 
 \includegraphics[width=.18\hsize,trim={0.2cm 0.2cm 0.3cm 0.3cm},clip]{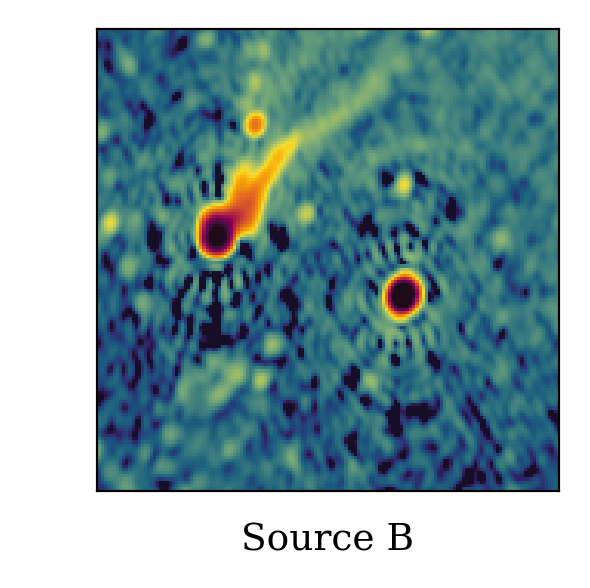} 
 \includegraphics[width=.18\hsize,trim={0.2cm 0.2cm 0.3cm 0.3cm},clip]{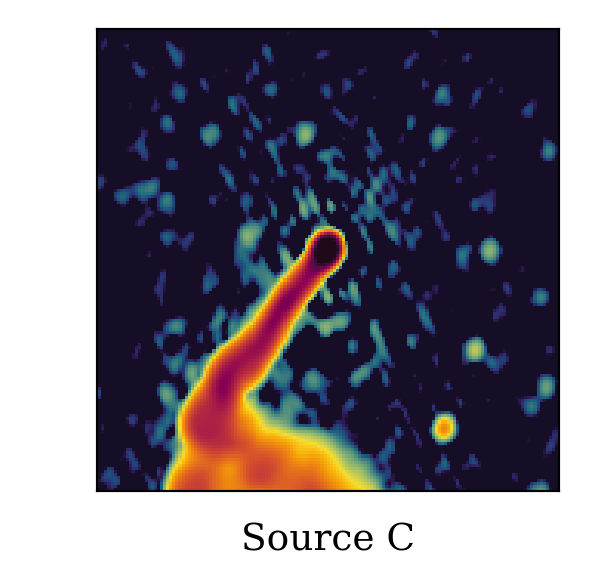} 
 \includegraphics[width=.18\hsize,trim={0.2cm 0.2cm 0.3cm 0.3cm},clip]{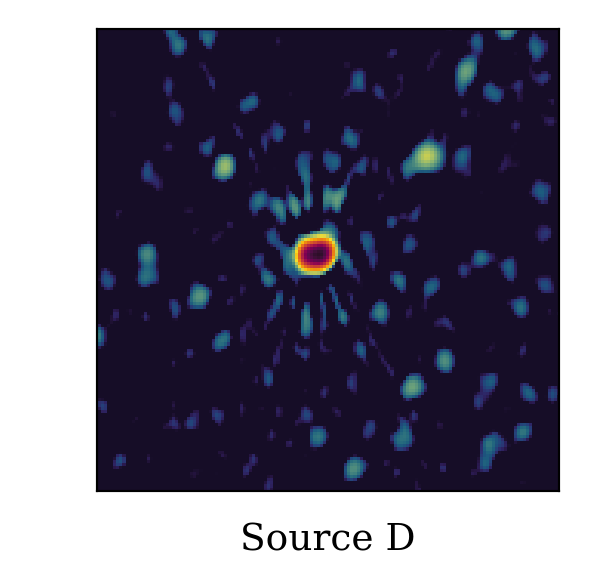} 
 \includegraphics[width=.18\hsize,trim={0.2cm 0.2cm 0.3cm 0.3cm},clip]{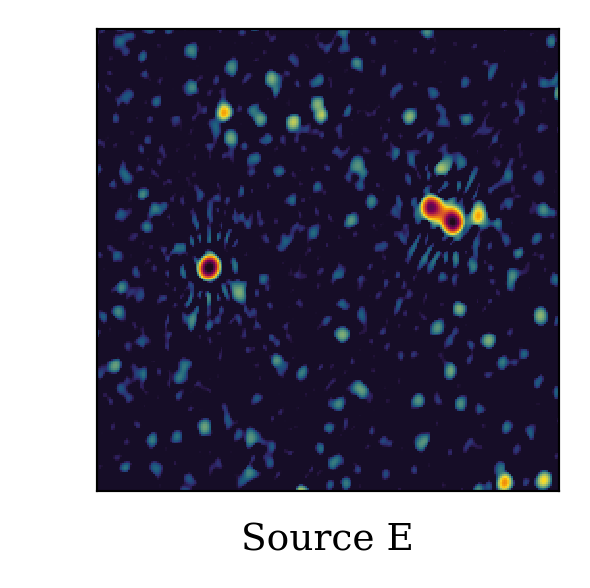} 
 \caption{Zoom-in images of sources A to E (\textit{left} to \textit{right}) for the images obtained with the SDP Continuum pipeline calibration (\textit{top panels}) and after direction-dependent calibration with \texttt{facetselfcal.py} (\textit{bottom panels}).}
 \label{fig:selfcal_sources}
\end{figure*}

As a first attempt to improve the calibration with respect to the images produce by the SDP Continuum pipeline (Fig.~\ref{fig:sdp_images}), we performed a self-calibration the entire \meerkat\ FoV. In this and in the following subsections, we shall use the UHF datasets (which were jointly calibrated) to showcase our results obtained using the \texttt{facetselfcal.py} pipeline described in \citet{vanweeren21}, which was originally developed for the (re)calibration of \lofar\ observations from \lotss. We adopted a similar procedure also for the calibration of the \meerkat\ L-band observations. We recall that \texttt{facetselfcal.py} uses DP3 to determine calibration solutions and \wsclean\ for imaging.  \\
\indent
The self-calibration step consisted of five rounds of ``scalarphase+complexgain'' calibration. The scalarphase aimed to solve for the polarization independent (Stokes I) phase correction on a fast timescale of 16~s. These solutions were pre-applied before solving for complexgain, for which we used a solution interval of 3~min, aiming to solve for slow gain variations using a function that depends on time, frequency and polarization. For both solves, we constrained the solutions to be smooth over frequency by convolving the solutions with a Gaussian kernel of 25 MHz using the \texttt{smoothnessconstraint} parameter in DP3. Outlying amplitude solutions were automatically flagged. Baselines shorter than 500$\lambda$ were ignored during calibration to filter out the most extended low-surface brightness emission from the cluster. During the calibration step, the addition of new columns in the MS files increases their size to $\sim$90~GB each. \\
\indent
For the imaging, we used the \texttt{wstacking} algorithm \citep{arras21, ye22} and enabled the multiscale multifrequency deconvolution option \citep{offringa17}, subdividing the bandwidth into 12 channels. A Briggs robustness weighting of $-0.5$ \citep{briggs95} was adopted. Data below 10$\lambda$ were excluded to avoid including auto-correlations which were not flagged\footnote{Formally, the shortest baselines in our observations are $40\lambda$ (UHF) and $62\lambda$ (L).}. The bright radio source Hydra A \citep[40.8 Jy at 1.4 GHz;][]{condon98}, located at about 3.3\deg\ from A754, was included in the image in order to remove its sidelobes affecting the pointing center. Both cleaning masks generated from restored images of previous self-calibration cycles and the auto-masking feature implemented in \wsclean\ were used to guide the deconvolution process. To improve the computational performance of the deconvolution and gridding of large images (with $\gtrsim$10$^8$ pixels) from large datasets (with $\gtrsim$10$^6$ visibilities), we enabled the parallel cleaning process (\texttt{-parallel-deconvolution}), which splits into subimages deconvolving them separately, and the baseline-dependent averaging (\texttt{-baseline-averaging}), which averages short baselines more than long baselines. \\
\indent
In Fig.~\ref{fig:selfcal_compare}a we show the image with the SDP Continuum pipeline calibration, with the additional Stokes V flagging and reimaged with \wsclean\ as described above, while in Fig.~\ref{fig:selfcal_compare}b we show the image obtained with \texttt{facetselfcal.py} in last self-calibration cycle. While more extended emission is recovered in the latter image, calibration artifacts are still present around some sources in the FoV. We also note that no appreciable difference/improvement was observed during the five self-calibration rounds.

\subsubsection{Extraction and calibration toward the target}\label{sec:cal_DI}

Because \meerkat\ FoV is large and our research focuses on A754 at the center of the field, in our second attempt we aimed to refine the calibration specifically toward this target. This was achieved with a step called ``extraction'' \citep{vanweeren21}, which consists of the subtraction in the visibility data of all sources located outside a user-defined squared region containing the region of interest. The model used for the subtraction of these sources is obtained from the last self-calibration iteration described in the previous section. The model visibility prediction was performed with \wsclean. Ideally, the extracting region should not be excessively large (more than, say, 0.5 deg $\times$ 0.5 deg) to limit possible direction-dependent distortions due to ionospheric effects. A small extracting region has also the advantage of allowing for fast reimaging in the subsequent analysis. Nonetheless, due to its low-$z$, A754 appears as a large target in the sky and it hosts diffuse emission of interest even at significant distance from the cluster center, such as the tailed source C and the diffuse emission southwest of it (\cf\ Fig.~\ref{fig:selfcal_compare}). For this reason, the extraction region we adopted for A754 is large: 1.6 deg $\times$ 1.6 deg. After subtraction, the \uv-data were phase-shifted with DP3 to the center of the extracting region. The shift of about 10.7 arcmin from the original (RA: 09$^h$09$^m$06.69$^s$, Dec.: $-$09\deg40\arcmin12.50\arcsec) to the new (RA: 09$^h$08$^m$57.80$^s$, Dec.: $-$09\deg50\arcmin39.55\arcsec) phase center is represented by the red arrow in the Fig.~\ref{fig:selfcal_compare}c. \\
\indent
The self-calibration of the extracted datasets was done as described in Section~\ref{sec:full_fov_cal}. Results are reported in Fig.~\ref{fig:selfcal_compare}c. In this specific case, the calibration does not improve with respect to the full FoV calibration performed before (\cf\ Fig.~\ref{fig:selfcal_compare}b), except in terms of computing speed, as the subtraction of the sources outside the region of interest implies that the new self-calibration rounds could be performed on images with 16 times less pixels. As anticipated, the reason of the lack of improvements is due to presence of direction-dependent effects that could not be corrected for during self-calibration. In general, the user focusing on small targets/regions may already find the calibration obtained after this extraction step satisfactory to carry out the scientific analysis. However, for larger targets/regions of interest and/or in the presence of significant direction-dependent effects, the procedure outlined in the following subsection is recommended.

\subsubsection{Direction-dependent calibration}\label{sec:cal_DD}

To address the challenge of correcting for direction-dependent effects, we incorporated in \texttt{facetselfcal.py} the capability to calibrate data across different directions (facets) of the FoV. This functionality was enabled by the recent introduction of the facet-based imaging mode in \wsclean\ (available from v3.0). An example of application of direction-dependent calibration with \wsclean\ was presented in \citet{dejong22} (but see also \citealt{sweijen22, ye23arx}), who analyzed \lofar\ observations at 144~MHz of the cluster pair A399-A401. The method used in \citet{dejong22} consisted in finding the solutions across multiple sources in the FoV, which were extracted and self-calibrated independently, and thus applying the solutions during the facet-based imaging. In the new implementation of \texttt{facetselfcal.py}, a full joint facet-calibration is performed. The direction-depended calibration was carried out on the same extracted datasets discussed Section~\ref{sec:cal_DI}. \\
\indent
As a first step, we need to define the directions where to find calibration solutions. These can be internally determined within \texttt{facetselfcal.py}, which makes use of the \texttt{LSMTool}\footnote{\url{https://git.astron.nl/RD/LSMTool}} package either to determine the directions containing a given target flux or to tesselate the FoV in a given number of facets. Alternatively, the user can provide a list of directions for the tesselation or even a pre-existing facet layout. After performing some rounds of self-calibration with different facets layouts, we adopted the one leading to best results. This is shown in Fig.~\ref{fig:selfcal_compare}d and comprises five directions. We used the same layout also for the calibration of the L-band observations. \\
\indent
After a first round of direction-independent imaging, calibration solutions were found for each facet. As before, the calibration was of ``scalarphase+complexgain'' type, and consisted in five iterations. However, in this case, the DP3 solves for the model data column of each facet (for the $i$-direction, \texttt{facetselfcal.py} calls these columns MODEL\_DATA\_DD$i$). The addition of new model data columns may significantly increase the data volume. In our case, for 5 directions, each MS file reached a size of $\sim$350~GB during the self-calibration step. Solutions for each solve type are collected in Hierarchical Data Format v5 (HDF5) files, that we merged into a single file (one per MS) using the \texttt{lofar\_helpers}\footnote{\url{https://github.com/jurjen93/lofar_helpers}} package \citep{dejong22}. This step is required to apply the facet solutions during direction-dependent imaging. \\
\indent
In Fig.~\ref{fig:selfcal_compare}d we show the image obtained in the last cycle of direction-dependent calibration. In general, we mention that the calibration converges after just two cycles and that only for source B there is a further improvement at the third iteration step. Overall, the direction-dependent calibration leads to an enhancement of the image quality (\cf\ Fig.~\ref{fig:selfcal_sources}). Although there is room for potential improvement (note some residual calibration artifacts around sources A, D, and E), we considered this calibration suitable for the scope of this work and did not attempt further refinement (note indeed that artifacts toward sources B and C, which are close to the diffuse emission of our interest, have been sensibly reduced).

\subsubsection{Final considerations}

Overall, we conclude that the \texttt{facetselfcal.py} pipeline, originally developed in the context of \lofar\ observations, can be successfully used also for the calibration of \meerkat\ data. However, the type of solves and solution intervals described here may need to be adapted for datasets with different limitations. Our tests performed on the UHF data of A754, for example, suggest that similar calibration results can be achieved by solving only for gain corrections, using solution intervals from a few minutes up to $\sim$1~h and \texttt{smoothnessconstraint} values in the range $\sim$10--100~MHz, with the faster ``scalarcomplexgain'' solve type. As these parameters may influence the stability of solutions and the memory footprint of the pipeline on the computing resource, we recommend some consideration when setting them. Depending on the scientific goals and the quality of the observations, the user can decide whether it is necessary to pass through the extraction step (which is advised in most cases to improve the flexibility in the reimaging and analysis) and/or a direction-dependent calibration. For A754, we shall use the data calibrated as described in Section~\ref{sec:cal_DD}.

\begin{figure}
 \centering
 \includegraphics[width=\hsize,trim={0cm 0cm 0cm 0cm},clip]{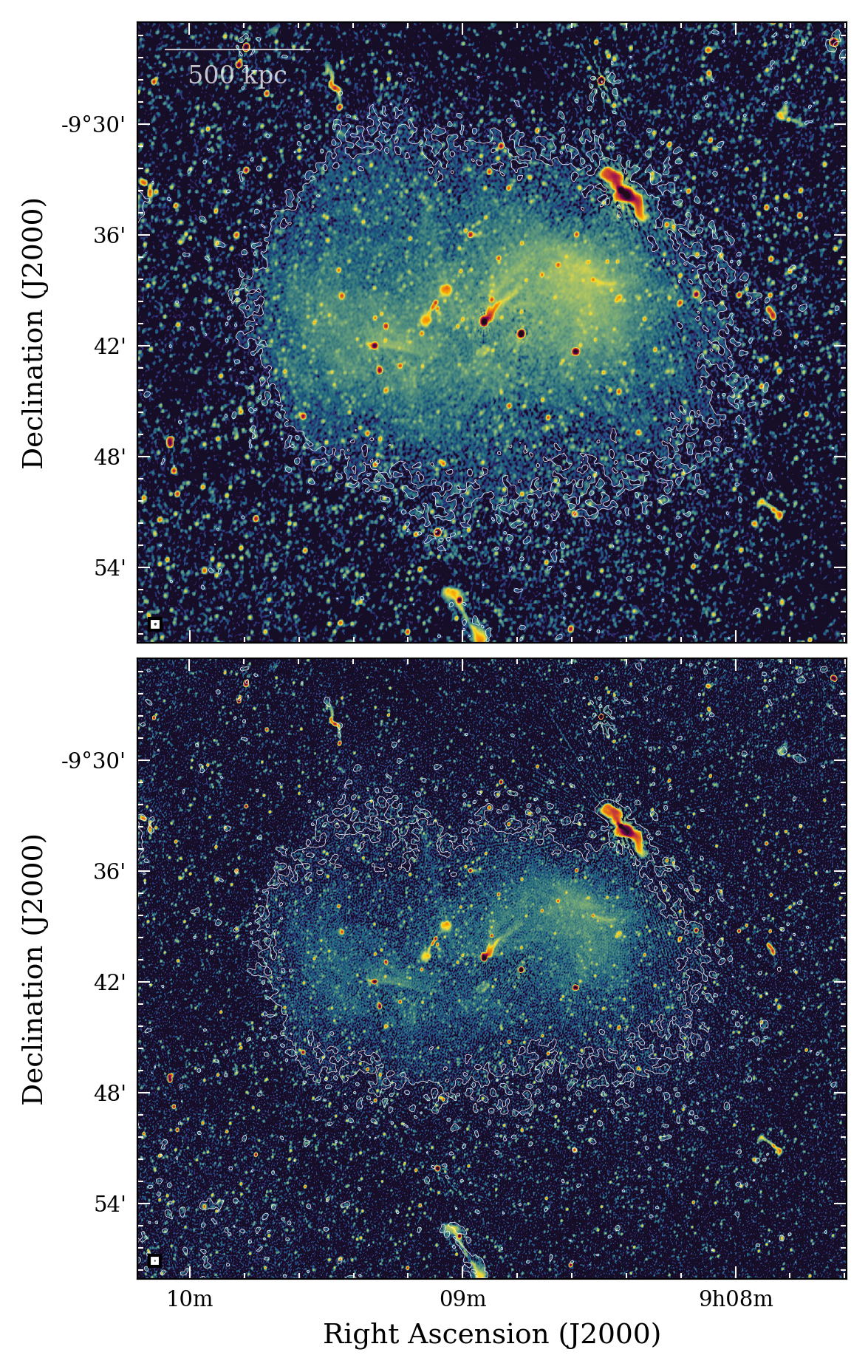}
 \caption{Radio halo recovered with \meerkat\ at 819 MHz (UHF, \textit{top panel}) and at 1.28 GHz (L, \textit{bottom panel}). The colors show high-resolution images while the contour represent the 3$\sigma$ emission from a lower resolution image with discrete sources subtracted (see Tab.~\ref{tab:imaging_par} for more details).}
 \label{fig:halo}
\end{figure}

\begin{figure}
 \centering
 \includegraphics[width=\hsize,trim={0cm 0cm 0cm 0cm},clip]{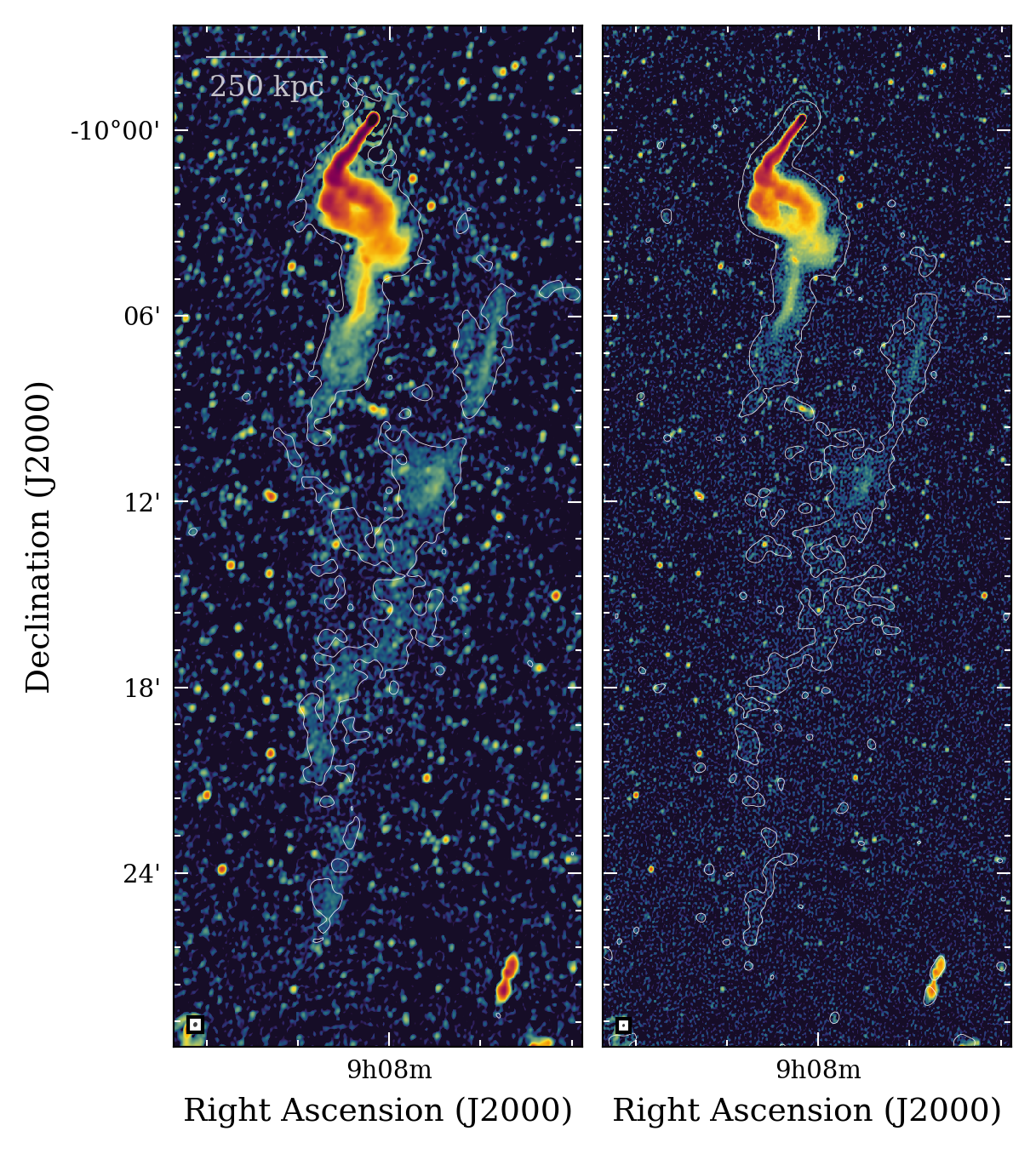}
 \caption{Radio relic recovered with \meerkat\ at 819 MHz (UHF, \textit{left panel}) and at 1.28 GHz (L, \textit{right panel}). The colors show high-resolution images while the contour represent the 3$\sigma$ emission from a lower resolution image with discrete sources subtracted (see Tab.~\ref{tab:imaging_par} for more details).}
 \label{fig:relic}
\end{figure}

\begin{table*}
 \centering
 \caption{Properties of the images reported in the paper.}\label{tab:imaging_par}
 \begin{tabular}{lcccccc} 
  \hline
  \hline
  Frequency & Robust & Taper    & Beam  & rms & Figure \\
  (MHz)     &        & (\arcsec) & ($\arcsec \times \arcsec$) & (\mujyb) & \\
  \hline
  \multirow{4}*{819}  & $-0.5$ & $-$ & $9.5\times7.9$ & 5.7 & \ref{fig:halo} and \ref{fig:relic} (colors) \\
                      & $-0.5$ & 15  & $16.1\times15.6$ & 8.3 & \ref{fig:halo} (contours$^\dagger$) \\
                      & $-0.5$ & 20  & $20.8\times20.3$ & 11.7 & \ref{fig:relic} (contours$^\dagger$) \\
                      %& $-0.5$ & 15 & $16.1\times15.6$ & 12.6 &  \\
                      %& $-0.5$ & 20 & $20.8\times20.3$ & 19.2 & \\
                      & $-0.5$ & 20  & $25.0\times25.0$ & 13.8 & \ref{fig:spix}$^\dagger$ \\
                      %& $-1.0$ & $-$ & $7.5\times6.1$ & 6.9 & \\
  \hline
  \multirow{5}*{1282} & $-0.5$ & $-$ & $6.1\times5.1$ & 3.1 & \ref{fig:halo} and \ref{fig:relic} (colors) \\
                      & $-0.5$ & 15  & $15.4\times15.2$ & 5.8 & \ref{fig:halo} (contours$^\dagger$) \\
                      & $-0.5$ & 20  & $20.4\times20.1$ & 7.8 & \ref{fig:relic} (contours$^\dagger$) \\
                      %& $-0.5$ & 15  & $15.4\times15.2$ & 6.2 & \\
                      %& $-0.5$ & 20  & $20.4\times20.1$ & 9.4 & \\
                      & $-0.5$ & 20  & $25.0\times25.0$ & 10.1 & \ref{fig:spix}$^\dagger$ \\
                      & $-1.0$ & $-$ & $4.9\times6.1$ & 5.4 & \ref{fig:leda} \\
  \hline
 \end{tabular}
 \tablefoot{Circular beams were obtained by smoothing the images with a Gaussian kernel to reach the desired resolution. Images indicated with a dagger symbol have discrete sources subtracted.}
\end{table*}

\subsection{Imaging}

After the last self-calibration iteration, the MS files produced by \texttt{facetselfcal.py} are ready for imaging. Conveniently, \texttt{facetselfcal.py} has an option to ``archive'' the calibrated datasets. It consists of creating new MS files where the CORRECTED\_DATA column obtained during self-calibration is copied into the DATA column of the new files (also in this case, data are compressed with \textsc{dysco}). In this way, the archived MS size goes back to $\sim$20~GB. In the case of direction-dependent imaging, the merged HDF5 files are also required. For reference, these were $\sim$18~GB for each dataset of A754. \\
\indent
The imaging was carried out similarly as done during the calibration, taking particular care of deconvolving the diffuse emission from the cluster. Primary beam correction was performed within \wsclean, which uses the \texttt{EveryBeam}\footnote{\url{https://git.astron.nl/RD/EveryBeam}} library. The images obtained with a robust value of $-0.5$ have a resolution and rms noise ($\sigma$) of 9.5 arcsec $\times$ 7.9 arcsec and $\sigma = 5.7$ \mujyb\ at 819~MHz (UHF) and 6.1 arcsec $\times$ 5.1 arcsec and $\sigma = 3.1$ \mujyb\ at 1.28~GHz (L). These images are shown in Figs.~\ref{fig:halo} and \ref{fig:relic} in colors. In these images, contours are derived from lower resolution images obtained using a Gaussian taper during imaging and where discrete sources were subtracted following the procedure outlined below. Our highest resolution image, that has a beam of 4.9 arcsec $\times$ 6.1 arcsec and rms noise of 5.4 \mujyb\ was obtained using a robust value of $-1.0$ on the L-band data. \\
\indent
To better measure the flux density of the diffuse cluster emission, we subtracted the contribution of contaminating discrete sources from the \uv-data. The first step of this procedure consists in constructing a sky model containing only the discrete sources that we want to subtract from the visibilities. This was achieved by generating high-resolution images using a robust parameter of $-1.0$ and filtering out the most extended emission by imaging baselines longer than 200$\lambda$. To ensure that the model included the emission associated with extended radio galaxies, we forced the multiscale clean to use only the scale corresponding to the synthesized beam and delta components. If we were not using the multiscale clean (namely, if we were using only delta components) part of faint emission associated with radio galaxies would not be captured by the model. Conversely, if we were not limiting the scales used by the multiscale clean, we would risk to include into the model part of the cluster low surface brightness radio emission. While this significantly helps to isolate and subtract the extended emission from radio galaxies in the image, part of the brightest regions of the cluster extended emission may still be picked up by the deconvolution process, and thus added to the model. Therefore, before predicting the visibilities, we carefully inspected the model images and manually removed components associated with the cluster extended emission. Additionally, we excluded from the model the clean components associated with the brightest and most extended radio galaxies (namely, those that will be discussed in Section~\ref{sec:radiogal}), as the subtraction of their complex emission would not be reliable \citep[see \eg][for a similar argument]{botteon22a2255}. The clean components of the model were then subtracted from the \uv-data and the residual visibilities were deconvolved to produce images with discrete sources subtracted. \\
\indent
The spectral analysis was performed on images with discrete sources subtracted obtained by adopting an inner \uv-cut of 62$\lambda$ ($\simeq$55.5 arcmin), which corresponds to the shortest common baseline of the UHF and L datasets, to ensure that both observations recover the same largest angular scale in the sky. Subsequently, the images were convolved to a common resolution of 25 arcsec $\times$ 25 arcsec as well as corrected for positional offsets and regridded to the same pixelation to create the spectral index map that will be discussed in Section~\ref{sec:radio_emission}. The corresponding error map is reported in Fig.~\ref{fig:spix_err}. \\
\indent
More details on the images presented in this work are given in Tab.~\ref{tab:imaging_par}. The errors on the flux densities reported in the paper take into account both the statistical and systematic uncertainly, which we assumed to be 5\% (L) and 15\% (UHF), following previous \meerkat\ results \citep{knowles21, knowles22, sikhosana24arx}.

\section{\xmm\ data reduction}

\subsection{Archival observations}

A754 was observed 13 times with \xmm. The four earlier observations took place in 2001 and 2002 (ObsIDs: 0112950301, 0112950401, 0136740101, 0136740201; PI: Turner) and covered the central region of the cluster. In 2008, two deep observations targeted the shock region (ObsIDs: 0556200101, 0556200501; PI: Leccardi) and one was off-set from the cluster (ObsID: 0556200301; PI: Leccardi). The six latest observations, conducted in 2019 and 2020, covered the cluster outskirts in various directions (ObsIDs: 0821270601, 0821270801, 0844050101, 0844050201, 0844050301, 0844050401; PI: Ghirardini). We downloaded these data from the \xmm\ Science Archive and processed them using the \sasE\ (\sas\ v16.1) and the \esasE\ \citep[\esas;][]{snowden08}.

\subsection{Data analysis and image preparation}

\begin{figure}
 \centering
 \includegraphics[width=\hsize,trim={0cm 0cm 0cm 0cm},clip]{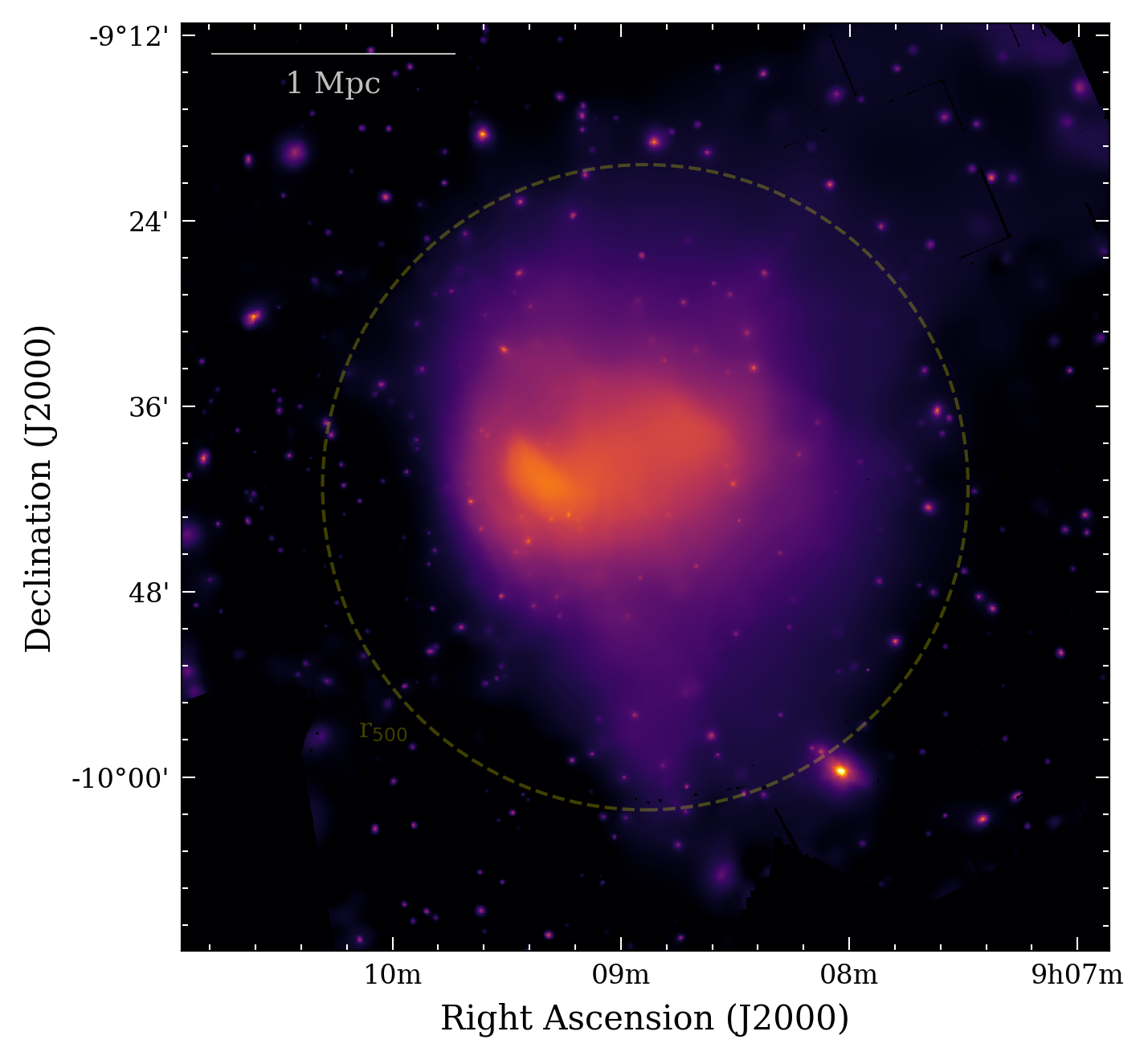}
 \caption{
 \xmm\ mosaic in the 0.4--7.0 keV energy band, adaptively smoothed and exposure corrected. The dashed circle denotes \rfive.}
 \label{fig:xmm_csmooth}
\end{figure}

The analysis followed the steps detailed in \citet{ghirardini19universal}. In particular, we extracted calibrated event files by running the standard event screening chains and then used the tools \texttt{mos-filter} and \texttt{pn-filter} to generate light curves for each detector (MOS1, MOS2, pn) mounted on the \epicE\ (\epic) to remove time intervals affected by soft-proton flares. For each detector, we thus extracted photon-count images, particle background, and exposure maps in different energy bands. The particle background was modeled using \texttt{mos-spectra} and \texttt{pn-spectra}, rescaling the filter-wheel-closed data available in the calibration database to match the observed count rate by measuring the high-energy count rate ratio in the unexposed corners of the three detectors and the closed-filter observations. Exposure images were produced with the \texttt{eexpmap} task, which accounts for vignetting, chip gaps, and dead pixels while computing the local effective exposure time. At the end, we mosaicked photon-count images, particle background, and exposure maps for each \epic\ detector and \obsid\ to obtain a large \xmm\ mosaic of A754. Narrow band images were used to produce thermodynamic maps of the ICM; we refer the reader to Section~\ref{sec:thermo_maps} for more details. A broad band image in the 0.4--7.0 keV energy band was instead used to produce the adaptively smoothed and exposure corrected mosaic reported in Fig.~\ref{fig:xmm_csmooth}. While summing the exposure map of each detector, we rescaled the pn exposure maps in MOS units by multiplying them for a factor representing the ratio of pn to MOS effective areas. This is an energy-dependent factor that was computed for each considered band, taking into account the different response of the 2008 observations, which employed the thick filter instead of the medium filter as the others. We note that this rescaling holds when narrow energy bands are used; for the 0.4--7.0 keV range, it represents the average correction factor across the broad band. For this reason, the image shown in Fig.~\ref{fig:xmm_csmooth} was used only for visualization purposes. The total exposure map given by the sum of MOS1, MOS2, and pn detectors, after soft protons cleaning procedure, is reported in Fig.~\ref{fig:xmm_exp}.

\subsection{Thermodynamic maps}\label{sec:thermo_maps}

We produced thermodynamic maps of the ICM from the \xmm\ data following the procedure outlined in \citet{niemiec23}, summarized as follows. We started by generating photon-count, exposures, and background mosaics in five energy bands (0.4--0.7, 0.7--1.2, 1.2--2.0, 2.0--4.0, and 4.0--7.0 keV) by combing the 13 ObsIDs available on the cluster. Temperature and the other thermodynamic quantities, were determined form the best-fit thermal plasma model, absorbed for the Galactic column density of hydrogen in the direction of A754 \citep[$\nh = 4.96 \times 10^{20}$ cm$^{-2}$;][]{hi4pi16}, derived using spectral model templates in the five considered bands folded with the \xmm\ response in \xspec\ \citep[see also section 7 in][]{jauzac16}. The metallicity was fixed to 0.3 solar. The sky background was modeled using the spectrum extracted in a cluster-free region from \obsid\ 0556200301, adopting a model that accounts for both a cosmic X-ray background (CXB) component and a foreground component \citep[\eg][]{kuntz00}. The CXB was modeled with an absorbed power-law with photon index $\Gamma = 1.46$ \citep{deluca04}. The foreground component was modeled with both a unabsorbed and an absorbed thermal plasma with solar metallicity: the former represents the local hot bubble with a fixed temperature of 0.11 keV, and the latter represents the Galactic halo, for which we found a best-fit temperature of 0.17 keV. We used \texttt{apec} \citep{smith01} as the thermal plasma model and \texttt{phabs} as the photoelectric absorption model, adopting the abundance table from \citet{asplund09rev} and the cross-sections from \citet{verner96}. \\
\indent
We used the five energy bands to evaluate the spectral energy distribution in adaptively circular binned regions. The radius of these regions, centered on each pixel of the mosaic not contaminated by discrete sources, was determined by accumulating counts in the 0.4--7.0 keV band until when the threshold of 2000 counts was reached. Before this step, all images were rebinned by a factor of two to increase the count statistics per pixel. The radii of the circular regions mostly range from 7.5 arcsec in the cluster brightest region to 60 arcsec in the outskirts (Fig.~\ref{fig:tmap_radius}), implying that pixels are correlated on these lengths. The temperature, pseudo-entropy and pseudo-pressure maps will be discussed in Section~\ref{sec:xray_emission}. The corresponding error maps are reported in Fig.~\ref{fig:tmap_err}.

\section{Radio emission from A754}\label{sec:radio_emission}

\begin{table*}
 \centering
 \caption{Properties of the radio halo and relic in A754.}\label{tab:halorelic}
 \begin{tabular}{lccccc} 
  \hline
  \hline
  Source & LLS   & $S_{0.82}$ & $S_{1.28}$ & $\alpha$ & $P_{1.4}$ \\
         & (Mpc) & (mJy) & (mJy) & & (\whz) \\
  \hline
  Halo  & 1.6 & $457.8\pm68.7$ & $223.7\pm11.2$ & $1.30\pm0.39$ & $(1.4 \pm 0.1) \times 10^{24}$ \\
  Relic & 1.6 & $21.8\pm3.4$ & $12.5\pm1.0$ & $1.24\pm0.39$ & $(8.0 \pm 0.7) \times 10^{22}$ \\
  \hline
 \end{tabular}
 \tablefoot{Reported quantities are: largest-linear size (LLS), flux densities at 819~MHz ($S_{0.82}$) and 1.28~GHz ($S_{1.28}$), spectral index derived from images with common inner \uv-cut ($\alpha$), and $k$-corrected radio power extrapolated at 1.4~GHz ($P_{1.4}$).}
\end{table*}

\begin{figure*}
 \centering
 \includegraphics[width=.33\hsize,trim={0cm 0cm 0cm 0cm},clip]{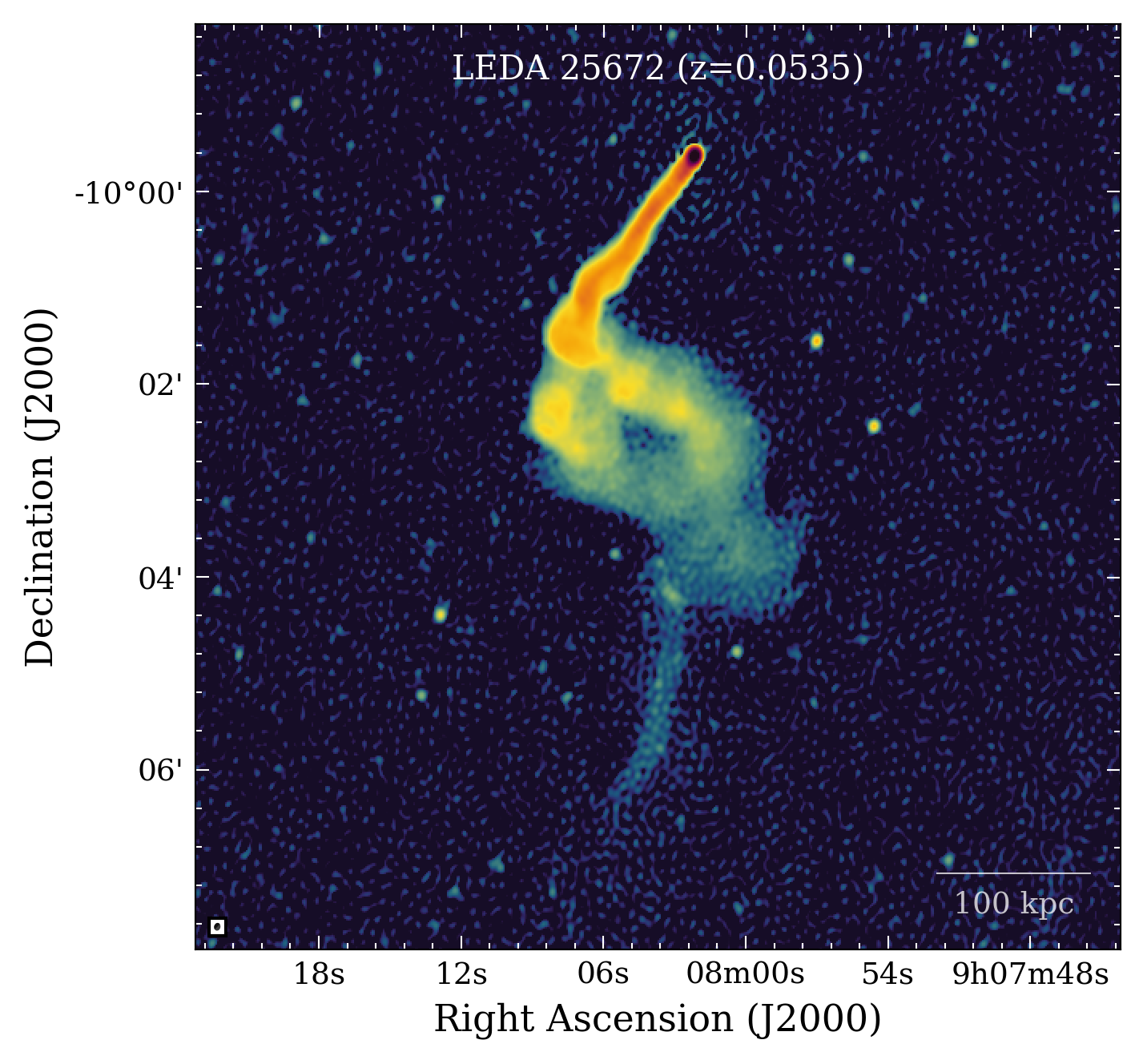}
 \includegraphics[width=.33\hsize,trim={0cm 0cm 0cm 0cm},clip]{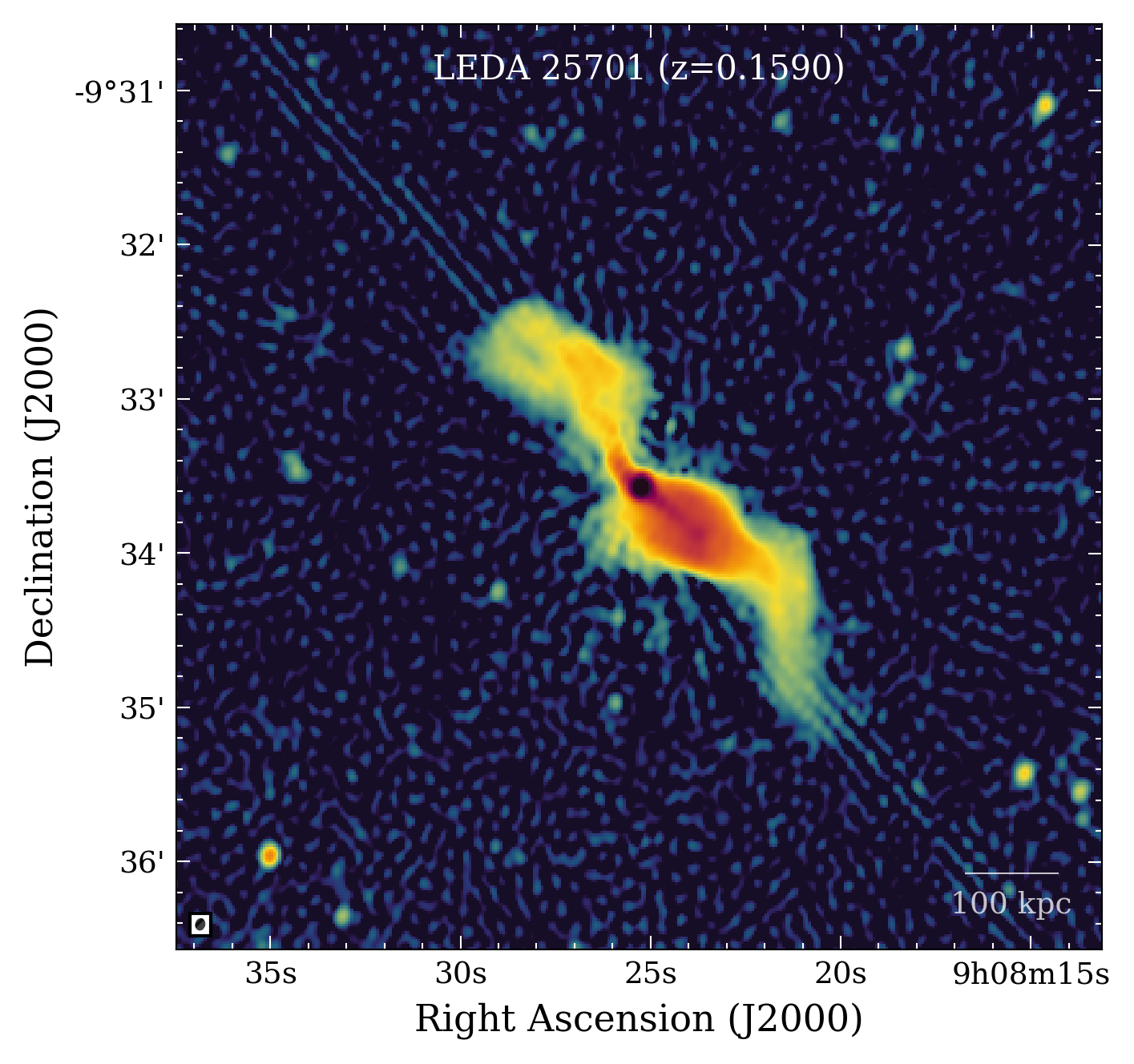}
 \includegraphics[width=.33\hsize,trim={0cm 0cm 0cm 0cm},clip]{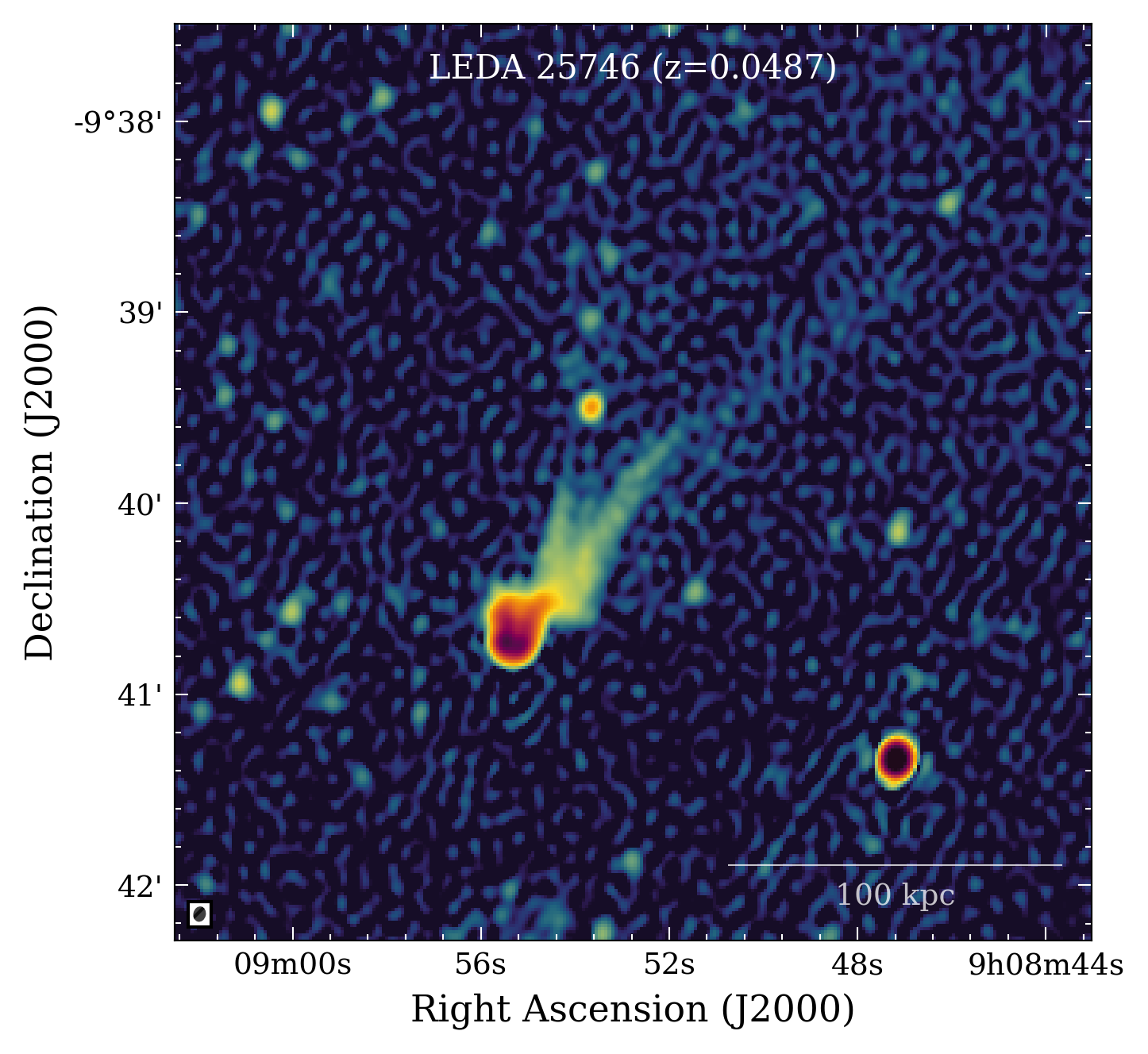}
 \includegraphics[width=.33\hsize,trim={0cm 0cm 0cm 0cm},clip]{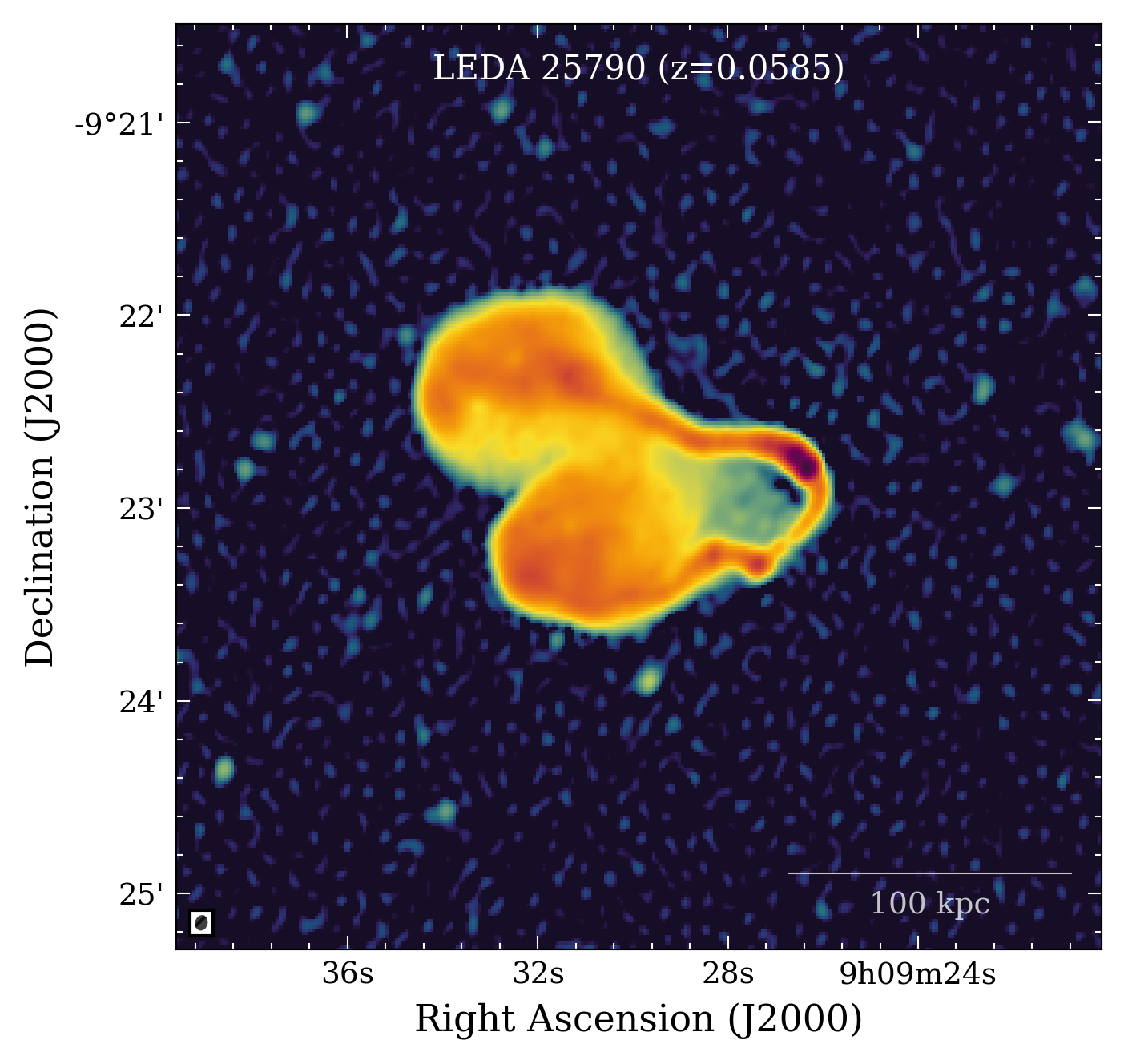}
 \includegraphics[width=.33\hsize,trim={0cm 0cm 0cm 0cm},clip]{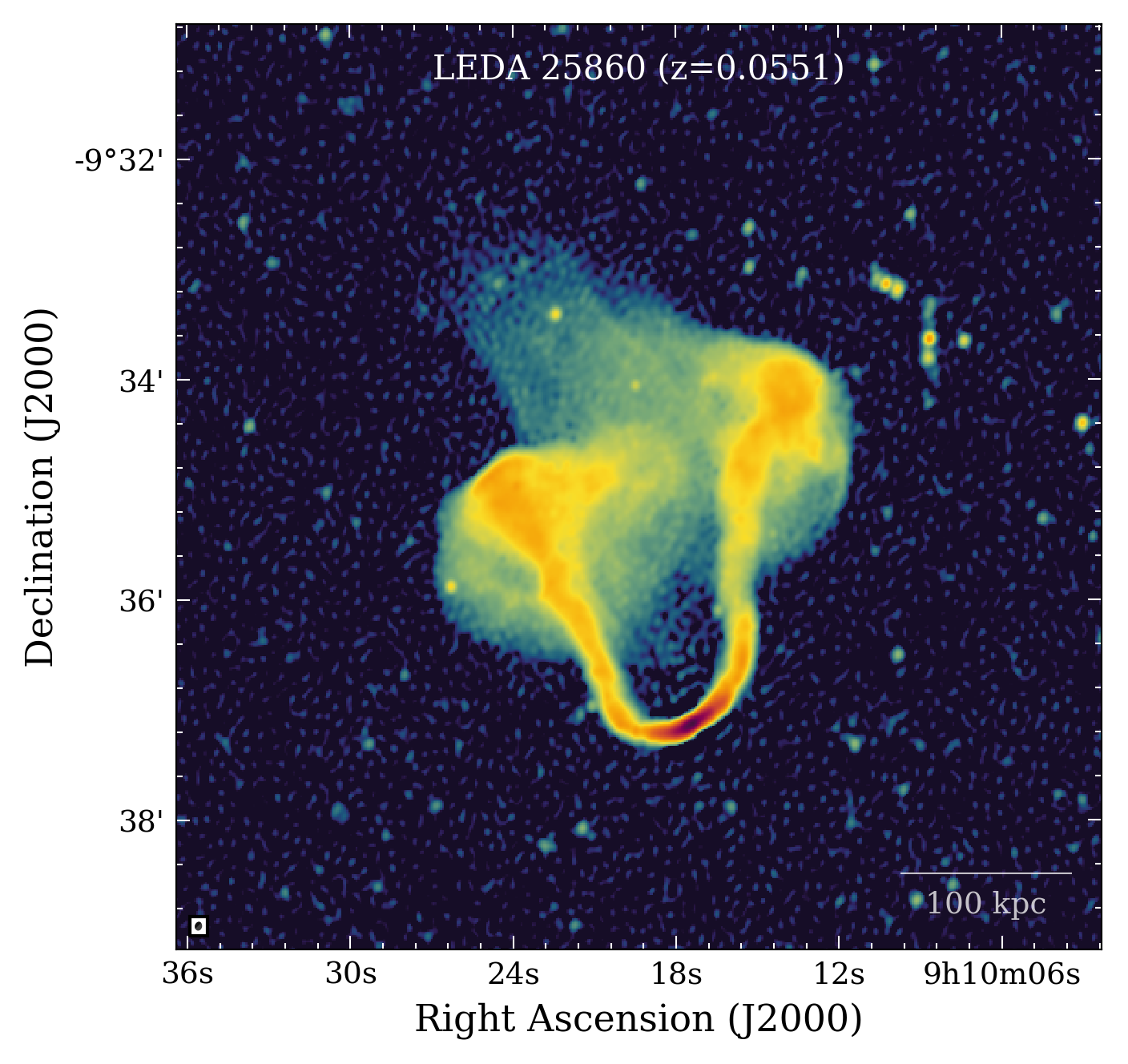}
 \caption{Extended radio galaxies from the highest resolution image at 1.28 GHz (L) we produced in this work (see Tab.~\ref{tab:imaging_par} for more details).}
 \label{fig:leda}
\end{figure*}

\begin{figure}
 \centering
 \includegraphics[width=\hsize,trim={0cm 0cm 0.3cm 0.28cm},clip]{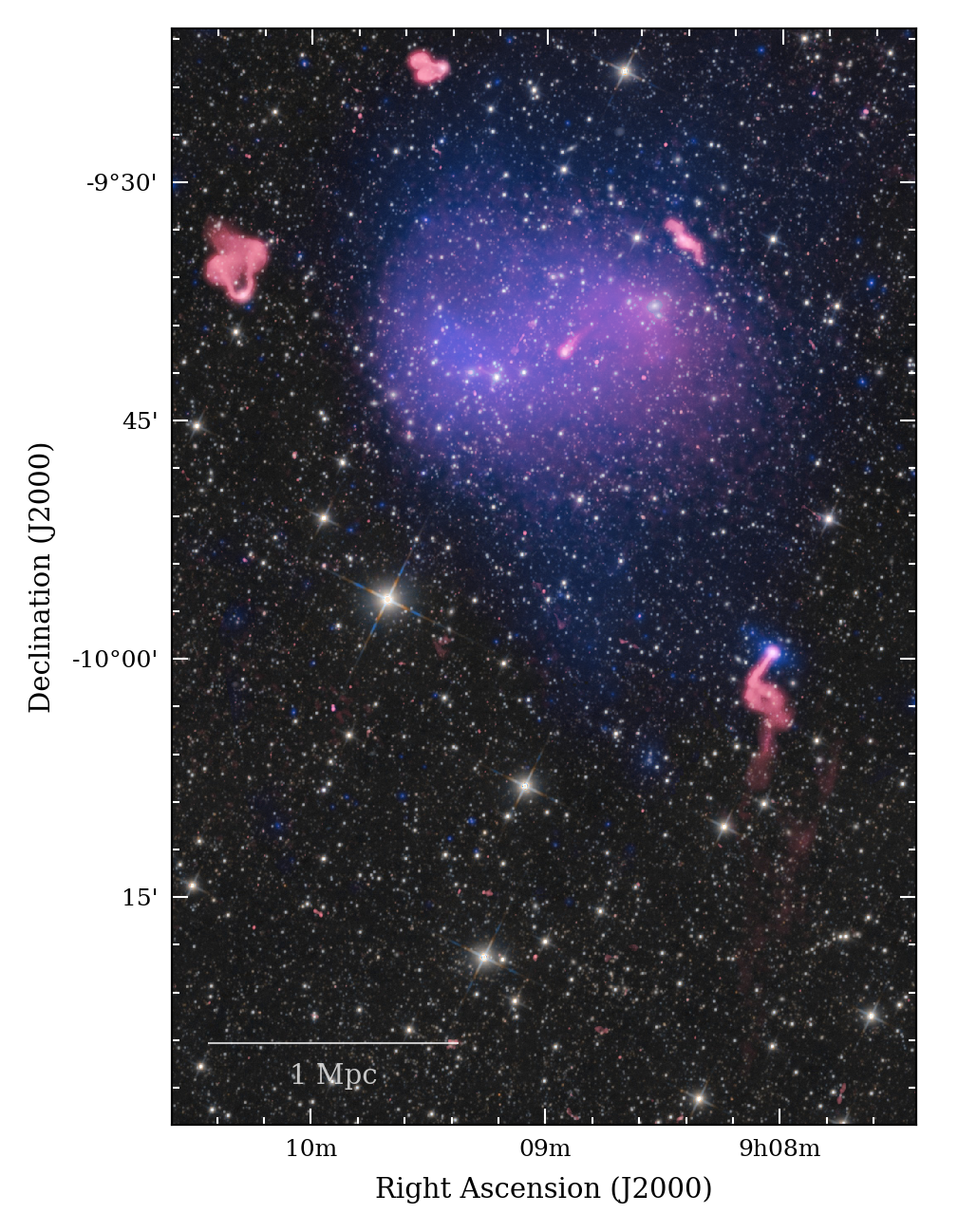}
 \caption{Composite multiwavelength image of A754. The red color denotes the radio emission detected with \meerkat\ while blue represents the X-ray emission recovered with \xmm. The background image is from NEOWISE \citep{mainzer14}.}
 \label{fig:composite}
\end{figure}

\begin{figure}
 \centering
 \includegraphics[width=\hsize,trim={0cm 0cm 2.3cm 0.7cm},clip]{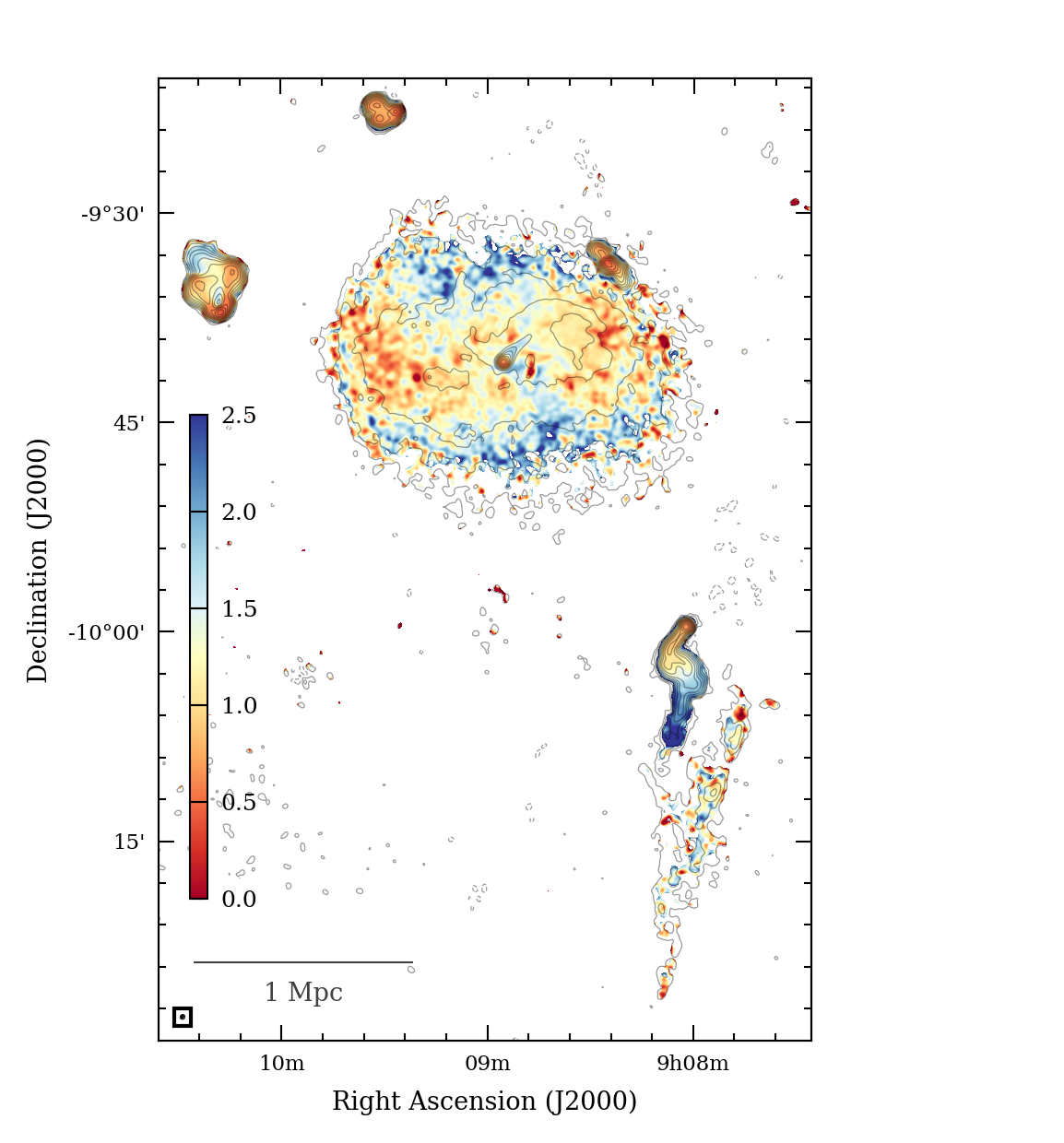}
 \caption{Spectral index map between 819 and 1282 MHz at a resolution of 25\arcsec\ $\times$ 25\arcsec. Pixels with values below 3$\sigma$ were blanked. Contours start the from the 3$\sigma$ level of the UHF image and are spaced by a factor of 2 (see Tab.~\ref{tab:imaging_par} for more details). The spectral index error map is reported in Fig.~\ref{fig:spix_err}.}
 \label{fig:spix}
\end{figure}

The \meerkat\ observations of A754 unveil the presence of diffuse emission in form of a radio halo in the central region of the cluster (Fig.~\ref{fig:halo}) and a radio relic in its southwest periphery (Fig.~\ref{fig:relic}), whose properties are summarized in Tab.~\ref{tab:halorelic}. Additionally, the images capture a number of bright radio galaxies exhibiting extended emission (Fig.~\ref{fig:leda}). Their position within the cluster can be inferred from Figs.~\ref{fig:selfcal_compare}d and \ref{fig:composite}. In the subsequent subsections, we discuss each type of source individually.

\subsection{Radio halo}

The radio halo in A754 is well-recovered by both our \meerkat\ UHF- and L-band high-resolution images of Fig.~\ref{fig:halo}. The emission is mostly elongated in the east-west direction, mirroring the primary merger axis, and exhibits a projected largest linear size of $\sim$1.6 Mpc. At lower frequency (Fig.~\ref{fig:halo}, top panel), its morphology appears more roundish because of the detection of faint emission in the north-south direction. The most striking feature of the diffuse emission is that its surface brightness rapidly decreases toward east. Although the detailed analysis of this region will be the subject of our subsequent paper (Botteon et al., in preparation), we anticipate here that this feature marks the edge of the radio halo emission, which is bounded by the X-ray detected shock front (Fig.~\ref{fig:composite}). Past observations with limited sensitivity and number of short baselines led to believe that the radio emission at the shock was detached from the central diffuse emission, resulting in its misclassification as a radio relic \citep{kale09, macario11}. The brightest region of the halo is located to the west, in-between the radio galaxies LEDA 25746 and LEDA 25701 discussed in Section~\ref{sec:radiogal}. \\
\indent
The integrated flux densities of the radio halo measured from the low-resolution images with discrete sources subtracted\footnote{As the radio galaxies LEDA 25746 and LEDA 25701 were not subtracted but are projected onto the radio halo, their emission was masked and replaced with the average surface brightness of the halo.} within an elliptical region encompassing the $\sim$3$\sigma$ contour of the UHF image are $S_{0.82} = 457.8 \pm 68.7$ mJy (UHF) and $S_{1.28} = 223.7 \pm 11.2$ mJy (L). The integrated spectral index derived from images with common inner \uv-cut is $\alpha = 1.30 \pm 0.39$. The $k$-corrected radio power extrapolated at 1.4 GHz is thus $P_{1.4} = (1.4\pm0.1) \times 10^{24}$ \whz, in line with the value expected from the known $P_{1.4}$--\mfive\ relation for radio halos \citep{cassano13, cuciti21b}. \\ 
\indent
The spectral index map of Fig.~\ref{fig:spix} shows that the radio halo does not have a uniform distribution of spectral index values. Flatter spectrum ($\alpha \lesssim 1.0$) emission is found on the west, in the anticipated post-shock region, and on the east, in the anticipated region with highest brightness (the typical spectral index error in these regions is $\sim$0.2, \cf\ Fig.~\ref{fig:spix_err}). These trace sites of efficient or more recent particle acceleration. Emission with steeper spectrum ($\alpha \gtrsim 1.5$) is instead detected in the north-south direction, perpendicularly with respect to the main cluster merger axis, where particle acceleration is less efficient or occurred long ago (the typical spectral index error in these regions is $\sim$0.4, \cf\ Fig.~\ref{fig:spix_err}).

\subsection{Radio relic}

New diffuse emission is unveiled in our \meerkat\ image in the southwest periphery of A754, at a projected distance of $\sim$38 arcmin $\simeq$ 2.4 Mpc from the cluster center (Figs.~\ref{fig:relic} and \ref{fig:composite}). This source is strongly elongated, with a projected largest-linear size of $\sim$1.6 Mpc and a 1:7 axis ratio. The emission has low surface brightness and is irregular, showing some brighter patches and a somewhat straight edge in the southwest direction while it blends with the emission from the radio galaxy LEDA 25672 (Section~\ref{sec:radiogal}) on the opposite side. A proper characterization of the source morphology, as well as of its flux density and spectral properties, is hampered by its peripheral position in the pointing, which limits the sensitivity of our images due to the primary beam attenuation. With this in mind, the flux densities of the diffuse source that we measured within a polygonal region encompassing the $\sim$3$\sigma$ contour of the UHF image are $S_{0.82} = 21.8 \pm 3.4$ mJy (UHF) and $S_{1.28} = 12.5 \pm 1.0$ mJy (L). The integrated spectral index derived from images with common inner \uv-cut is $\alpha = 1.24 \pm 0.39$. The $k$-corrected radio power extrapolated at 1.4 GHz is thus $P_{1.4} = (8.0\pm0.7) \times 10^{22}$ \whz. \\
\indent
We classify this emission as a radio relic because of its peripheral location, elongated morphology, surface brightness apparently declining toward the cluster center, and steep integrated spectrum. Due to the limitations of the images mentioned above, it is not possible to study the resolved spectral properties of the emission (\cf\ Fig.~\ref{fig:spix}). The presence of a spectral steepening toward the cluster would further support the interpretation of the source as a relic. Nevertheless, the investigation of this additional property is postponed until better data become available.

\subsection{Extended radio galaxies}\label{sec:radiogal}

\begin{figure*}
 \centering
 \includegraphics[width=.33\hsize,trim={0.3cm 0.2cm 0.3cm 0.2cm},clip]{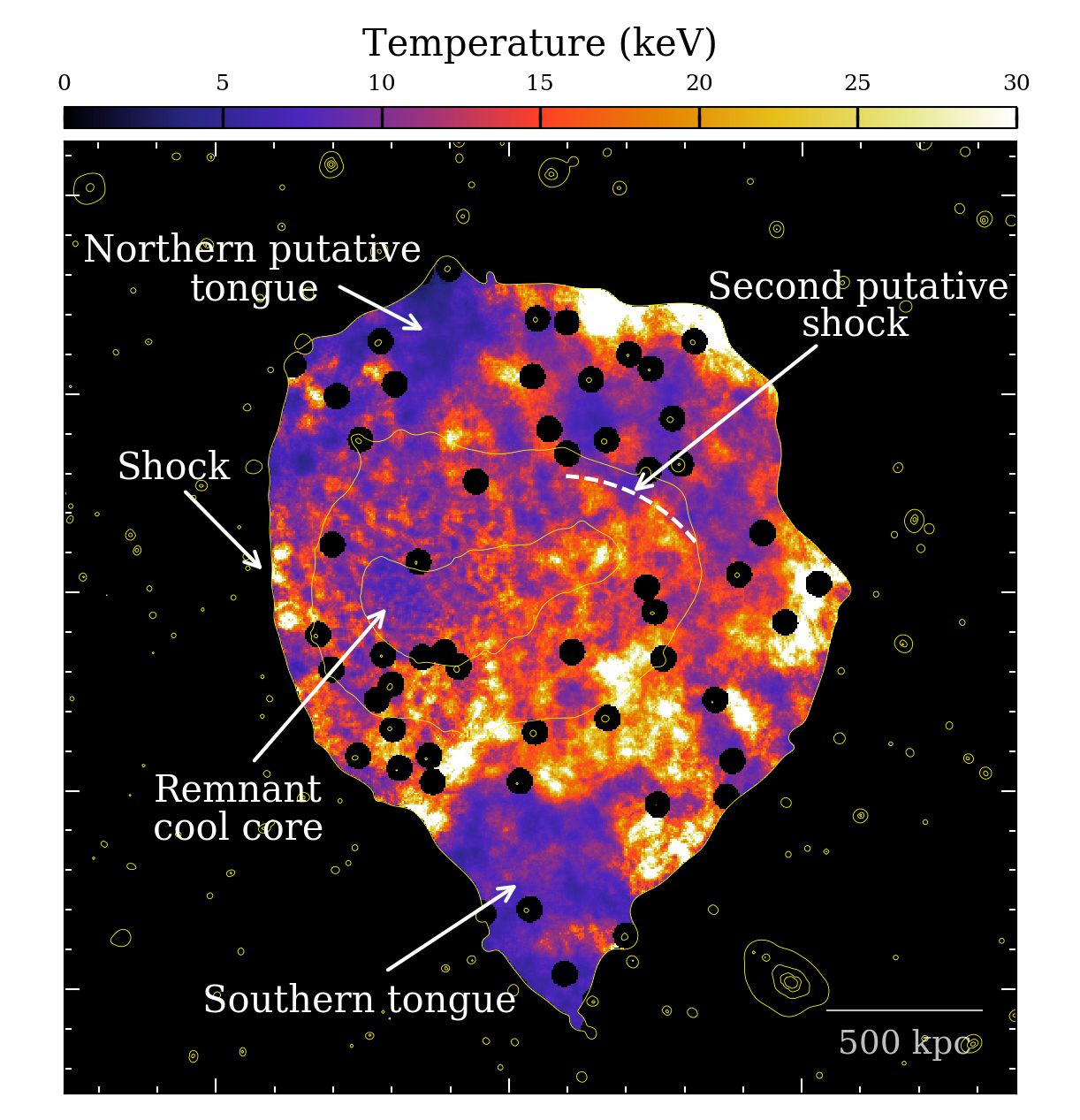}
 \includegraphics[width=.33\hsize,trim={0.3cm 0.2cm 0.3cm 0.2cm},clip]{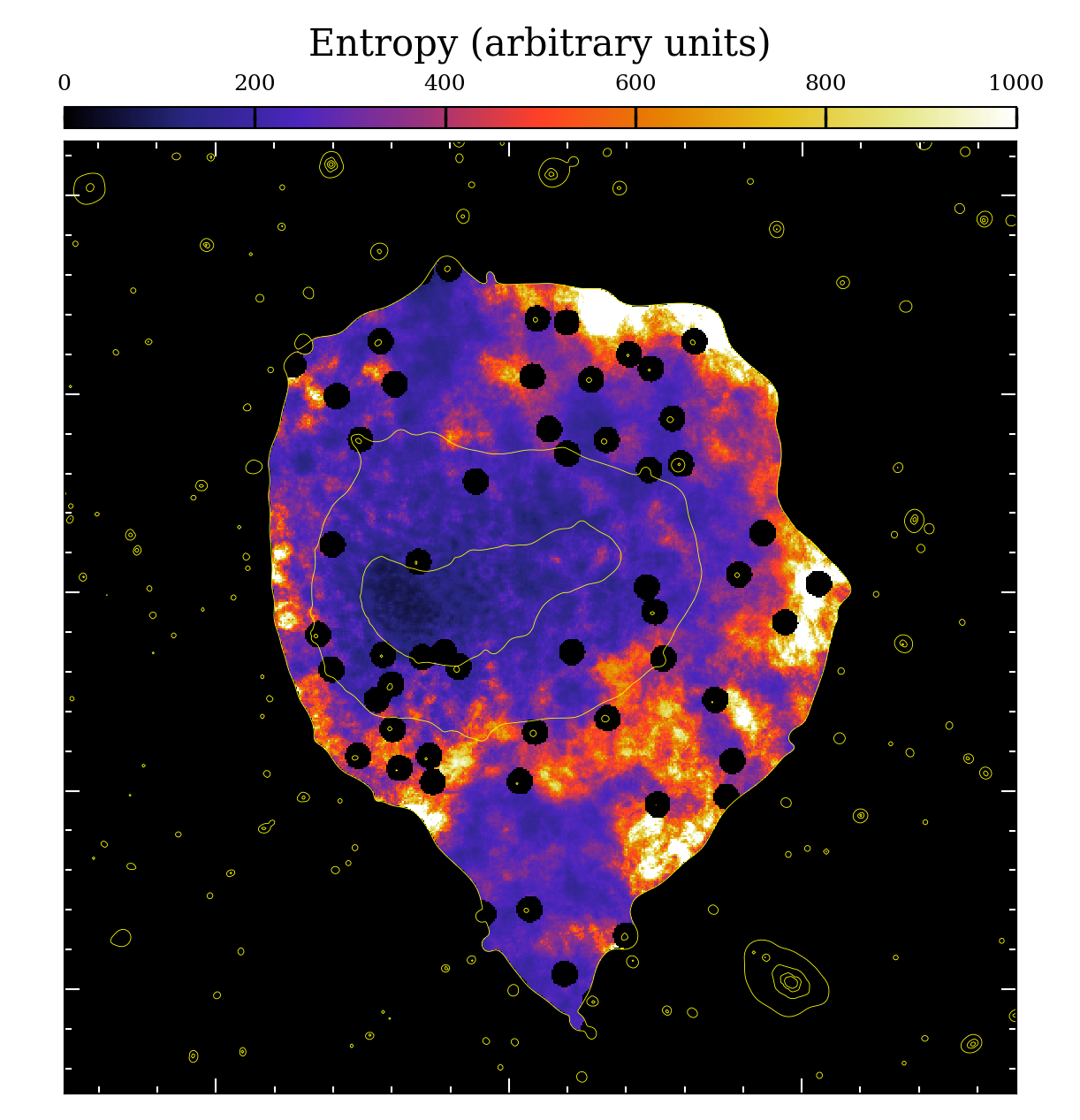}
 \includegraphics[width=.33\hsize,trim={0.3cm 0.2cm 0.3cm 0.2cm},clip]{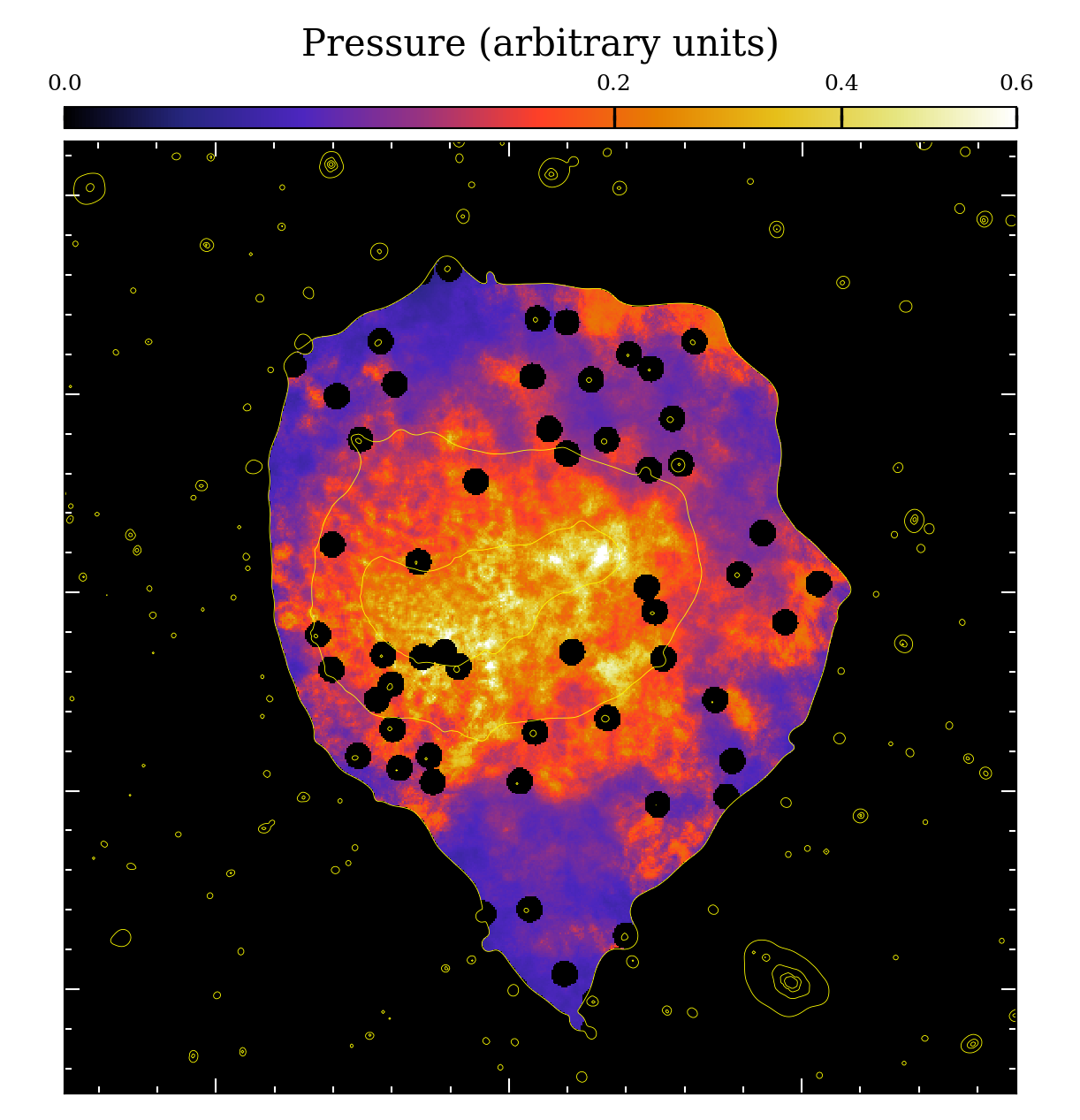}
 \caption{Thermodynamic maps of the ICM with contours from the adaptively smoothed \xmm\ image of Fig.~\ref{fig:xmm_csmooth} reported in yellow. Blanked circular regions indicate masked area because of the presence of point sources.}
 \label{fig:tmap}
\end{figure*}

In Fig.~\ref{fig:leda} we show high-resolution images of the five most prominent extended radio galaxies in the proximity of A754. In the following, we briefly comment on them adopting their Lyon-Meudon Extragalactic Database \citep[LEDA;][]{paturel89} number. All the physical lengths quoted below should be intended as projected sizes. \\
\indent
LEDA 25672 ($z=0.0535$) is a tailed radio galaxy located $\sim$23 arcmin southwest to cluster center. The radio emission keeps a collimated and bright structure up to $\sim$150 kpc from its core, then it bends and bifurcates roughly by 90 degree becoming more diffuse. An additional, fainter, isolated filament of emission characterized by $\alpha \gtrsim 2.0$ (\cf. Fig.~\ref{fig:spix}) extends for a further $\sim$140 kpc to the southwest. At lower resolution, lower surface brightness emission is recovered around the isolated filament, blending with the newly discovered radio relic (Fig.~\ref{fig:relic}). \\
\indent
LEDA 25701 ($z=0.1590$) is a bright background radio galaxy, unrelated to A754, projected at $\sim$10 arcmin to the northwest of the cluster center. It shows a double lobe structure, extending for $\sim$590 kpc. \\
\indent
LEDA 25746 ($z=0.0487$) is projected onto the central region of A754, and is embedded into the radio halo emission. This is a wide-angle tailed radio galaxy, whose tails appear to bifurcate at $\sim$50 kpc from its core. The radio galaxy emission extends at least for a distance of 150 kpc before blending with the radio halo (Fig.~\ref{fig:halo}). The difference between the galaxy line-of-sight velocity and that of A754 is $\Delta v_{\rm los} \simeq -1680$ km s$^{-1}$, larger than the velocity dispersion of cluster galaxies (\cf\ Tab.~\ref{tab:a754_summary}), suggesting the presence of a significant peculiar motion. \\
\indent
LEDA 25790 ($z=0.0585$) is a bright wide-angle tailed radio galaxy found $\sim$19 arcmin north to A754. The bent jets are well resolved in our image, showing the presence of emission knots, before inflating the radio lobes which lead to a total source size of $\sim$150 kpc. As before, the line-of-sight relative velocity between the galaxy and A754 of $\Delta v_{\rm los} \simeq 1260$ km s$^{-1}$ suggests the presence of a significant peculiar motion. \\
\indent
LEDA 25860 ($z=0.0551$) is another bright radio galaxy, placed $\sim$20 arcmin east to A754, showing prominent jets separated by a wide angle which inflate radio lobes. The source has a largest extension of $\sim$320 kpc.

\section{X-ray emission from Abell 754}\label{sec:xray_emission}

The large \xmm\ mosaic of Fig.~\ref{fig:xmm_csmooth} demonstrates the highly disturbed dynamical state of A754. The cluster has a remnant cool core on the west side which follows the bow shock induced by the cluster merger \citep{macario11}. On the opposite side, the X-ray surface brightness possibly shows another rapid surface brightness decrement, which may trace a counter shock. In addition to the east-west features, the X-ray emission shows an elongated ``tongue'' of low surface brigthness emission extending south to a distance of 1.6 Mpc from the cluster center. In order to have a better understanding of the processes undergoing in the thermal gas during the merger, we produced thermodynamic maps of the ICM as described in Section~\ref{sec:thermo_maps}. These maps, shown in Fig.~\ref{fig:tmap}, highlight a number of interesting features, which we have labeled in the left-hand panel and will comment on below. \\
\indent
The cluster is hot and exhibits a complex temperature distribution, as previously indicated by temperature maps obtained with \asca\ \citep{henriksen96a754}, \chandra\ \citep{markevitch03, henriksen04}, and \xmm\ \citep{henry04, lagana10spiral, lagana19maps}. The brightest ICM region is coincident with the remnant cool-core disrupted during the cluster merger which, as expected, exhibits lower temperature and entropy than the surrounding gas. The bow-shaped border on the east of the map delineates the location of the known shock front studied in \citet{macario11}, where temperature, entropy, and pressure increase due to the shock heating. The rise in thermodynamic values compared to the upstream gas is not very evident in our maps because they are truncated at the position of the shock. While the west region of the cluster has been less investigated, past studies pointed out the presence of high temperature values \citep[see \eg][]{henriksen96a754, henriksen04, henry04, inoue16}, which are confirmed by our analysis. In particular, \citet{inoue16} proposed that this portion of the cluster may trace a shock-heated region where the plasma is a nonequilibrium ionization state. Our thermodynamic maps are in line with a scenario where a shock front is propagating toward the northwestern cluster outskirts. We have indicated the position of this second putative shock in our temperature map. The presence of a shock front is further supported by both the X-ray and radio emission, which exhibit a rapid decrement/compression in this region (Figs.~\ref{fig:halo} and \ref{fig:xmm_csmooth}). In addition, the synchrotron radiation here shows a flatter spectrum than the surroundings (Fig.~\ref{fig:spix}), which can also be interpreted in a scenario where particles are (re)accelerated by the passage of a shock front. The southern tongue of low surface brightness emission is characterized by low temperature/entropy/pressure. A similar feature with comparable thermodynamic characteristics, which we have labeled as northern putative tongue, is observed in the opposite direction. \\
\indent
In Section~\ref{sec:merger_scenario} we shall use the features discuss above to propose a merger scenario for A754.

\section{Discussion}\label{sec:discussion}

A754 is a prototypical merging cluster at low-$z$ which hosts nonthermal diffuse radio sources on Mpc-scale. Thanks to the high sensitivity provided by \meerkat, these sources have been recovered with unprecedented detail. In particular, the new \meerkat\ images show that the emission located at the position of the shock front and previously claimed to be a radio relic \citep{kale09, macario11} actually marks the edge of the radio halo. This is not the only case where a shock front is bounding the emission from a halo \citep[see \eg][and references therein]{markevitch10arx, vanweeren19rev}. Recently, other radio edges, not necessary at the border of the radio emission, have been reported in a number of clusters observed with \meerkat\ \citep{botteon23}. \\
\indent
The cluster Abell 520 is probably the most extensively studied case of radio halo edge-shock connection \citep{markevitch05, wang18a520, hoang19a520}. The analysis of such a region and, especially, of the jump in radio emissivity at the front, allows to discriminate the origin of the edge; that is, whether relativistic particles are reaccelerated or adiabatically compressed by the shock. A similar discussion for the shock in A754 was also presented by \citet{macario11}. However, in both cases, determining the physical processes responsible for generating the radio-emitting electrons has remained inconclusive. In our forthcoming work, we will perform a multiband study of the shock front in A754 (Botteon et al., in preparation). \\
\indent
Additionally, the \meerkat\ images have unveiled a new extended radio source $\sim$2.4 Mpc southwest from the cluster center, which we have classified as a radio relic. Relics are the tracers of particles (re)accelerated by shock fronts propagating through the outskirts of clusters during mergers \citep[\eg][]{finoguenov10a3667, bourdin13, botteon16a115, ge19a1367, sarkar24}. The proximity of the radio galaxy LEDA 25672 (Fig.~\ref{fig:relic}) gives a tantalizing indication of ongoing cosmic ray seeding, recalling the well-known case of Abell 3411-3412 \citep{vanweeren17a3411}. Nonetheless, as we will discuss more in detail in Section~\ref{sec:elusive}, the direct acceleration of electrons from the thermal pool is not ruled out in this case. While the distance of the relic in A754 is considerable ($\sim$1.8$\rfive$), it is not unique. Other relics exist at comparable or even greater distances from their respective cluster, including those in Abell 2255 \citep{pizzo08, botteon22a2255}, ClG 0217+70 \citep{brown11multiple, hoang21clg0217}, PLCKESZ G287.0+32.9 \citep{bagchi11, bonafede14reacc}, Ant Cluster \citep{botteon21ant}, Coma Cluster \citep{bonafede22}, and Bullet Cluster \citep{sikhosana23}. Particularly noteworthy is the latter case, which appears as a close analog to A754. Indeed, in both clusters, the peripheral diffuse emission appears in an unusual location relative to the ICM and to the east-west merger axis that one can infer at first glance from the location of the bow shocks found in the two systems. \\
\indent
In simple head-on collisions, pairs of shocks propagating in opposite directions develop, but the scenario becomes more complex if the merger occurs with a nonzero impact parameter \citep[\eg][]{evrard90, roettiger93, ricker01}. Moreover, recent numerical simulations have indicated that even in nearly head-on collisions, equatorial shocks moving perpendicularly to the merger axis can form \citep[\eg][]{ha18shocks, lee20a115, zhang21splashback}. Based on the analysis of \rosat\ and \asca\ data, \citet{henriksen96a754} suggested that A754 is undergoing a slighly off-axis collision. This scenario was later corroborated by tailored numerical simulations \citep{roettiger98}. In the following subsection, we use the results obtained in this paper to refine the merger scenario of A754.

\begin{figure*}
 \centering
 \includegraphics[width=\hsize,trim={0cm 0cm 0cm 0cm},clip]{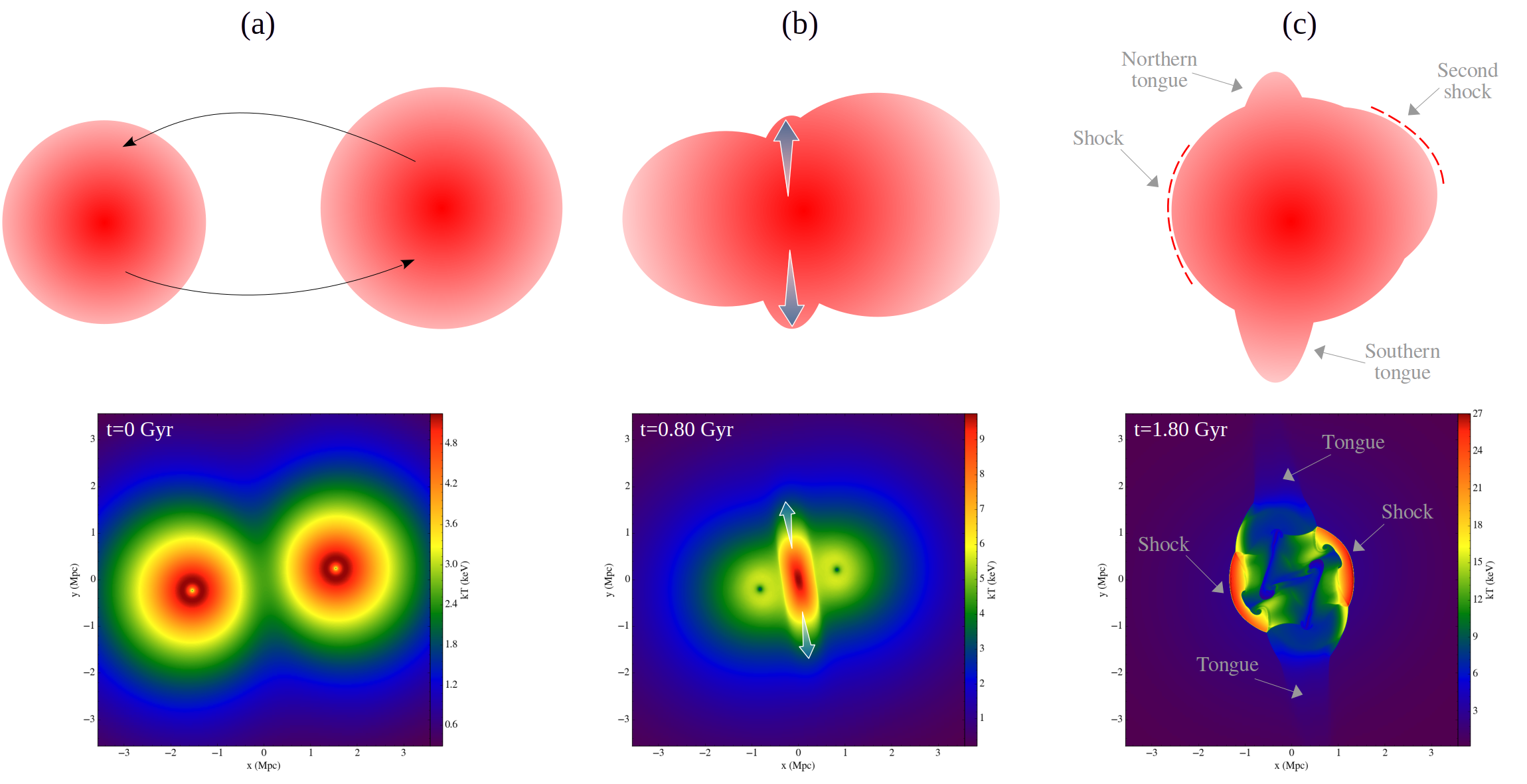}
 \caption{\textit{Top panels}: Schematic representation of the proposed merger scenario for A754. (\textit{a}) Two clusters undergo a slightly off-axis collision nearly in the plane of the sky. (\textit{b}) As clusters approach, the gas in-between is squeezed and propagates in equatorial directions. (\textit{c}) In the current configuration of A754, tongues of cold and low entropy plasma develop and shock fronts propagate in opposite directions roughly along the merger axis. \textit{Bottom panels}: temperature slices (for the $z$ projection) from an idealized simulation by \citet{zuhone18catalog} of a 1:1 merger with an impact parameter of 500 kpc taken at different times and used to support our sketched merger scenario of A754.}
 \label{fig:sketch}
\end{figure*}

\begin{figure*}
 \centering
 \includegraphics[width=\hsize,trim={0.0cm 0.0cm 0.0cm 0.0cm},clip]{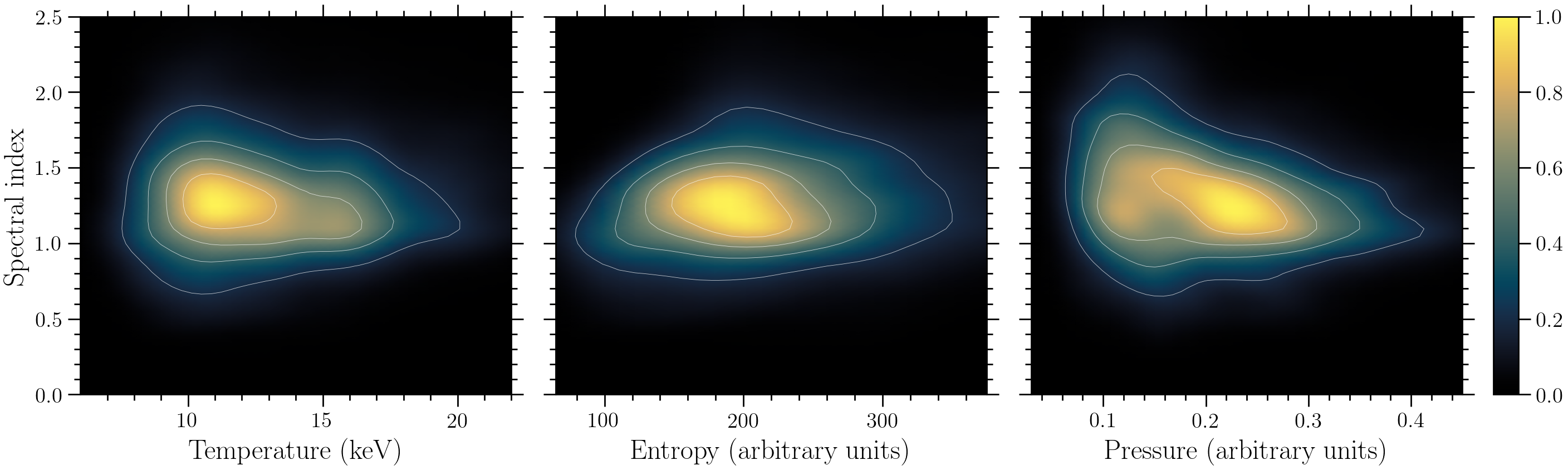}
 \caption{Normalized 2D KDE distributions between spectral index and thermodynamic properties of the ICM, with isocontours drawn at the same levels (0.2, 0.4, 0.6, 0.8).}
 \label{fig:pixpix}
\end{figure*}

\subsection{Merger scenario}\label{sec:merger_scenario}

The presence of the sharp discontinuity (shock/radio halo edge) in the east suggests that at least part of the merger motion is occurring along this direction. Based on the thermodynamic maps shown in Fig.~\ref{fig:tmap}, we suggested the existence of a second shock likely propagating toward the northwest. Non-planar shocks (\ie\ shocks that are not moving in diametrically opposite directions) are distinctive of mergers between clusters with unequal masses colliding with a nonzero impact parameter. In such cases, the shock driven by the less massive cluster curves after the pericenter passage, sweeping through the cluster outskirts \citep{ricker01}. The other interesting features highlighted by our thermodynamic maps are the two tongues of low temperature/entropy/pressure gas developing in the north and south directions, with the latter being more prominent. We interpret these structures as flows originating form the gas in-between the colliding clusters that is being squeezed and expelled toward perpendicular directions, accelerating down the pressure gradient \citep[see \eg][for early reports of these features in numerical simulations]{ricker98, ritchie02}. Also the asymmetry of the tongues can be interpreted as a non head-on collision and/or unequal mass ratio. \\
\indent
In light of these results, we propose that A754 is undergoing a major merger as sketched in the top panels of Fig.~\ref{fig:sketch}, where two clusters collide with a small impact parameter. The schematic representation delineates how the observed features developed during the merger (gray labels). To support our sketched scenario, we inspected the Galaxy Cluster Merger Catalog \citep{zuhone18catalog}. This catalog contains the results from idealized binary cluster mergers exploring a parameter space with different mass ratios and/or impact parameters. In the bottom panels of Fig.~\ref{fig:sketch}, we report temperature slices for snapshots at different epochs for the merger with mass ratio 1:1 and impact parameter of 500 kpc, where structures similar to those observed in A754 can be identified (gray labels). We remark that the aim of this comparison is not to closely replicate the case of A754\footnote{For reference, the values of the mass ratio and impact parameter in the numerical model of A754 by \citet{roettiger98} were 2.5:1 and 120 kpc, respectively.}, but to help understanding the merger scenario and observed features. In this context, the new relic in the southwest may trace a shock driven by the outflowing gas of the southern tongue. For a sequence illustrating the launch of a shock in that direction by these motions, see, for example, figure 12 in \citet{zhang21splashback}. \\
\indent
To conclude, the results from the radio and X-ray analysis of the \meerkat\ and \xmm\ data reported in this paper are in line with a major merger between two galaxy clusters undergoing a slightly off-axis collision. Whilst similar features observed in A754 can be identified also in idealized binary merger simulations, we note that in reality departures from these simulations are naturally expected given the cosmological accretion inflows and mergers with clusters containing substructures.

\subsection{Correlating thermodynamic and spectral index maps}

A connection between thermal and nonthermal components in the ICM is expected in the formation models for radio halos \citep[\eg][]{brunetti14rev}. Such a connection could be reflected as correlations between the local spectral index of the synchroton radiation and the local thermodynamic properties (temperature, entropy, and pressure) of the thermal gas. Following the argument of \citet{botteon20a2255}, in a simplistic scenario where the spectral index reflects the turbulent reacceleration efficiency and the thermal quantities represent the level of heating and mixing in the gas, one may expect regions with flatter spectral index to correlate with regions with higher values of these properties. However, such investigations have so far been conducted only in a handful of clusters due to the limited availability of suitable X-ray and multifrequency radio data required for this kind of analysis. The search for a correlation between spectral index and temperature was performed in the following systems: Bullet Cluster \citep{shimwell14}, A521 \citep{santra24}, A2142 \citep{riseley24}, A2255 \citep{botteon20a2255}, and A2744 \citep{orru07, pearce17, rajpurohit21a2744}. The findings from these studies are inconclusive, showing mild to no correlations between spectral index and temperature. In A2255, there was also an investigation of the relations between spectral index and entropy and pressure, revealing a tendency for flatter spectrum emission to be located in higher entropy regions. \\
\indent
The high quality of the \meerkat\ and \xmm\ observations of A754 allows us to investigate these possible correlations by using an approach which takes advantage of the large angular extent of the cluster and of the method we used to produce the thermodynamic maps of Fig.~\ref{fig:tmap}. Our approach consists of correlating the thermodynamic and spectral index maps on a pixel-to-pixel basis and reporting the results as 2D kernel density estimate (KDE) plots, with data weighted by the inverted quadrature sum of the fractional pixel errors. Clearly, the pixels in our maps are correlated, and this method is not suitable to perform a proper statistical analysis (\eg\ linear regression). Nonetheless, its advantage is that it condenses the spatial information of two quantities in a single plot, allowing to search for common (sub)structures in the distribution, in addition to provide information on the general trend of the plotted quantities. \\
\indent
In order to correlate the different maps, images were first aligned and regridded to the same pixelation. Pixels covering contaminating X-ray point sources (\ie\ the circular regions in Fig.~\ref{fig:tmap}) and the extended radio galaxies that were not subtracted in the \uv-plane (\ie\ LEDA 25746 and LEDA 25701) were excluded from the analysis. The resulting 2D KDE plots, normalized for the maximum value of each distribution, obtained from the values of 44,435 pixels, are shown in Fig.~\ref{fig:pixpix}. While no clear trends can be identified between spectral index and temperature or entropy, evidence for a possible anti-correlation with the pressure is reported. This may indicate that in regions with higher gas pressure, which may be in higher disturbed state, the acceleration of nonthermal electrons is more efficient, leading to generation of radio emission with flatter synchrotron spectrum. However, even if a trend may be naively interpreted, we note that the scenario is more complex as (i) these thermodynamic quantities are not good proxies of the turbulence in the ICM, (ii) the timescales of gas heating/cooling and particle acceleration/cooling are different, and (iii) the degree of plasma collisionality, albeit poorly understood, likely affects the acceleration rate. The presence of these relations needs to be validated by further observations and numerical simulations.

\begin{figure}
 \centering
 \includegraphics[width=\hsize,trim={0.0cm 0.0cm 0.0cm 0.0cm},clip]{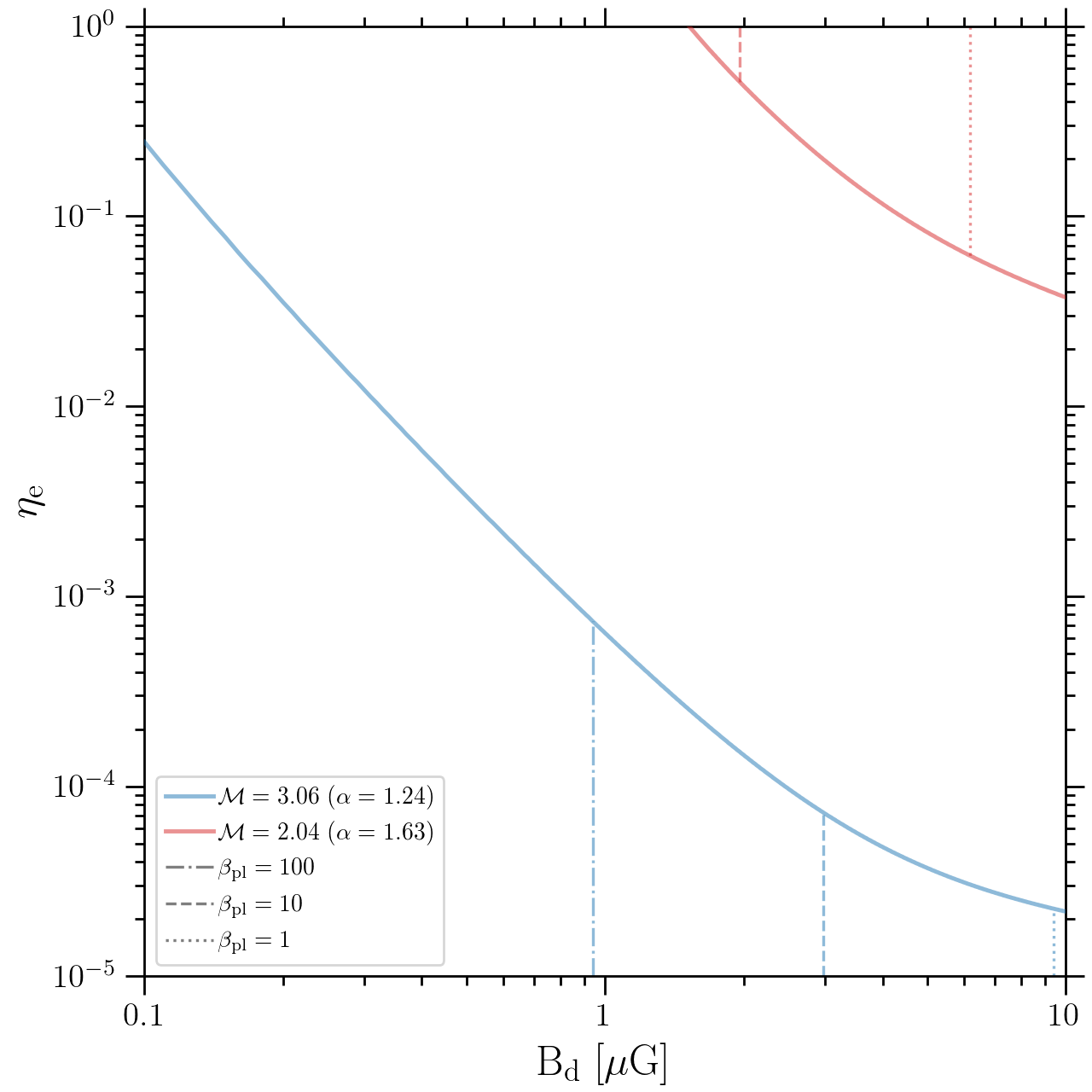}
 \caption{Electron acceleration efficiency that is required to reproduce the radio luminosity of the radio relic in A754 as a function of the downstream magnetic field. Vertical lines denote magnetic field strengths corresponding to different $\beta_{\rm pl}$ values.}
 \label{fig:eff}
\end{figure}

\subsection{A weak radio relic powered by direct acceleration of thermal pool electrons?}\label{sec:elusive}

The radio power of $P_{1.4} = (8.0\pm0.7) \times 10^{22}$ \whz\ makes the relic in A754 one of the least powerful detected at high-frequencies. This adds to the growing number of low-power radio relics being discovered lately, especially due to sensitive observations at low frequencies, where steep spectrum sources appear brighter \citep[\eg][]{locatelli20, bonafede21, botteon21ant, botteon22dr2, duchesne22, schellenberger22, riseley22a3266, jones23, chatterjee24, rajpurohit24}. Whilst the relic--shock connection is supported by a number of radio/X-ray observations, particle acceleration of thermal pool electrons via diffusive shock acceleration (DSA) is known to be a rather inefficient process for weak shocks ($\mach \lesssim 3$) such as those typically found in galaxy clusters \citep{kang02large, kang05}. This, combined with the high radio luminosity observed in radio relics, implies untenable acceleration efficiencies of thermal electrons, suggesting that other mechanisms (\eg\ reacceleration of supra-thermal seed electrons) play a role in most relics \citep{botteon20efficiency}. It is therefore interesting to investigate the so-called acceleration efficiency problem for a low-power radio relic such as that in A754, whose large extent suggests that a significant kinetic energy flux is dissipated through the shock surface supporting the nonthermal luminosity of the relativistic electrons radiating in the radio band. \\
\indent
We use the formalism of \citet{botteon20efficiency} to derive the electron acceleration efficiency ($\eta_{\rm e}$) required to reproduce the luminosity of the radio relic as a function of the (unknown) downstream magnetic field strength ($B_{\rm d}$). The computation requires the parameters of the relic (\ie\ power and spectral index) and shock (\ie\ density, temperature, and surface area). The former are provided by the \meerkat\ observations (Tab.~\ref{tab:halorelic}). The latter cannot be determined with current X-ray observations as the relic is located outside the FoV of the \xmm\ mosaic. Therefore, we assume a shock surface area of $800^2 \pi$ kpc$^2$, where 800~kpc is the semi-axis of the relic emission, and the density and temperature expected from the universal thermodynamic profiles of \citet{ghirardini19universal}, that we consider as unshocked (\ie\ upstream) values. At the distance of the relic in A754 ($\sim$2.4 Mpc $\simeq $ 1.8\rfive), these are $n_{\rm e} \sim 3\times10^{-5}$ cm$^{-3}$ and $kT \sim 3.2$ keV. We use the classic DSA formula to compute the shock Mach number from the integrated synchroton spectral index \citep{blandford87rev} and the Rankine-Hugoniot jump conditions to estimate the downstream density and temperature \citep{landau59}. \\
\indent
Our results reported in Fig.~\ref{fig:eff} indicate that a shock with Mach number $\mach = 3.06$, corresponding to the central value of our spectral index estimate $\alpha = 1.24$, can reproduce the luminosity of the radio relic with electron acceleration efficiencies $\eta_{\rm e} < 10^{-3}$ for magnetic field strengths $B_{\rm d} > 0.8$ \muG. Although the acceleration mechanisms at weak shocks and the magnetic fields in cluster outskirts are still poorly constrained, the values reported here appear reasonable (for $B_{\rm d} \sim 1$ \muG, the corresponding thermal-to-magnetic-pressure ratio is $\beta_{\rm pl} \sim 100$) and significantly more favorable than those obtained by \citet{botteon20efficiency} for a sample of much more powerful radio relics. This suggests that DSA of thermal electrons represents a viable scenario in A754. We note, however, that if we assume the worst-case scenario allowed by our measurements, \ie\ adopting the upper bound of our spectral index measurement ($\alpha =  1.63$) which leads to $\mach = 2.04$ under DSA assumptions, the acceleration efficiency required to reproduce the luminosity of the radio relic is significantly higher: $\eta_{\rm e} < 10^{-1}$ for large magnetic field strengths $B_{\rm d} > 4.4$ \muG, which in turn imply $\beta_{\rm pl} < 2$ (\cf\ Fig.~\ref{fig:eff}). Efficiencies of the order of $10^{-1}$ are generally associated to protons in strong ($\mach \sim 10^{3}$) supernova shocks \citep[\eg][]{morlino12}, and thus appear untenable for electrons in weak cluster shocks. Should the radio relic have such a steep spectrum, DSA of thermal electrons would be disfavored compared to a scenario where seed electrons provided by the nearby LEDA 25672 are reaccelerated at the shock, as anticipated in Section~\ref{sec:discussion}. Therefore, it will be crucial to better constrain the spectral index of the emission and search for the hypothetical $\mach \sim 3$ shock with future follow-up observations. \\
\indent
Overall, observations with the new generation of radio interferometers present the tantalizing possibility of uncovering the population of weak relics anticipated by recent numerical work \citep[\eg][]{bruggen20, zhou22, lee24}. These relics may be powered by the direct acceleration of thermal pool electrons \citep[\eg][]{locatelli20} and could have been missed by previous instruments due to their faint emission.

\section{Conclusions}

In this paper, we have presented new \meerkat\ data and archival \xmm\ observations of the prototypical major cluster merger A754. The main outcomes of our work, that is the first in a series dedicated to this target, are outlined below. \\
\indent
From a technical point of view, we tested a new strategy to calibrate \meerkat\ observations, adopting the \texttt{facetselfcal.py} pipeline originally developed to calibrate \lofar\ data. This method probed to be effective also for \meerkat\ UHF- and L-band observations. Noteworthy features of this scheme include the possibility to improve the calibration toward a specific target of interest through the extraction step, which also enables faster calibration and subsequent reimaging, and to correct for direction-dependent effects across the FoV. This approach allowed us to produce highly sensitive images of A754 at 819 MHz and 1.28 GHz, providing the most detailed view yet of the cluster nonthermal emission. \\ 
\indent
From a scientific point of view, our results can be summarized as follows.

\begin{itemize}
 \item We revised the nature of the diffuse radio emission associated with the well-known shock front detected a few arcminutes east of the cluster core. While this emission was previously classified as a radio relic, our new images reveal that it actually marks the edge of the radio halo, which is bounded by the underlying shock. The integrated spectral index of the halo between 819 MHz and 1.28 GHz is $\alpha = 1.30\pm0.39$, implying a radio power at 1.4 GHz of $(1.4 \pm 0.1) \times 10^{24}$ \whz. This places it within the known $P_{1.4}$--\mfive\ relation for radio halos. The resolved spectral index map shows regions of flatter spectrum emission in the east-west direction and steeper spectrum emission in the north-south direction, likely reflecting different efficiencies in particle acceleration along and perpendicular to the main cluster merger axis.
 
 \item We discovered a new, elongated, peripheral emission at $\sim$2.4 Mpc $\simeq $1.8\rfive\ from the cluster center, which we classified as a radio relic.  This emission is detected at both 819 MHz and 1.28 GHz, with an integrated spectral index between these frequencies of $\alpha = 1.24 \pm 0.39$. The radio power at 1.4 GHz of $(8.0 \pm 0.7) \times 10^{22}$ \whz\ makes it one of the least powerful radio relics observed to date. We found that DSA of thermal pool electrons represents a viable scenario to explain its origin, unlike what is generally found for more powerful radio relics. Still,  we noted the presence of a radio galaxy (LEDA 25672) likely connected to the relic, which may naturally provide supra-thermal electrons, further reducing the electron acceleration efficiency required to explain the relic radio luminosity. 
 
 \item We produced a large \xmm\ mosaic that allowed us to study the properties of the thermal gas. In particular, we found that the X-ray emission from the ICM shows a prominent elongation toward south, which we dubbed ``tongue''. Thermodynamic maps suggest that this tongue is constituted of gas with low temperature/entropy/pressure, and that it has a possible counterpart in the northern direction. Signs of shock-heated gas are found in the western region of the cluster, in line with previous X-ray analyses and with the flatter spectrum emission of the radio halo observed in this region. 
 
 \item We interpreted the observed radio and X-ray features in the context of a major cluster-cluster collision with a small impact parameter. Aided by a catalog of idealized simulations of cluster mergers, we propose that in the current phase the gas in-between the colliding clusters is squeezed and forced to outflow in perpendicular directions creating the southern tongue and its putative northern counterpart. We speculate that the newly discovered relic may trace a shock driven by this perpendicularly outflowing gas. 

 \item We investigated possible correlations between spectral index of the radio emission and thermodynamic properties of the X-ray emitting gas. A possible anti-correlation between spectral index and pressure was noted. This may suggest a scenario where higher pressure regions trace regions with higher acceleration efficiencies, leading to synchrotron emission with flatter spectrum. Further investigation is needed to confirm these relations (or lack thereof) and understand their implications for the interplay between thermal and nonthermal components in the ICM.
\end{itemize}

New generation radio interferometers combined with novel calibration techniques are allowing us to obtain radio images of cluster extended emission with unparalleled quality. In combination with complementary X-ray data, these are crucial for unraveling the physics of galaxy cluster mergers. A754, as the prototype of a major cluster-cluster collision, motivates additional work exploiting the new \meerkat\ data.

\begin{acknowledgements}
We thank the anonymous referee for the comments on the manuscript.
We thank Andr\'{e} Offringa for responsive support and bug fixes with \wsclean\ and Frits Sweijen for creating the composite image displayed in Fig.~\ref{fig:composite}.
A.B. acknowledges financial support from the European Union - Next Generation EU.
R.J.vW. acknowledges support from the ERC Starting Grant ClusterWeb 804208.
R.K. acknowledges the support of the Department of Atomic Energy, Government of India, under project no. 12-R\&D-TFR-5.02-0700 and the SERB Women Excellence Award WEA/2021/000008.
Basic research in radio astronomy at the Naval Research Laboratory is supported by 6.1 Base funding.
The \meerkat\ telescope is operated by the South African Radio Astronomy Observatory, which is a facility of the National Research Foundation, an agency of the Department of Science and Innovation.
The scientific results reported in this article are based in part on observations obtained with \xmm, an ESA science mission with instruments and contributions directly funded by ESA Member States and NASA.
This work made use of data from the Galaxy Cluster Merger Catalog (\url{http://gcmc.hub.yt}).
This research made use of the following \textsc{python} packages: \texttt{APLpy} \citep{robitaille12}, \texttt{astropy} \citep{astropy22}, \texttt{CMasher} \citep{vandervelden20}, \texttt{iminuit} \citep{dembinski24}, \texttt{matplotlib} \citep{hunter07}, \texttt{numpy} \citep{vanderwalt11}, and \texttt{scipy} \citep{virtanen20}.
\end{acknowledgements}


\bibliographystyle{aa}
\bibliography{library.bib}

\begin{appendix}

\FloatBarrier
\section{Ancillary images}\label{sec:appendix}

\begin{figure}[h!]
 \centering
 \includegraphics[width=\hsize,trim={0cm 0cm 2.3cm 0.7cm},clip]{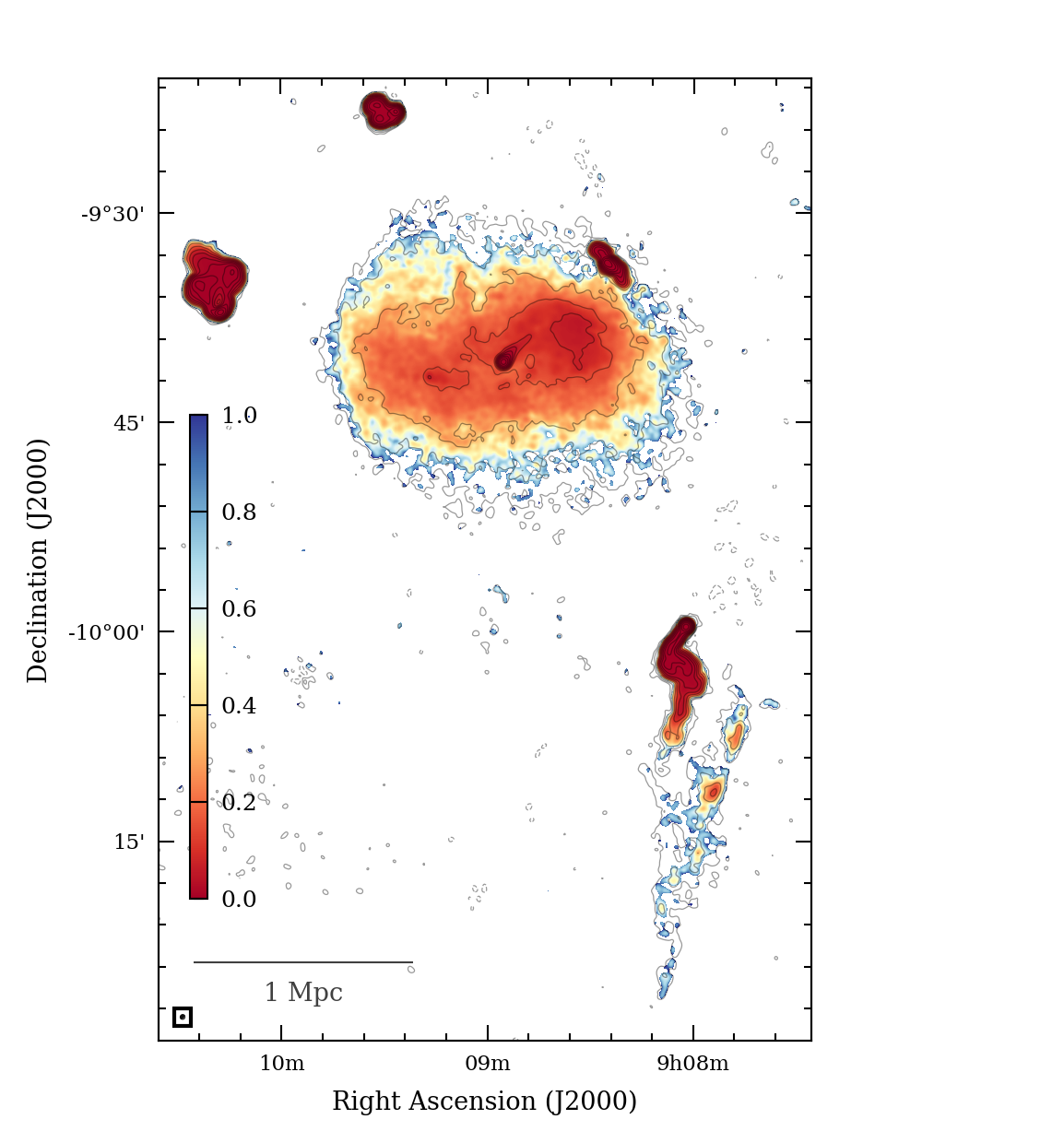}
 \caption{Spectral index error map of Fig.~\ref{fig:spix}.}
 \label{fig:spix_err}
\end{figure}

\begin{figure}
 \centering
 \includegraphics[width=\hsize,trim={0.2cm 0.2cm 0.3cm 0.2cm},clip]{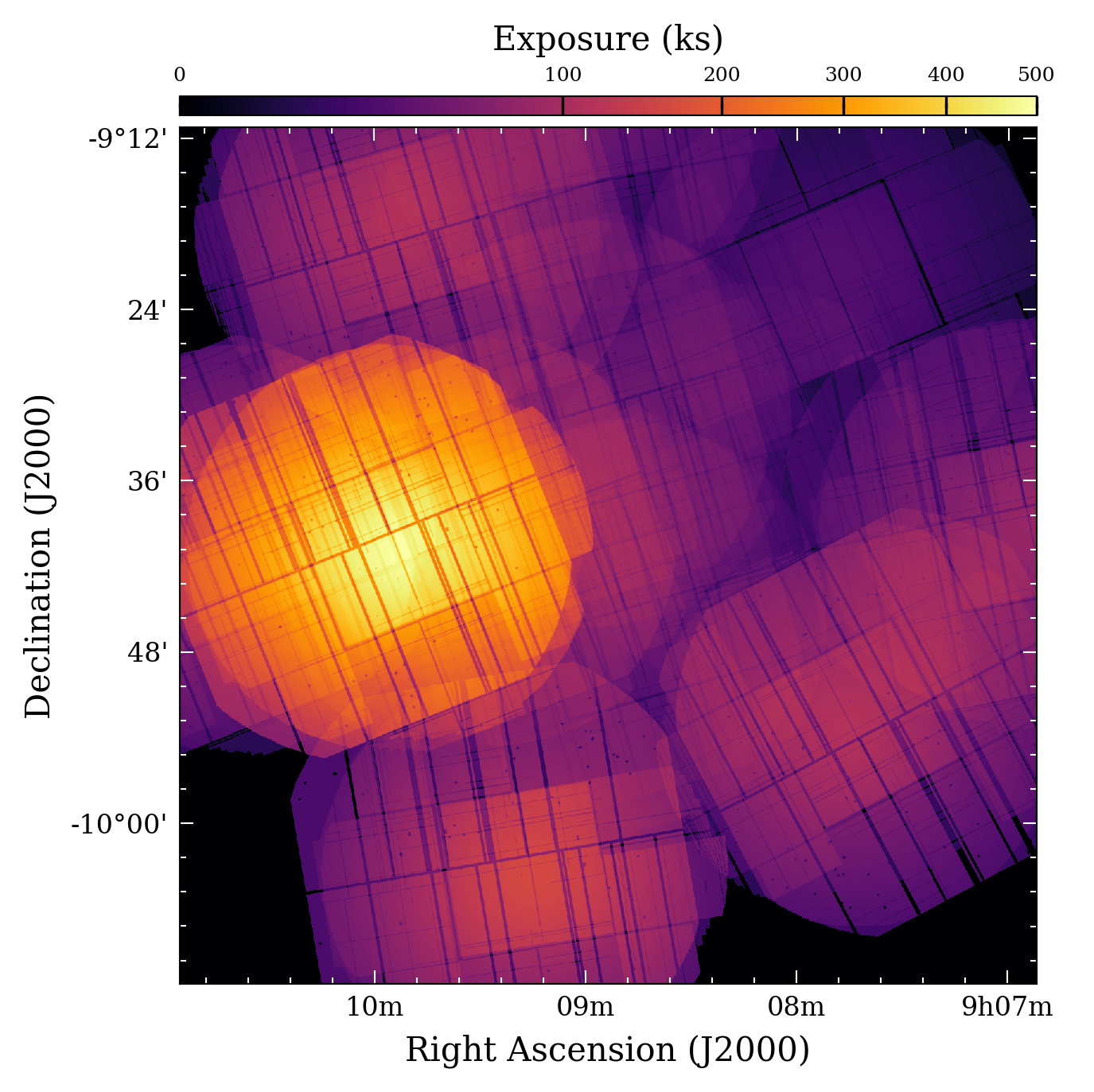}
 \caption{Exposure map of the \xmm\ mosaic of A754. It shows the \epic\ net exposure in MOS units.}
 \label{fig:xmm_exp}
\end{figure}

\begin{figure}
 \centering
 \includegraphics[width=\hsize,trim={0.3cm 0.2cm 0.3cm 0.2cm},clip]{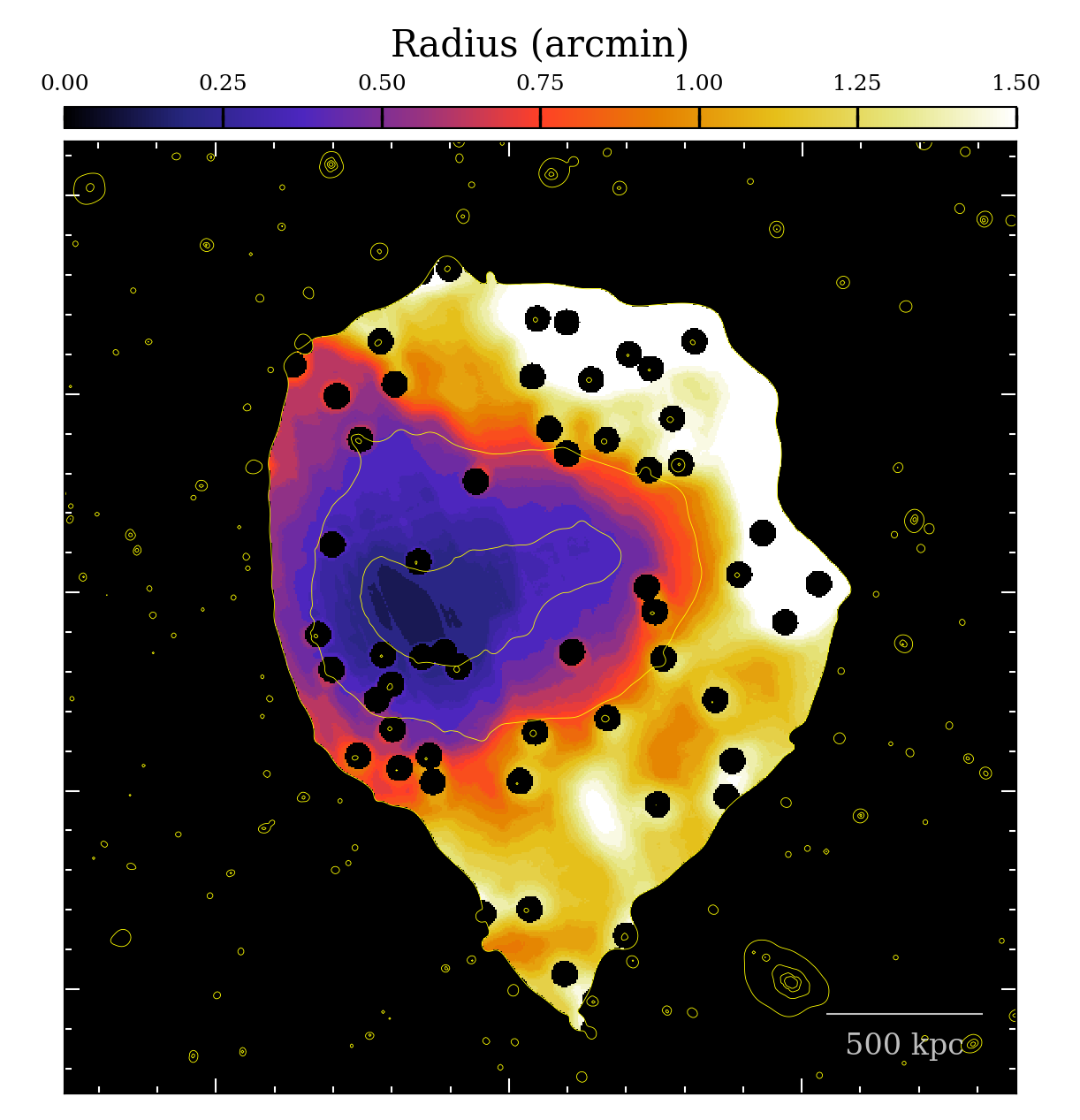}
 \caption{Map showing the radii of the adaptively binned circular regions.}
 \label{fig:tmap_radius}
\end{figure}

\begin{figure*}
 \centering
 \includegraphics[width=.33\hsize,trim={0.3cm 0.2cm 0.3cm 0.2cm},clip]{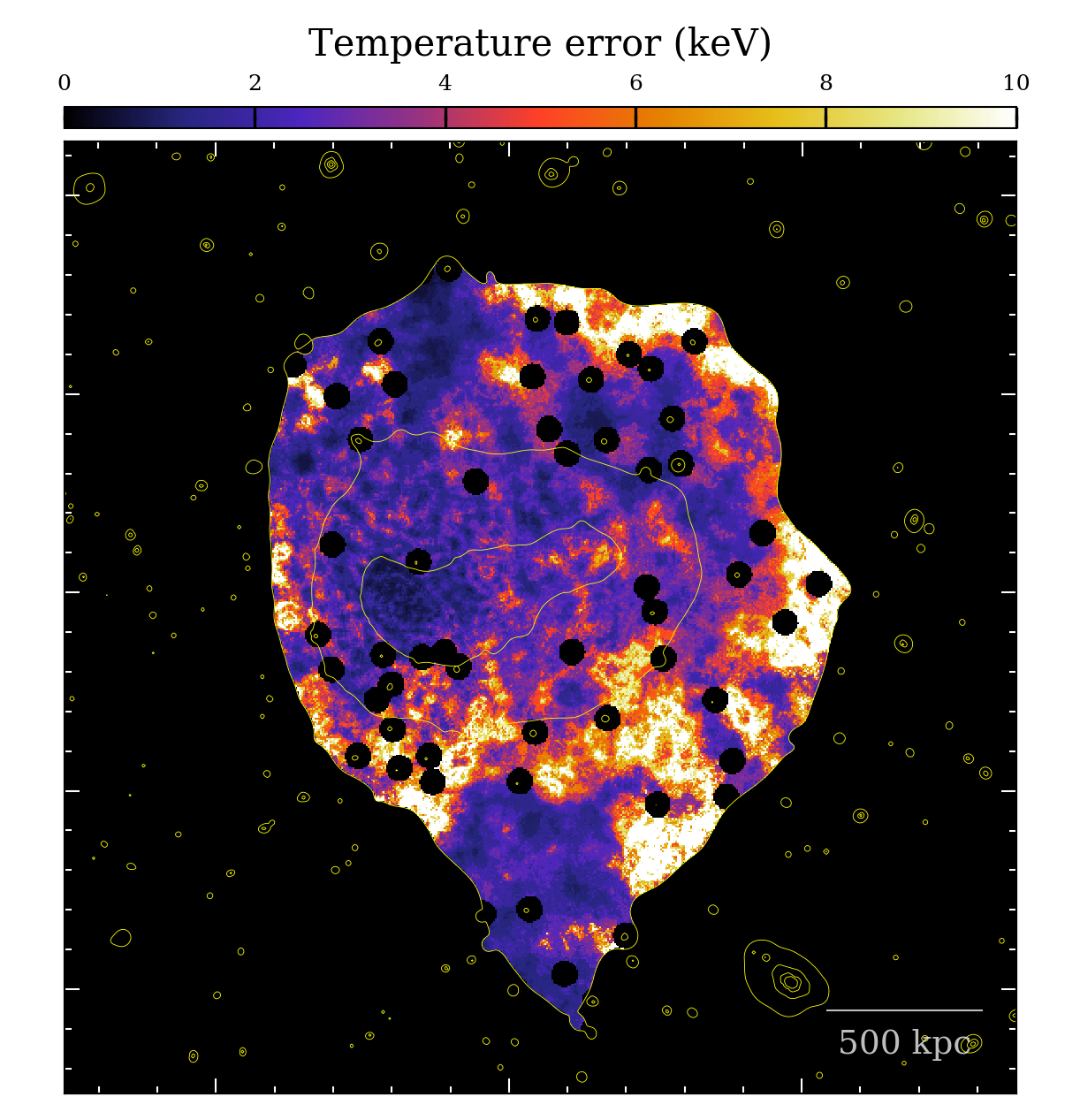}
 \includegraphics[width=.33\hsize,trim={0.3cm 0.2cm 0.3cm 0.2cm},clip]{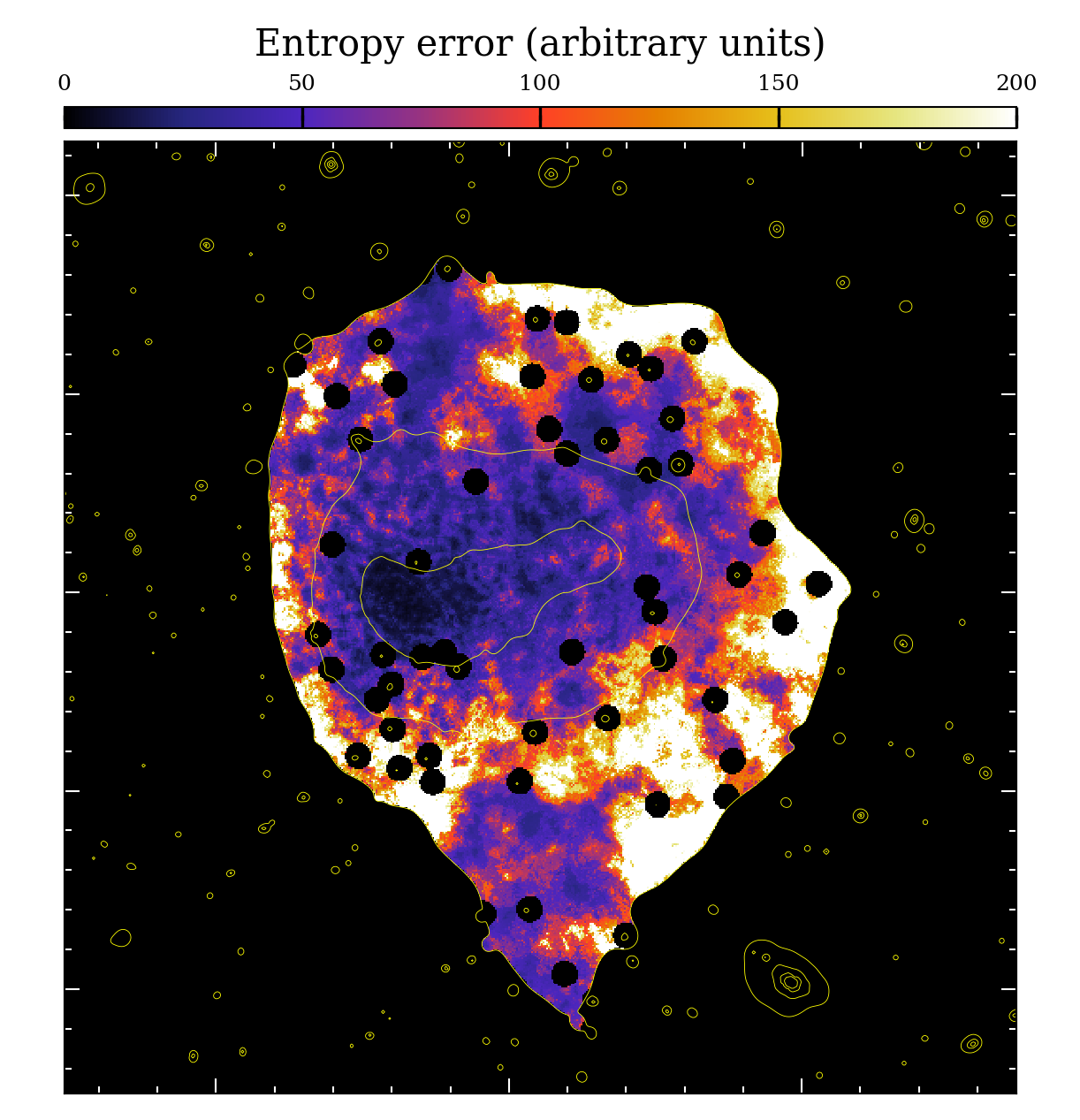}
 \includegraphics[width=.33\hsize,trim={0.3cm 0.2cm 0.3cm 0.2cm},clip]{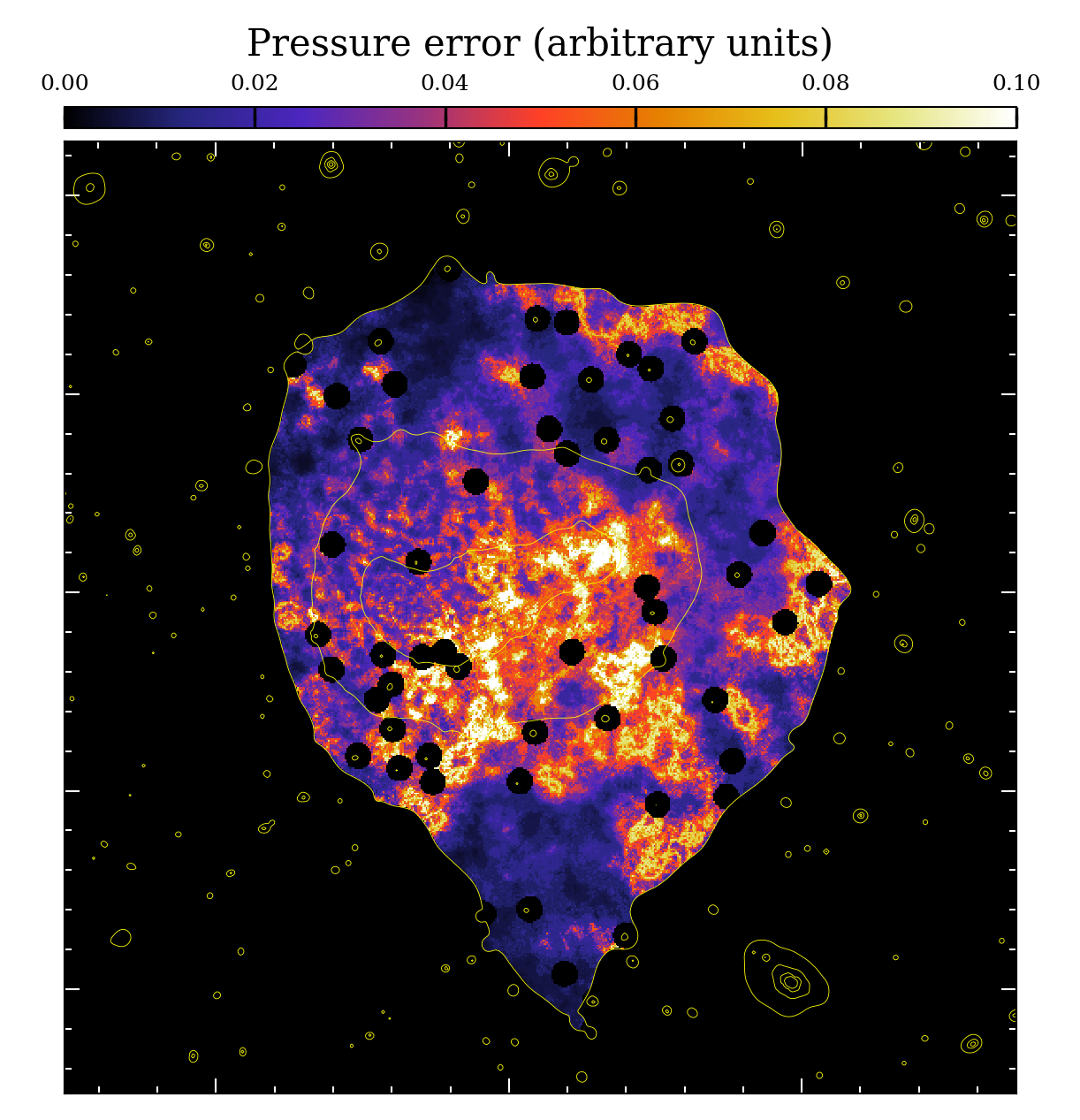}
 \caption{Thermodynamic error maps of Fig.~\ref{fig:tmap}.}
 \label{fig:tmap_err}
\end{figure*}

\end{appendix}

\end{document}